\documentclass[]{jfm}

\makeatletter
\def\absfooterflag{}
\def\pagelimitfooter{}
\makeatother

\usepackage{graphicx}
\usepackage{newtxtext}
\usepackage{newtxmath}
\usepackage{natbib}
\usepackage{hyperref}
\hypersetup{
    colorlinks = true,
    urlcolor   = blue,
    citecolor  = blue,
}

\newcommand{\RomanNumeralCaps}[1]
\linenumbers

\usepackage{color}

\usepackage{bm}
\usepackage[per-mode=symbol]{siunitx}
\DeclareSIUnit\rpm{rpm}

\newcommand{\strain}{\bm{S}}

\newcommand{\strainhat}{\hat{\bm{S}}}
\newcommand{\straincoh}{\widetilde{\bm{S}}}

\newcommand{\qfull}{\mathbf{q}}
\newcommand{\ufull}{\mathbf{u}}
\newcommand{\ai}{\mathbf{a}_i}

\newcommand{\qmean}{\overline{\mathbf{q}}}

\newcommand{\kmean}{\overline{k}}
\newcommand{\epsmean}{\overline{\varepsilon}}
\newcommand{\nutmean}{\overline{\nu}_t}

\newcommand{\qcoh}{\widetilde{\mathbf{q}}}
\newcommand{\ucoh}{\widetilde{\mathbf{u}}}
\newcommand{\pcoh}{\widetilde{p}}
\newcommand{\kcoh}{\widetilde{k}}
\newcommand{\epscoh}{\widetilde{\varepsilon}}
\newcommand{\nutcoh}{\widetilde{\nu}_t}

\newcommand{\qturb}{\mathbf{q}''}
\newcommand{\uturb}{\mathbf{u}''}

\newcommand{\qphase}{\langle \mathbf{q} \rangle}

\newcommand{\kphase}{\langle k \rangle}
\newcommand{\epsphase}{\langle \varepsilon \rangle}
\newcommand{\nutphase}{\langle \nu_t \rangle}

\newcommand{\qhat}{\hat{\mathbf{q}}}
\newcommand{\uhat}{\hat{\mathbf{u}}}

\newcommand{\khat}{\hat{k}}
\newcommand{\epshat}{\hat{\varepsilon}}

\newcommand{\nut}{\nu_t}
\newcommand{\nuthat}{\hat{\nu}_t}
\newcommand{\eps}{\varepsilon}

\newcommand{\imag}{\mathrm{i}}

\usepackage{tikz}
\usetikzlibrary{calc,arrows.meta,arrows,3d}
\definecolor{sketch_gray}{RGB}{225,225,225}
\definecolor{paraview_red}{RGB}{158,43,43}
\definecolor{paraview_red_light}{RGB}{213,101,101}
\definecolor{guidevanes_red}{RGB}{208,73,71}
\definecolor{runner_blue}{RGB}{14,110,232}
\definecolor{oj}{RGB}{221,102,29}
\definecolor{gn}{RGB}{22,174,51}

\usepackage{floatrow}
\usepackage[labelformat=simple]{subfig}
\makeatletter
\renewcommand{\p@subfigure}{} 
\makeatother

\title{Impact of perturbed eddy-viscosity modeling on stability and shape sensitivity of the hydro-turbine vortex rope using linearized Reynolds-averaged Navier--Stokes equations}

\author{J. S. Müller\aff{1}
  \corresp{\email{jens.mueller@tu-berlin.de}},
  S. J. Knechtel\aff{1}
 \and K. Oberleithner\aff{1}}

\affiliation{\aff{1}Laboratory for Flow Instabilities and Dynamics, Technische Universität Berlin, Müller-Breslau-Straße 8, 10623 Berlin, Germany}

\begin{document}
\maketitle

\begin{abstract}

This study investigates the influence of a perturbed eddy-viscosity model on linear stability and shape sensitivity of the global vortex rope mode arising in a hydro-turbine flow under fully turbulent conditions. The framework is based on the Reynolds-averaged Navier--Stokes equations with a standard $k$--$\eps$ turbulence closure, linearized around a base-flow state. This base state is tuned to match the vortex-rope bifurcation predicted from three-dimensional unsteady simulations. The shape sensitivity of the global mode is derived, accounting for perturbations of both the base flow and the linear operator. We show that although the perturbed eddy-viscosity model has only a marginal effect on the eigenvalues and eigenmodes of interest, it substantially alters the resulting shape sensitivities. These differences arise primarily through the base-flow contribution to the total sensitivity, which dominates the sensitivity to shape deformations. Although both models identify coherent-velocity production and advection as the leading contributors, the linearized model captures additional mechanisms associated with eddy-viscosity perturbations. Comparison with experiments shows that only the perturbed eddy-viscosity model reproduces the correct trends in shape sensitivity, whereas the frozen model fails to do so. These findings highlight the importance of consistently linearizing turbulence models for sensitivity-based control of turbulent global instabilities.

\end{abstract}

\begin{keywords}
Authors should not enter keywords on the manuscript, as these must be chosen by the author during the online submission process \ldots
\end{keywords}

\section{Introduction} \label{sec:intro} 
The mitigation and control of coherent structures in fluid flows driven by global instabilities is an ongoing and important topic in current research \citep{Sipp2010,Camarri2015,Taira2020}. Such instabilities are often responsible for increased drag, increased noise, or severe structural vibrations. Thus, their understanding and eventual control are of high practical relevance. A physics-based framework for addressing these phenomena is provided by linear stability analysis (LSA), which describes the dynamics of coherent structures driven by linear modal mechanisms. Furthermore, adjoint-based sensitivity analysis allows for the efficient computation of gradients of eigenvalues with respect to control inputs, such as base-flow modifications or periodic forcing. These gradients reveal where an active or passive flow control has the largest impact on the instability dynamics in terms of frequency and growth rate. In practice, this framework can be used either qualitatively, e.g., to guide optimal actuator placement, or quantitatively, as the basis for gradient-based optimization of shape design.

In the remainder of the introduction, we will review previous studies with respect to (1) LSA-based sensitivity analyses for flow control, (2) LSA-based shape sensitivity frameworks, (3) LSA application in turbulent flows, and (4) flow control in hydro-turbine flows, before we define (5) the research objective of this work.

\subsection{Linear stability and sensitivity analysis for flow control}
LSA and sensitivity analysis have been widely applied to a range of technical configurations under both laminar and turbulent conditions. Numerous works have used sensitivity analysis in a qualitative fashion to identify regions of intrinsic feedback, often referred to as the `wavemaker' \citep[e.g.][]{Giannetti2007,Qadri2013}, or to detect regions where periodic forcing is most effective for active control \citep[e.g.][]{Tammisola2016,Kaiser2018,Mueller2020}. Sensitivity analysis has also been used quantitatively to predict where and how a small control device would change the frequency and growth rate of a global mode \citep[e.g.][]{Marquet2008,Meliga2012b}. Other quantitative applications that exploit gradient information, such as shape optimization to stabilize the flow, have been used mainly in laminar conditions, e.g., in incompressible airfoil flows to delay laminar-turbulent transition \citep{Amoignon2006}; in hydrofoil flows \citep{Mueller2024} and in cylinder flows \citep{Heuveline2009,Tammisola2017,Kiriyama2018,Boujo2019,Brewster2020} to suppress global wake instability; in sudden expansion flows \citep{Nakazawa2016} and contraction flows \citep{Wang2019} to suppress asymmetric global modes; and in microfluidic channel flows to enhance decay rates of perturbations \citep{Kungurtsev2019}. They have rarely been used in turbulent conditions, with exceptions such as cylinder flows to suppress the wake mode \citep{Brewster2019} or transonic airfoil flows to suppress the buffet instability \citep{Martinez-Cava2020}.

\subsection{Shape sensitivity derivation and application}
There are several methods for deriving shape sensitivities. In the continuous framework (differentiate-then-discretize), two main formulations can be distinguished. The first is the classical shape calculus approach, in which the so-called shape derivative is derived analytically in an infinite-dimensional space using the Hadamard form \citep{Grinfeld2010,Delfour2011}. This yields a boundary integral expression of the variation of a given cost functional, such as the eigenvalue, that quantifies the local influence of infinitesimal shape deformations and can be subsequently evaluated numerically after discretization. This approach is typically formulated within an optimal control framework via Lagrangian functionals that are used to define the cost function and to derive the equations of the dual or adjoint variable. Its application to the Navier--Stokes equations in general was pioneered in \cite{Pironneau1974} and has found its way into an LSA-specific context in various studies \citep{Nakazawa2016,Kiriyama2018,Kungurtsev2019,Brewster2020}. A second continuous approach is based on perturbation theory \citep{Hinch1991,Luchini2014}. Here, infinitesimal boundary deformations are introduced and the governing equations are expanded in a perturbation series, allowing first- or arbitrary-order \citep{Knechtel2024} variations of the cost function to be expressed analytically in terms of the unperturbed state and adjoint variables. This formulation has been successfully adapted for LSA-based shape sensitivity and optimization problems to evaluate second-order shape sensitivities \citep{Tammisola2017,Boujo2019}. In contrast, the discrete approach (discretize-then-differentiate) formulates the governing equations directly in matrix form, such that discrete shape gradients are obtained as sensitivities with respect to mesh node displacements through Lagrangian functions and the corresponding Jacobian or Hessian matrices. This framework inherently embeds boundary conditions in the discrete operator, and thus circumvents the explicit evaluation of boundary integrals. Discrete adjoint-based formulations of shape sensitivity in LSA have been successfully used in \cite{Browne2014,Wang2019,Martinez-Cava2020,Mueller2024}.

\subsection{Various eddy-viscosity approaches for application in turbulent flows}
When LSA and shape sensitivity are applied in turbulent flows, additional challenges emerge. Linearization about both the turbulent base flow and the time-averaged mean flow requires a closure for the fluctuating component of the turbulent Reynolds stresses due to the passage of a coherent structure, which is essentially the turbulence modeling problem in the context of LSA \citep{Reynolds1972,Reau2002}. The most prevalent method is the linear Boussinesq eddy viscosity model, where two main approaches exist: (i) the frozen eddy-viscosity model, in which only the mean eddy viscosity is retained and eddy-viscosity fluctuations are neglected, and (ii) the perturbed eddy-viscosity model, which explicitly accounts for perturbations of turbulence model quantities. The frozen model is frequently employed, often without a strong justification or quantitative validation, even though it may neglect relevant mechanisms. This is mainly due to the fact that the turbulent mean-flow fields are often obtained using an approach that is not based on Reynolds-averaged Navier--Stokes (RANS) equations, such as direct numerical simulations, large-eddy simulations, or experimental measurements. Thus, turbulence model equations are not available \textit{a priori}, making it difficult to infer the linearized behavior of the eddy viscosity. Instead, the (frozen) eddy viscosity is computed \textit{a posteriori} from the data. The determination of the eddy viscosity can be done algebraically using a direct or least-squares fit approach when the Reynolds stress and shear rate tensor are known \citep[e.g.][]{Reau2002,Tammisola2016,Rukes2016,Mueller2024b}; by calibrating an algebraic mixing length model \citep[e.g.][]{Viola2014,Fan2024}; by using semi-empirical functions \citep[e.g.][]{Morra2019,Symon2023}; by energy budget considerations \citep[e.g.][]{Kuhn2022,VonSaldern2024}; or by optimization and data assimilation \citep[e.g.][]{Pickering2021,Mons2024,VonSaldern2024}. When mean fields are computed with a RANS approach or data-assimilated to a set of RANS and turbulence model equations, the inclusion of a linearized turbulence model and, thus, fluctuating eddy viscosity is more straightforward \citep[e.g.][]{Crouch2007,Meliga2012b,Brewster2019,Martinez-Cava2020,Sarras2024}. There is no conclusive consensus on whether a perturbed eddy-viscosity model is required or not, since the relevance of the perturbed eddy-viscosity model depends on the specific application and flow conditions. For example, a perturbed eddy-viscosity model has been found to be destabilizing in a transonic airfoil flow \citep{Crouch2007}; stabilizing and with significant changes to the frequency in stalled airfoil flows \citep{Mons2024,Sarras2024}; stabilizing but with virtually no effect on frequency and mode shape in a deep cavity flow \citep{Mettot2014}; and to have almost no influence on sensitivity to steady and unsteady forcing in a cylinder wake flow \citep{Mettot2014a}. In this work, we continue the discussion on the impact of perturbed eddy-viscosity modeling on LSA and shape sensitivity specifically, and we aim to discuss the differences compared to a frozen eddy-viscosity model.

\subsection{Flow control in hydro-turbine flows}
For the eddy-viscosity modeling discussion, we apply the framework to a fully turbulent hydropower application. Specifically, we investigate the shape sensitivities of the dominant global mode in a Francis turbine draft tube operated at part load. This configuration is particularly relevant, as modern hydropower plants are increasingly forced to operate away from their best efficiency point in order to balance fluctuations in the power grid caused by renewable energy sources. At part load, the so-called vortex rope develops: a nonlinearly saturated global instability at limit cycle, which induces large-scale pressure pulsations and poses a significant hazard to safe operation \citep{Valentin2017,Goyal2018a}. The vortex rope is also known as the helical or spiral vortex breakdown, or the precessing vortex core \citep{LuccaNegro2001}. Many strategies for vortex-rope control in draft tubes have been investigated \citep{Kumar2021}, including active flow control measures \citep{Bosioc2012,Muntean2014,Javadi2017,Holmstrom2021,Shiraghaee2024}, passive control devices \citep{Kurokawa2010,Tanasa2019,Susan-Resiga2021,Holmstrom2022,Urban2022}, and shape modifications \citep{Li2021,Lueckoff2022,Khullar2022,Zhou2024}. However, only a few studies have considered the problem through the lens of LSA \citep{Pasche2017,Mueller2022b,Mitrut2022,Seifi2023} or sensitivity-based optimization \citep{Pasche2019}, and all of these works rely on a frozen eddy-viscosity model.

\subsection{Outline}
The objective of the present paper is to derive, validate and discuss shape sensitivities of the vortex rope instability in a fully turbulent flow, with a particular focus on the role of turbulence modeling. We juxtapose the frozen and the perturbed eddy-viscosity models in order to identify and discuss the physical mechanisms that are neglected in the frozen formulation and those that are added through the non-frozen model. To this end, we evaluate their influence on eigenvalues, eigenmodes, and shape sensitivities. The following research questions guide our study: (i) What is the influence of the perturbed eddy-viscosity model on the eigenvalues, eigenmodes, and sensitivities? (ii) Which mechanisms are omitted in the frozen eddy-viscosity model and which are captured by the perturbed eddy-viscosity model? (iii) How valid are the sensitivities obtained from RANS-based LSA compared to empirically determined values from experimental data?

By addressing these questions, the present work contributes to the systematic development of sensitivity-based design and control in turbulent flows. The Francis turbine vortex rope serves as a representative and practically relevant case study, while the methodological insights are expected to be of more general relevance for the application of LSA and sensitivity analysis in turbulent flows.

\section{Linear stability and shape sensitivity theory} \label{sec:theory}
The fundamental starting point for all our physics-based analyses is the incompressible Navier--Stokes momentum and continuity equations, reading
\begin{equation}
    \breve{\mathcal{B}} \frac{\partial \qfull}{\partial t} + \mathcal{N}(\qfull) = \mathbf{0} ,
    \label{eq:NSE_conti}
\end{equation}
with
\begin{equation}
    \breve{\mathcal{B}} =
    \begin{bmatrix}
        \mathcal{I} & 0\\
        \mathbf{0} & 0
    \end{bmatrix} ,
\end{equation}
\begin{equation}
    \mathcal{N}(\qfull) =
    \begin{bmatrix}
        \mathcal{N}_\ufull(\qfull)\\
        \mathcal{N}_p(\qfull)
    \end{bmatrix} ,
\end{equation}
and the components
\begin{subequations}
    \begin{equation}
        \mathcal{N}_\ufull(\qfull) = \left( \ufull \cdot \nabla \right) \ufull + \frac{1}{\rho} \nabla p - \nabla \cdot \left( 2\nu \strain \right) ,
    \end{equation}
    \begin{equation}
        \mathcal{N}_p(\qfull) = \nabla \cdot \ufull ,
    \end{equation}
\end{subequations}
where $\mathcal{N}$ is the nonlinear Navier--Stokes operator, $\breve{\mathcal{B}}$ is a restriction operator, $\mathcal{I}$ is the identity operator, $\ufull$ is the velocity vector, $p$ is the pressure, $\rho$ is the density, $\nu$ is the molecular kinematic viscosity, and $\strain(\ufull) = 1/2 \left( \nabla \ufull + \nabla \ufull^\mathrm{T} \right)$ is the strain-rate tensor.

The basis for modeling coherent structures is the triple decomposition \citep{Reynolds1972},
\begin{equation}
    \qfull = \qmean + \qcoh + \qturb ,
    \label{eq:triple_decomposition}
\end{equation}
which decomposes any general flow quantity $\qfull$ into a time-averaged mean part $\qmean$, a coherent fluctuation part $\qcoh$ and an incoherent turbulent-stochastic part $\qturb$. In the following, we derive the unsteady RANS (URANS) and RANS equations and stick closely to the derivations outlined in previous works such as \cite{Reynolds1972,Meliga2016,Sarras2024}.

\subsection{Unsteady Reynolds-averaged Navier--Stokes equations and turbulent base-flow states}
To obtain the URANS equations, we substitute the triple decomposition~\eqref{eq:triple_decomposition} into the incompressible Navier--Stokes equations~\eqref{eq:NSE_conti}, and perform an ensemble-averaging operation with the properties $\qphase = \qmean + \qcoh$ and $\langle \qturb \rangle = 0$, which leads to the URANS equations,
\begin{equation}
    \breve{\mathcal{B}} \frac{\partial \qphase}{\partial t} + \mathcal{N}(\qphase) =
    \begin{bmatrix}
        -\nabla \cdot \langle \uturb \otimes \uturb \rangle\\
        0
    \end{bmatrix},
    \label{eq:URANS_conti_pavg}
\end{equation}
with $\qfull = [\ufull, p]^\mathrm{T}$.
The right-hand side term is the ensemble-averaged turbulent Reynolds stress term, representing the nonlinear interaction between the ensemble-averaged field and the turbulent field. To solve this set of equations, this term needs to be modeled, which is done here with the classical Boussinesq eddy viscosity assumption, which models the anisotropic part of the ensemble-averaged Reynolds stress tensor, assuming that the tensor is aligned with the strain-rate tensor so that
\begin{equation}
    \langle \uturb \otimes \uturb \rangle = - 2\nutphase \langle \strain \rangle + \frac{2}{3} \kphase \mathcal{I} ,
    \label{eq:boussinesq}
\end{equation}
where $\nutphase = \nutmean + \nutcoh$ is the ensemble-averaged eddy viscosity and $\kphase = 1/2 \langle \uturb\cdot\uturb \rangle = 1/2 \overline{\uturb\cdot\uturb} + 1/2 \widetilde{\uturb \cdot \uturb} = \overline{k} + \kcoh$ is the ensemble-averaged turbulent kinetic energy. We use the Boussinesq approach~\eqref{eq:boussinesq} for the URANS equations~\eqref{eq:URANS_conti_pavg} and complement with a $k$--$\eps$ turbulence model in its standard form \citep{Launder1974} to compute the unknown eddy viscosity $\nutphase$ and turbulent kinetic energy $\kphase$. We redefine $\qfull = [\ufull, p, k, \eps]^\mathrm{T}$ and obtain
\begin{equation}
    \mathcal{B} \frac{\partial \qphase}{\partial t} + \mathcal{R}(\qphase) = \mathbf{0} ,
    \label{eq:URANS_conti_pavg_boussinesq}
\end{equation}
with
\begin{equation}
    \mathcal{B} =
    \begin{bmatrix}
        \mathcal{I} & 0 & 0 & 0\\
        \mathbf{0} & 0 & 0 & 0\\
        \mathbf{0} & 0 & 1 & 0\\
        \mathbf{0} & 0 & 0 & 1
    \end{bmatrix} ,
\end{equation}
\begin{equation}
    \mathcal{R}(\qphase) =
    \begin{bmatrix}
        \mathcal{R}_\ufull(\qphase)\\
        \mathcal{R}_p(\qphase)\\
        \mathcal{R}_k(\qphase)\\
        \mathcal{R}_\eps(\qphase)
    \end{bmatrix} ,
\end{equation}
where $\mathcal{R}$ is the nonlinear RANS operator with the components
\begin{subequations}
    \begin{align}
        &\mathcal{R}_\ufull(\qfull) = \left( \ufull \cdot \nabla \right) \ufull + \frac{1}{\rho} \nabla p - \nabla \cdot \left( 2(\nu + \nut) \strain \right) + \frac{2}{3} \nabla k , \\
        &\mathcal{R}_p(\qfull) = \nabla \cdot \ufull , \\
        &\mathcal{R}_k(\qfull) = \ufull \cdot \nabla k - \nabla \cdot \left( \left( \nu + \frac{\nut}{\sigma_k} \right) \nabla k \right) - 2 \nut \strain : \strain + \eps , \\
        &\mathcal{R}_\eps(\qfull) = \ufull \cdot \nabla \eps - \nabla \cdot \left( \left( \nu + \frac{\nut}{\sigma_\eps} \right) \nabla \eps \right) - 2 C_1 \frac{\eps}{k} \nut \strain : \strain + C_2 \frac{\eps^2}{k} .
    \end{align}
    \label{eq:RANS_operators}
\end{subequations}
%
%
In the $k$--$\eps$ turbulence model, the ensemble-averaged eddy viscosity is defined with
\begin{equation}
    \nutphase = C_\mu \frac{\kphase^2}{\epsphase} .
    \label{eq:nut_pavg}
\end{equation}
The empirical coefficients are chosen at their default values \citep{Launder1974}, namely $C_\mu = 0.09$, $C_1 = 1.44$, $C_2 = 1.92$, $\sigma_k = 1$, and $\sigma_\eps = 1.3$.

The solutions of Equation~\eqref{eq:URANS_conti_pavg_boussinesq} can be unsteady and converge to a stable time-periodic orbit of an attracting limit cycle. Here, `stable' is meant in a dynamical systems sense, not in an eigenvalue-based linear stability sense. The solutions can also be steady and a fixed point with $\partial \qphase / \partial t = 0$ and $\qcoh = \mathbf{0}$. Equation~\eqref{eq:URANS_conti_pavg_boussinesq} then reduces to the steady RANS equations,
\begin{equation}
    \mathcal{R}(\qfull_b) = \mathbf{0} ,
    \label{eq:RANS_conti_k_eps}
\end{equation}
with $\qfull_b$ being a turbulent base flow. Note that the terminology of a `turbulent base flow' is chosen here in analogy to a laminar base flow being a fixed point solution of the Navier--Stokes equations. By that, the terminology may differ from the more common terminology in the literature, in which a solution of the RANS equations is often termed turbulent mean flow. Also note that the turbulent base flow phenomenologically still contains fluctuations in the form of stochastic `background' turbulence, which is, however, modeled in the RANS framework. The actual resolved fluctuations in the form of coherent structures are absent in the steady RANS equations.

\subsection{Linearized Reynolds-averaged Navier--Stokes equations}
The fixed points solutions of Equation~\eqref{eq:RANS_conti_k_eps} can be linearly stable or unstable to perturbations, and the stability of these solutions can be determined using the LSA. In order to derive the LSA equations, we first define the linearized RANS equations. The RANS operator~\eqref{eq:RANS_operators} linearized around a base flow $\qfull_b$ is
\begin{equation}
    \mathcal{L}(\qfull_b)\qcoh = \left. \frac{\partial \mathcal{R}(\qfull)}{\partial \qfull} \right|_{\qfull=\qfull_b} \qcoh = 
    \begin{bmatrix}
        \mathcal{L}_\ufull(\qfull_b)\qcoh\\
        \mathcal{L}_p(\qfull_b)\qcoh\\
        \mathcal{L}_k(\qfull_b)\qcoh\\
        \mathcal{L}_\eps(\qfull_b)\qcoh
    \end{bmatrix}
\end{equation}
where $\qcoh$ is a coherent fluctuation on which the linear Jacobian operator $\mathcal{L}$ acts. We further linearize the $k$--$\eps$ URANS equations~\eqref{eq:URANS_conti_pavg_boussinesq} accordingly, with $\qphase = \qfull_b + \qcoh$ and neglecting second- and higher-order terms, leading to
\begin{equation}
    \mathcal{B} \frac{\partial \qcoh}{\partial t} + \mathcal{L}(\qfull_b)\qcoh = \mathbf{0} ,
    \label{eq:LURANS}
\end{equation}
with the components of the linearized RANS operator being
\begin{subequations}
    \begin{align}
        \mathcal{L}_\ufull(\qfull)\qcoh = &\left( \ufull \cdot \nabla \right) \ucoh + \left( \ucoh \cdot \nabla \right) \ufull + \frac{1}{\rho} \nabla \pcoh - \nabla \cdot \left( 2(\nu + \nut) \straincoh \right) - \nabla \cdot \left( 2\nutcoh \strain \right) + \frac{2}{3} \nabla \kcoh , \\
        \mathcal{L}_p(\qfull)\qcoh = &\nabla \cdot \ucoh , \\
        \begin{split}
            \mathcal{L}_k(\qfull)\qcoh = &\ufull \cdot \nabla \kcoh + k \cdot \nabla \ucoh - \nabla \cdot \left( \left( \nu + \frac{\nut}{\sigma_k} \right) \nabla \kcoh \right) - \nabla \cdot \left( \frac{\nutcoh}{\sigma_k} \nabla k \right)\\
            &- 2 \nutcoh \strain : \strain - 4 \nut \strain : \straincoh + \epscoh , 
        \end{split}\\
        \begin{split}
            \mathcal{L}_\eps(\qfull)\qcoh = &\ufull \cdot \nabla \epscoh + \ucoh \cdot \nabla \eps - \nabla \cdot \left( \left( \nu + \frac{\nut}{\sigma_\eps} \right) \nabla \epscoh \right) - \nabla \cdot \left( \frac{\nutcoh}{\sigma_\eps} \nabla \eps \right)\\
            &- 2 C_1 \left( \frac{\epscoh}{k} \nut \strain : \strain - \frac{\eps}{k^2} \kcoh \nut \strain : \strain + \frac{\eps}{k} \nutcoh \strain : \strain + \frac{2\eps}{k} \nut \strain : \straincoh \right)  + C_2 \left( \frac{2\eps}{k}\epscoh - \frac{\eps^2}{k^2}\kcoh \right) .
        \end{split}
    \end{align}
    \label{eq:LRANS_operator}
\end{subequations}
The coherent fluctuation of the eddy viscosity is obtained by linearizing Equation~\eqref{eq:nut_pavg},
\begin{equation}
    \nutcoh(\qfull) = C_\mu \left( \frac{2k}{\eps}\kcoh - \frac{k^2}{\eps^2}\epscoh \right) .
    \label{eq:nutcoh}
\end{equation}

\subsection{Linear stability analysis} \label{sec:LSA_theory}
In the following, we develop the stability and sensitivity equations for the perturbed eddy-viscosity model. The frozen eddy-viscosity model, with which the linearized model is compared, will be introduced later in §\ref{sec:frozen_vs_linearized_theory}.

The linear stability equations build upon the linearized RANS equations~\eqref{eq:LURANS}. We assume the azimuthal and temporal coordinate to be homogeneous and introduce a normal mode ansatz
\begin{equation}
    \qcoh(x,r,\theta,t) = \qhat(x,r) \mathrm{e}^{\imag (m\theta - \lambda t)} + \mathrm{c.c.} ,
    \label{eq:normal_mode_ansatz}
\end{equation}
where we use a cylindrical coordinate system with $x$, $r$ and $\theta$ being the axial, radial, and azimuthal coordinates, respectively. In Equation~\eqref{eq:normal_mode_ansatz}, $\qhat$ is the spatial eigenmode shape, $m \in \mathbb{Z}$ the azimuthal wavenumber, and $\lambda \in \mathbb{C}$ the complex eigenvalue, with $\omega = \Re(\lambda)$ being the angular frequency and $\sigma = \Im(\lambda)$ being the growth rate. The $\mathrm{c.c.}$ denotes the complex conjugate. Substituting Equation~\eqref{eq:normal_mode_ansatz} into Equation~\eqref{eq:LURANS}, rearranging and dropping the complex conjugate results in the generalized linear stability eigenvalue problem,
\begin{equation}
    \left( \mathcal{L}(\qfull_b,m) -\imag \lambda \mathcal{B} \right) \qhat = \mathbf{0} ,
\end{equation}
where the $m$ in the modified linear operator $\mathcal{L}(\qfull_b,m)$ indicates that the spatial derivatives in the homogeneous $\theta$-direction due to the $\nabla$ operator have been performed analytically. The expressions of the modified $\nabla$ operator are given in Appendix~\ref{app:operators_nabla}. 

Practically, the eigenvalue problem is solved in discretized form. In this study, a spatial discretization based on the finite element method is employed. Using this discretization, the linear operators are transformed into matrices that contain spatial weighting coefficients that depend on the underlying computational mesh. The state variables, which are originally defined as continuous functions, are, therefore, represented as vectors. We adopt a matrix--vector notation which transforms the formulation to finite-dimensional space. The discretized eigenvalue problem becomes
\begin{equation}
    \left( \mathbf{L}(\qfull_b,m) - \lambda \mathbf{B} \right) \qhat = \mathbf{0} .
    \label{eq:egv_problem_discretized}
\end{equation}
where $\mathbf{L}(\qfull_b,m)$ and $\mathbf{B}$ correspond to the discretized forms of the operators $\mathcal{L}(\qfull_b,m)$ and $\imag \mathcal{B}$. The solution of the eigenvalue problem~\eqref{eq:egv_problem_discretized} provides us with a spectrum of eigenvalues and eigenmodes. The frequency $\omega$ of each eigenmode determines how the mode oscillates. The growth rate $\sigma$ determines whether the base flow is stable ($\sigma < 0$), neutrally stable ($\sigma = 0$) or unstable ($\sigma > 0$), which, respectively, corresponds to an asymptotic decay, indifference, or exponential growth of the amplitude. 

The eigenvalue problem~\eqref{eq:egv_problem_discretized} can also be solved in its adjoint form to compute the adjoint eigenmodes $\qhat^\dagger$,
\begin{equation}
    (\qhat^\dagger)^\mathrm{H} \left( \mathbf{L}(\qfull_b,m) - \lambda \mathbf{B} \right) = \mathbf{0} ,
    \label{eq:egv_adjoint_problem_discretized}
\end{equation}
where $(\cdot)^\mathrm{H}$ is the Hermitian transpose.

\subsection{Shape sensitivity analysis} \label{sec:shape_sensitivity_theory}
We follow up with a brief derivation of the shape sensitivity using perturbation theory \citep{Luchini2014,Tammisola2017}. Mathematically, we use a discrete shape deformation approach, similar to \cite{Martinez-Cava2020}, by considering surface mesh deformations that are embedded in a finite-element framework.

Let us only consider shapes that are parameterized, and let $\ai = [(a_i)_x, (a_i)_r]^T$ be the $i$-th shape parameter containing the coordinate components $x$ and $r$ of the parameter. Although not explicitly written for brevity in the following derivations, note that the base flow $\qfull_b(\ai)$, the eigenvalue $\lambda(\ai)$, and the eigenmodes $\qhat(\ai)$ and $\qhat(\ai)^\dagger$ are functions of the shape parameter $\ai$. The discretized linear operator $\mathbf{L} = \mathbf{L}(\qfull_b(\ai), \ai, m)$ depends both indirectly (through the dependence on $\qfull_b$) and directly on the shape parameter $\ai$. Furthermore, the discretized restriction operator $\mathbf{B}(\ai)$ contains spatial weighting coefficients and directly depends on the shape parameter $\ai$. Thus, its derivative with respect to $\ai$ will appear as an additional term in the eigenvalue sensitivity.

Following \cite{Luchini2014}, but with additional terms due to the shape sensitivity, at first order we can differentiate the discretized eigenvalue problem~\eqref{eq:egv_problem_discretized} with respect to $\ai$,
\begin{equation}
    \frac{\mathrm{d} [\left( \mathbf{L}(\qfull_b,m) - \lambda \mathbf{B} \right) \qhat]}{\mathrm{d} \ai} = \mathbf{0} ,
\end{equation}
where the zero on the right-hand side applies since the equality in \eqref{eq:egv_problem_discretized} holds for any perturbation of the eigenvalue problem. We carry out the differentiation operation, leaving us with
\begin{equation}
    \left( \frac{\mathrm{d} \mathbf{L}(\qfull_b,m)}{\mathrm{d} \ai} - \lambda \frac{\mathrm{d} \mathbf{B}}{\mathrm{d} \ai} \right) \qhat - \frac{\mathrm{d} \lambda}{\mathrm{d} \ai} \mathbf{B} \qhat + (\mathbf{L}(\qfull_b,m) - \lambda \mathbf{B}) \frac{\mathrm{d} \qhat}{\mathrm{d} \ai} = \mathbf{0} .
\end{equation}
The third term on the right-hand side can be eliminated by premultiplying the equation with the adjoint eigenmode $(\qhat^\dagger)^\mathrm{H}$ and acknowledging that $(\qhat^\dagger)^\mathrm{H} \left( \mathbf{L}(\qfull_b,m) - \lambda \mathbf{B} \right) = \mathbf{0}$ (see Equation~\eqref{eq:egv_adjoint_problem_discretized}). Furthermore, we introduce the normalization condition $(\qhat^\dagger)^\mathrm{H} \mathbf{B} \qhat = 1$. We can then rearrange and obtain the eigenvalue sensitivity with respect to a shape deformation,
\begin{equation}
    \frac{\mathrm{d}\lambda}{\mathrm{d} \ai} = (\qhat^\dagger)^\mathrm{H} \left( \frac{\mathrm{d} \mathbf{L}(\qfull_b,m)}{\mathrm{d} \ai} - \lambda \frac{\mathrm{d} \mathbf{B}}{\mathrm{d} \ai} \right) \qhat ,
    \label{eq:shape_sensitivity}
\end{equation}
where Equation~\eqref{eq:shape_sensitivity} is a vector equation such that $\mathrm{d}\lambda/\mathrm{d}\ai = [\mathrm{d}\lambda/\mathrm{d}(a_i)_x, \mathrm{d}\lambda/\mathrm{d}(a_i)_y]^T$ contains the sensitivity of the eigenvalue with respect to changes in the $i$th shape parameter in $x$ and $r$ direction.

The total derivative of the discretized restriction operator $\mathbf{B}$ in Equation~\eqref{eq:shape_sensitivity} is approximated with a finite difference approach,
\begin{equation}
    \frac{\mathrm{d} \mathbf{B}(\ai)}{\mathrm{d} \ai} \approx \frac{ \mathbf{B}(\ai+\epsilon\mathbf{e}_j) - \mathbf{B} (\ai)}{\epsilon} \, ,
    \label{eq:B_operator_finite_difference}
\end{equation}
with $\epsilon$ being a very small perturbation of the unit basis vectors $\mathbf{e}_j$ in $x$ and $r$ direction. The total derivative of the linearized RANS operator $\mathbf{L}$ in Equation~\eqref{eq:shape_sensitivity} is computed as follows,
\begin{equation}
    \frac{\mathrm{d}\mathbf{L}(\qfull_b,m)}{\mathrm{d}\ai} = \frac{\partial\mathbf{L}(\qfull_b,m)}{\partial \ai} + \mathbf{H}(\qfull_b,m) \frac{\mathrm{d}\qfull_b}{\mathrm{d}\ai} \, ,
    \label{eq:LNSE_operator_total_derivative}
\end{equation}
where the first term on the right-hand side is computed with a finite difference in the same way as in Equation~\eqref{eq:B_operator_finite_difference}. The second term in Equation~\eqref{eq:LNSE_operator_total_derivative} contains the discretized form $\mathbf{H}(\qfull_b,m) = \partial \mathbf{L}(\qfull_b,m) / \partial \qfull_b$ of the Hessian operator $\mathcal{H}$, whose definition is given in Appendix~\ref{app:operators_hessian}. Note that in this work, the Hessian operator is a bilinear operator that acts on two different inputs.

In this work, we do not solve a single adjoint equation system, but instead compute the total derivative of the base flow contained in Equation~\eqref{eq:LNSE_operator_total_derivative}, namely the base-flow sensitivity, directly for each $\ai$ with
\begin{equation}
    \mathbf{L}(\qfull_b,m) \frac{\mathrm{d}\qfull_b}{\mathrm{d}\ai} = - \frac{\partial \mathbf{R}(\qfull_b)}{\partial \ai} \, ,
    \label{eq:base_flow_sensitivity}
\end{equation}
where $\mathbf{R}$ is the discretized RANS operator of Equation~\eqref{eq:RANS_conti_k_eps}. The right-hand side is again approximated with a finite difference approach as shown in Equation~\eqref{eq:B_operator_finite_difference}. Compared to computing an adjoint base flow, this direct approach does not pose a significant drawback in terms of computational costs. Even if the number of parameters is large, the preconditioner for the matrix $\mathbf{L}$, once computed, can be reused for all subsequent linear solvings~\citep{Knechtel2024}.

\subsection{Shape sensitivity decomposition for identifying physical mechanisms} \label{sec:shape_sensitivity_decomposition_theory}
The shape sensitivity~\eqref{eq:shape_sensitivity} can be further decomposed into individual contributions that correspond to different physical mechanisms. This will help us in interpreting the results later on. We will closely follow the terminology and approach of \cite{Brewster2020}.

We first substitute Equation~\eqref{eq:LNSE_operator_total_derivative} in Equation~\eqref{eq:shape_sensitivity} and rearrange, resulting in
\begin{equation}
    \frac{\mathrm{d}\lambda}{\mathrm{d} \ai} = (\qhat^\dagger)^\mathrm{H} \left( \frac{\partial\mathbf{L}(\qfull_b,m)}{\partial \ai} - \lambda \frac{\mathrm{d} \mathbf{B}}{\mathrm{d} \ai} \right) \qhat + (\qhat^\dagger)^\mathrm{H} \left( \mathbf{H}(\qfull_b,m) \frac{\mathrm{d}\qfull_b}{\mathrm{d}\ai} \right) \qhat .
    \label{eq:shape_sensitivity_decomposed}
\end{equation}
The first term of Equation~\eqref{eq:shape_sensitivity_decomposed} is the feedback contribution of the shape sensitivity. It quantifies the eigenvalue change due to a change in the intrinsic feedback mechanism encoded in the linearized RANS operator (assuming that the base flow does not change),
\begin{equation}
    \left( \frac{\mathrm{d}\lambda}{\mathrm{d} \ai} \right)_\textrm{feed} = (\qhat^\dagger)^\mathrm{H} \left( \frac{\partial\mathbf{L}(\qfull_b,m)}{\partial \ai} - \lambda \frac{\mathrm{d} \mathbf{B}}{\mathrm{d} \ai} \right) \qhat .
    \label{eq:shape_sensitivity_feedback}
\end{equation}
The second term of Equation~\eqref{eq:shape_sensitivity_decomposed} is the base-flow contribution of the shape sensitivity. It quantifies the eigenvalue change due to a change in the base flow (assuming that the linearized RANS operator does not change),
\begin{equation}
    \left( \frac{\mathrm{d}\lambda}{\mathrm{d} \ai} \right)_\textrm{base}  = (\qhat^\dagger)^\mathrm{H} \left( \mathbf{H}(\qfull_b,m) \frac{\mathrm{d}\qfull_b}{\mathrm{d}\ai} \right) \qhat .
    \label{eq:shape_sensitivity_baseflow}
\end{equation}
Thus, the shape sensitivity is split into these two overarching mechanisms of feedback change and base-flow change.

The base-flow contribution in Equation~\eqref{eq:shape_sensitivity_baseflow} can be even further decomposed into individual physical mechanisms encoded in the Hessian operator $\mathbf{H}$. These physical mechanisms are specified in detail in Appendix~\ref{app:operators_hessian}. It is also insightful to decompose Equation~\eqref{eq:shape_sensitivity_baseflow} into two products such that
\begin{equation}
    \left( \frac{\mathrm{d}\lambda}{\mathrm{d} \ai} \right)_\textrm{base}  = \frac{\partial \lambda}{\partial \qfull_b} \frac{\mathrm{d}\qfull_b}{\mathrm{d}\ai}
    \label{eq:shape_sensitivity_baseflow_decomposed}
\end{equation}
with
\begin{equation}
    \frac{\partial \lambda}{\partial \qfull_b} = (\qhat^\dagger)^\mathrm{H} \mathbf{H}(\qfull_b,m) \qhat .
    \label{eq:shape_sensitivity_with_respect_to_base_flow}
\end{equation}
This decomposition is possible due to the commutative nature of the Hessian operator according to Schwarz's theorem, allowing us to swap the last two factors in Equation~\eqref{eq:shape_sensitivity_baseflow}. With the decomposition in Equation~\eqref{eq:shape_sensitivity_with_respect_to_base_flow}, we can inspect the optimal changes in the eigenvalue due to changes in the base flow, $\partial \lambda / \partial \qfull_b$, and isolate them from the actual base-flow changes due to shape deformation, the base-flow sensitivity $\mathrm{d}\qfull_b / \mathrm{d}\ai$.

Although the eddy viscosity is not part of the base state vector $\qfull_b$, changes to the eddy viscosity and derivatives with respect to the eddy viscosity can still be easily computed using the algebraic relation
\begin{equation}
    \nu_{t,b} = C_\mu \frac{k_b^2}{\eps_b}
    \label{eq:nut_baseflow}
\end{equation}
and the chain rule. The eigenvalue change due to a change in base-flow eddy viscosity is
\begin{subequations}
    \begin{align}
        \frac{\partial \lambda}{\partial \nu_{t,b}} &= \frac{\partial \lambda}{\partial k_b} \frac{\partial k_b}{\partial \nu_{t,b}} + \frac{\partial \lambda}{\partial \eps_b} \frac{\partial \eps_b}{\partial \nu_{t,b}} \, , \\
        \frac{\partial k_b}{\partial \nu_{t,b}} &= \frac{1}{2} \sqrt{\frac{\eps_b}{C_\mu \nu_{t,b}}} \, , \\
        \frac{\partial \eps_b}{\partial \nu_{t,b}} &= -C_\mu \frac{k_b^2}{\nu_{t,b}^2}
    \end{align}
    \label{eq:egv_sensitivity_due_to_nut}
\end{subequations}
and the base-flow sensitivity of the eddy viscosity is
\begin{subequations}
    \begin{align}
        \frac{\mathrm{d} \nu_{t,b}}{\mathrm{d} \ai} &= \frac{\partial \nu_{t,b}}{\partial k_b} \frac{\mathrm{d} k_b}{\mathrm{d} \ai} + \frac{\partial \nu_{t,b}}{\partial \eps_b} \frac{\mathrm{d} \eps_b}{\mathrm{d} \ai} \, , \\
        \frac{\partial \nu_{t,b}}{\partial k_b} &= 2 C_\mu \frac{k_b}{\eps_b} \, , \\
        \frac{\partial \nu_{t,b}}{\partial \eps_b} &= - C_\mu \frac{k_b^2}{\eps_b^2} \, .
    \end{align}
    \label{eq:base_flow_sensitivity_nut}
\end{subequations}

\subsection{Frozen versus perturbed eddy-viscosity model} \label{sec:frozen_vs_linearized_theory}
In the previous sections, we introduced the eigenvalue problem and shape sensitivity with the corresponding operators, in which we always assumed the eddy viscosity, turbulent kinetic energy, and dissipation to be fluctuating. In the frozen eddy-viscosity model, we neglect these coherent fluctuations, i.e. $\nutcoh = 0$, $\kcoh = 0$ and $\epscoh = 0$. In order to compare the results, we need to define the equations and operators accordingly for the frozen model. In essence, this is done by discarding the equations for $\kcoh$ and $\epscoh$ in \eqref{eq:LURANS} and the terms that involve fluctuations of $\nutcoh$ and $\kcoh$ in the momentum equations, and setting $\qfull = [\ufull, p]^T$. In the following, we denote the operators of the frozen eddy-viscosity model with $(\breve{\cdot})$.

The discretized direct eigenvalue problem with frozen eddy viscosity is defined as
\begin{equation}
    ( \mathbf{\breve{L}}(\qfull_b,m) - \lambda \breve{\mathbf{B}} ) \qhat = \mathbf{0} ,
    \label{eq:egv_problem_discretized_frozen}
\end{equation}
and the discretized adjoint eigenvalue problem with frozen eddy viscosity is defined as
\begin{equation}
    (\qhat^\dagger)^\mathrm{H} ( \mathbf{\breve{L}}(\qfull_b,m) - \lambda \breve{\mathbf{B}} ) = \mathbf{0} ,
    \label{eq:egv_adjoint_problem_discretized_frozen}
\end{equation}
where $\mathbf{\breve{L}}(\qfull_b,m)$ is the discretized RANS operator of $\mathcal{\breve{L}}(\qfull_b,m)$, where
\begin{equation}
    \breve{\mathcal{L}}(\qfull_b)\qcoh = \left. \frac{\partial \breve{\mathcal{R}}(\qfull)}{\partial \qfull} \right|_{\qfull=\qfull_b} \qcoh = 
    \begin{bmatrix}
        \breve{\mathcal{L}}_\ufull(\qfull_b)\qcoh\\
        \breve{\mathcal{L}}_p(\qfull_b)\qcoh ,
    \end{bmatrix}
\end{equation}
with
\begin{subequations}
    \begin{equation}
        \mathcal{\breve{L}}_\ufull(\qfull_b,m)\qcoh = \left( \ufull_b \cdot \nabla \right) \ucoh + \left( \ucoh \cdot \nabla \right) \ufull_b + \frac{1}{\rho} \nabla \pcoh - \nabla \cdot \left( 2(\nu + \nu_{t,b}) \straincoh \right) ,
    \end{equation}
    \begin{equation}
        \mathcal{\breve{L}}_p(\qfull)\qcoh = \nabla \cdot \ucoh .
    \end{equation}
    \label{eq:LRANS_operator_frozen}
\end{subequations}
The shape sensitivity is defined as
\begin{equation}
    \frac{\mathrm{d}\lambda}{\mathrm{d} \ai} = (\qhat^\dagger)^\mathrm{H} \left( \frac{\mathrm{d} \mathbf{\breve{L}}(\qfull_b,m)}{\mathrm{d} \ai} - \lambda \frac{\mathrm{d} \breve{\mathbf{B}}}{\mathrm{d} \ai} \right) \qhat ,
    \label{eq:shape_sensitivity_frozen}
\end{equation}
where the momentum component of the RANS operator contained in the total derivative of the linearized RANS operator is adapted to
\begin{equation}
    \mathcal{\breve{R}}_\ufull(\qfull_b) = \left( \ufull_b \cdot \nabla \right) \ufull_b + \frac{1}{\rho} \nabla p_b - \nabla \cdot \left( 2(\nu + \nu_{t,b}) \strain_b \right) ,
\end{equation}
and the Hessian operator is adapted as given in Appendix~\ref{app:operators_hessian}.

\section{Francis turbine model and numerical setup} \label{sec:setup}
The configuration investigated in this work is derived from the well-established Francis-99 hydro turbine, a benchmark case of a high-head Francis turbine widely used in academic and industrial research \citep{Trivedi2016}. The Francis turbine model is designed to emulate the draft-tube flow that develops downstream of the Francis-99 runner, thereby capturing the relevant velocity distributions and vortex rope dynamics without resorting to a full-scale turbine. The operating conditions are generated by combining a stationary guide vane stage with a rotating runner, following the approaches described in \cite{Susan-Resiga2007,Litvinov2018}. This setup reproduces a broad range of draft-tube flow conditions, including the bifurcation behavior associated with the vortex rope instability, while excluding the upstream hydraulic duct, which is not of interest here. Previous works have also shown that cavitation and two-phase effects do not significantly alter the global characteristics of the vortex rope, allowing the present single-phase approach using air \citep[e.g.][]{Skripkin2022}.

In this work, the LSA and shape sensitivity analysis are based on base-flow fields computed with RANS simulations. The RANS setup comprises a simplified configuration of the Francis turbine model that omits the upstream guide vanes and runner blades, retaining only the draft tube and runner crown geometry. This reduction allows the analysis to focus entirely on the flow physics relevant to the vortex rope dynamics. In order to reproduce realistic inlet conditions representative of the Francis-99 turbine operating conditions, precursor URANS simulations were performed, which were validated against experimental measurements. From these simulations, the temporally and azimuthally averaged flow profiles at the interface between the runner outlet and the draft tube inlet were extracted and applied as boundary conditions for the RANS computations. Additional details on the precursor URANS simulations, the inlet profiles, and the experimental measurements are provided in Appendix~\ref{app:3d_urans_experiments}.

\begin{figure}[tbp]
    \centering
    \begin{tikzpicture}
        \begin{scope}
            \node[anchor=south west,inner sep=0] (image) at (0,0) {\includegraphics[width=0.75\columnwidth]{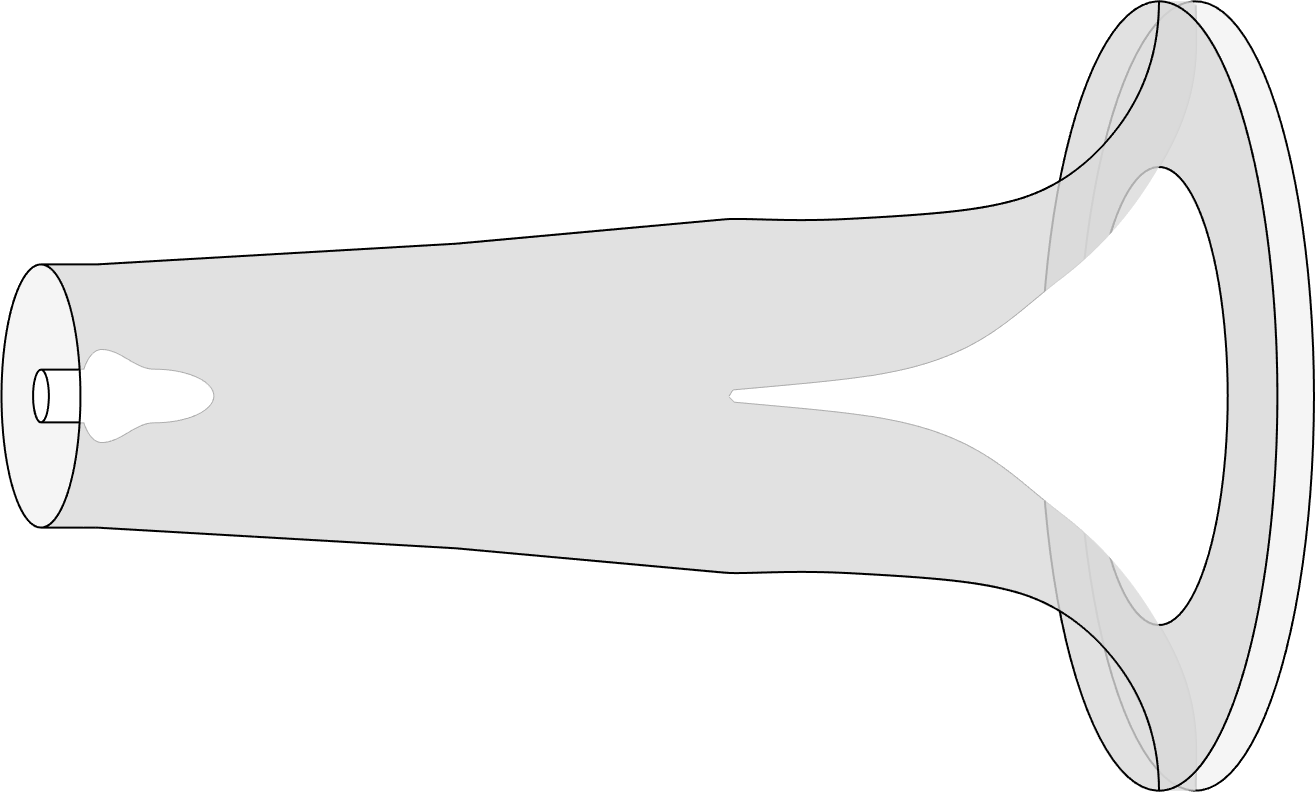}};
            \begin{scope}[x={(image.south east)},y={(image.north west)}]

                \draw [thin,dash dot,black] (0.22,0.5) -- ++(0.68,0);

                \draw [semithick, black] (0.032,0.5) -- ++(0.0255,0);
                \draw [-latex, semithick, black] (0.066,0.5) -- ++(0.11,0)
                node [right,black] {\footnotesize $x$};
                \draw [-latex, semithick, black] (0.032,0.5) -- ++(0,0.24)
                node [left,black] {\footnotesize $y$, $r$};
                \draw [-latex, semithick, black] (0.032,0.5) -- ++(0.1,-0.07)
                node [left,below,black] {\footnotesize $z$};   
                \draw[stealth-,semithick,black] (0.026,0.58) arc
                    [
                    start angle=110,
                    end angle=330,
                    x radius=0.017,
                    y radius =0.085
                    ] coordinate[pos=0.7] (mid);
                    \node[below=-1.5pt,semithick,black] at (mid) {\footnotesize $\theta$};

                \draw [latex-latex, very thin, black] (0.032-0.06,0.335) -- ++(0,0.33)
                node [left,midway,black] {\scriptsize $D = \SI{100}{\mm}$};
                \draw [-, thin, black] (0.032,0.335) -- ++(-0.07,0);
                \draw [-, thin, black] (0.032,0.665) -- ++(-0.07,0);

                \draw [-latex, thick,paraview_red] (0.24,0.61) to[out=0,in=180] (0.42,0.375) to[out=-180+180,in=180] (0.60,0.63);
                \draw[sketch_gray,fill=sketch_gray] (0.506,0.5) circle (0.012);
                
                \draw [-latex, thick,paraview_red] (0.24,0.39) to[out=0,in=180] (0.42,0.625) to[out=-180+180,in=180] (0.60,0.37);
                \draw[sketch_gray,fill=sketch_gray] (0.327,0.5) circle (0.012);

                \draw [thick,paraview_red] (0.24,0.61) to[out=0,in=180] (0.42,0.375); 

                \draw [-latex, thick,paraview_red] (0.62,0.63) to[out=0,in=200] (0.74,0.66) to[out=-180+200,in=240] (0.86,0.8) to[out=-180+240,in=270] (0.892,0.955) to[out=-180+270,in=270] (0.892,0.975);

                \draw [-latex, thick,paraview_red] (0.62,0.37) to[out=0,in=160] (0.74,0.34) to[out=-180+160,in=120] (0.86,0.2) to[out=-180+120,in=90] (0.892,0.045) to[out=-180+90,in=90] (0.892,0.025);

                \draw [latex-, thick, black] (0.033,0.335) -- ++(0,-0.135)
                node [below,black] {\footnotesize Inlet};
                
                \draw [latex-, thick, black] (0.155,0.48) -- ++(0,-0.28)
                node [below,black] {\footnotesize Centerbody};
                
                \draw [latex-, thick, black] (0.4,0.305) -- ++(0,-0.105)
                node [below,black] {\footnotesize Draft tube};
                
                \draw [latex-, thick, black] (0.66,0.27) -- ++(0,-0.07)
                node [below,black] {\footnotesize Moody tube};

                \draw [latex-, thick, black] (0.892,0.004) -- ++(0,-0.07)
                node [below,black] {\footnotesize Outlet};
            \end{scope}
        \end{scope}
    \end{tikzpicture}
    \caption{Geometrical domain of the RANS simulation and LSA setup.}
    \label{fig:domain_setup}
\end{figure}
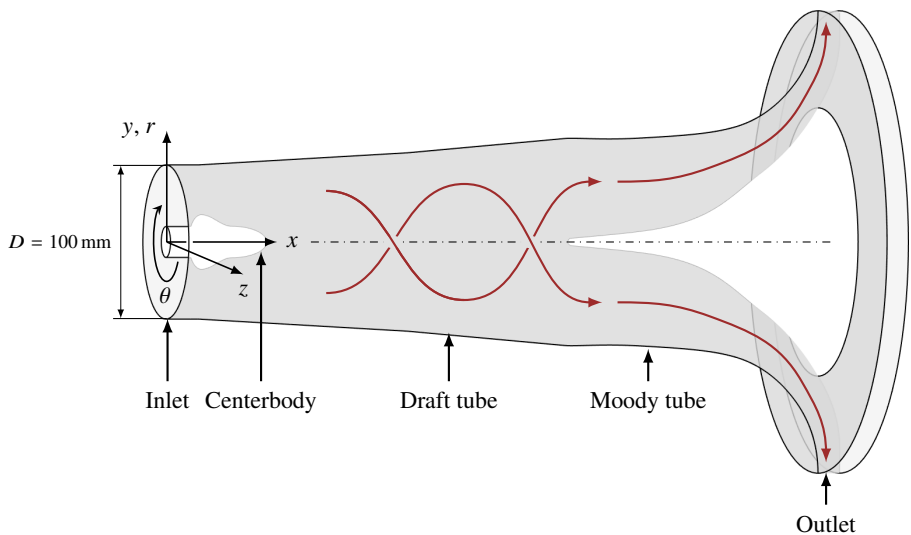

Geometrically, the setup consists of a conical draft tube with an inlet diameter of $D = \SI{100}{\mm}$ and an initial opening angle of $6.6^\circ$, as shown in Figure~\ref{fig:domain_setup}. A cylindrical coordinate system is used with $x$, $r$, and $\theta$ being the axial, radial, and azimuthal coordinates, respectively. For convenience, we will consider streamwise sections at $\theta = 0$ throughout this work and use the radial coordinate $r$ and the transverse coordinate $y$ interchangeably. At the inlet, an annular section comprises an axisymmetric centerbody composed of a cylindrical base and a smooth, free-form-like extension downstream, representing the runner crown. The smooth centerbody geometry is not part of the original Francis-99 design \citep[see][]{Lueckoff2022}. Downstream of the initial draft tube, a so-called Moody tube is attached, designed so that the streamwise discharge velocity is conserved similarly to that of the corresponding Francis-99 elbow draft tube. This ensures well-defined outlet boundary conditions in both the experiments and the simulations.  A wide range of part-load conditions is investigated, spanning normalized flow rates from $Q^* = Q/Q_\mathrm{BEP} = 0.50$ to $0.76$, where the flow rate at the turbine's best efficiency point (BEP) is $Q_\mathrm{BEP} = \SI{175}{m^3/h}$ \citep{Litvinov2018}. The corresponding bulk velocity of the range of part-load conditions varies between \SI{3.8}{\m\per\s} and \SI{5.7}{\m\per\s}, which translates to Reynolds numbers in the range of $Re = \numrange{25000}{38000}$.

\subsection{Numerical setup for the RANS simulations}
The simulations are performed using the open-source computational fluid dynamics framework \textit{OpenFOAM}. The code employs an unstructured finite-volume discretization on a cell-centered collocated mesh. The incompressible, steady-state RANS equations~\eqref{eq:RANS_conti_k_eps} are solved in an axisymmetric formulation where $\partial/\partial\theta = 0$. Turbulence closure is provided by the standard $k$–$\eps$ model \citep{Launder1974}. Spatial gradients are evaluated using a least-squares scheme. Convective fluxes are computed via the classical Gauss's theorem for surface integration. A second-order central differencing scheme is employed for the velocity $\mathbf{u}$, while $k$ and $\eps$ are discretized using a second-order total variation diminishing scheme with a Sweby flux limiter~\citep{Sweby1984}. Diffusive terms are approximated using a second-order central differencing scheme with a non-orthogonality correction. A linear scheme is used for the interpolation of quantities between the cell faces and the cell-center values. Pressure-velocity coupling is achieved through an under-relaxed SIMPLE algorithm \citep{VanDoormaal1984}. The discretized equations are solved iteratively: the velocity, turbulent kinetic energy, and dissipation rate equations are solved using a Gauss--Seidel method, while the pressure equation is handled by a geometric–algebraic multigrid solver. Convergence is assumed when all residuals exhibit a sustained plateau over several iterations. Although the exact levels depend on the flow rate, residuals typically drop below $10^{-8}$ for $\mathbf{u}$, $k$ and $\eps$, and below $10^{-6}$ for the pressure equation, in many cases reaching even lower values.

At the inlet, inhomogeneous Dirichlet boundary conditions are imposed for all velocity and turbulence quantities, where the profiles are prescribed based on temporally and azimuthally averaged data extracted from precursor URANS simulations, with further details provided in Appendix~\ref{app:3d_urans_experiments}. The pressure is set to a homogeneous Neumann condition at the inlet. At the outlet, homogeneous Neumann conditions are applied for all flow variables, except for the pressure, for which a homogeneous Dirichlet condition is used to fix the arbitrary reference pressure level. No-slip boundary conditions are enforced at all solid walls for the velocity, accompanied by a homogeneous Neumann condition for the pressure. For the turbulence quantities, $k$ is set to a homogeneous Dirichlet condition, while $\eps$ is assigned a homogeneous Neumann condition. Additional tests with a standard low-Reynolds-number wall function showed no discernible differences when compared to the homogeneous Neumann approach.

\begin{figure}
    \centering
    \begin{minipage}[b]{0.9\textwidth}
            \centering
            \begin{tikzpicture}
                \begin{scope}
                    \node[anchor=south west,inner sep=0] (image) at (0,0) {\includegraphics[width=0.96\textwidth]{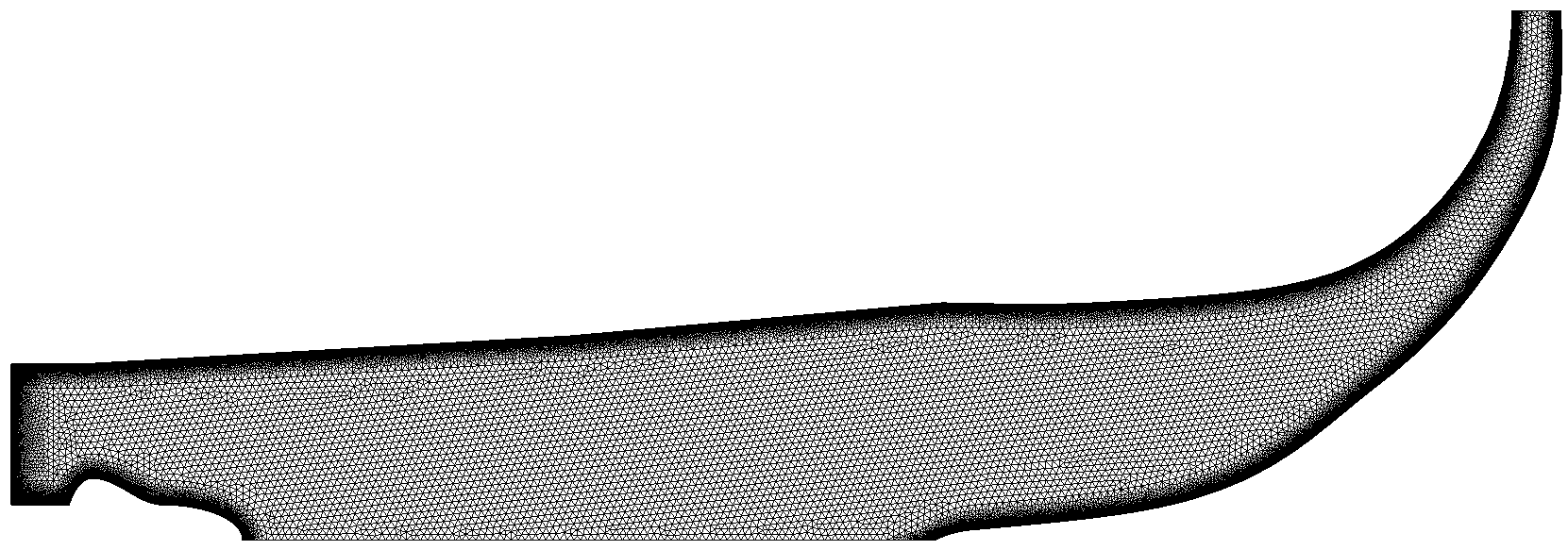}};
                    \begin{scope}[x={(image.south east)},y={(image.north west)}]
                        \draw[draw=paraview_red_light, thick] (0.0,0.0) rectangle
                        ++(0.20,0.38);
                        \draw[draw=paraview_red, thick] (0.038,0.07) rectangle
                        ++(0.05,0.11);
                    \end{scope}
                \end{scope}
            \end{tikzpicture}
            \vspace{0.1cm}
            \begin{tikzpicture}
                \begin{scope}
                    \node[anchor=south west,inner sep=0] (image) at (0,0) {\includegraphics[height=0.267\textwidth]{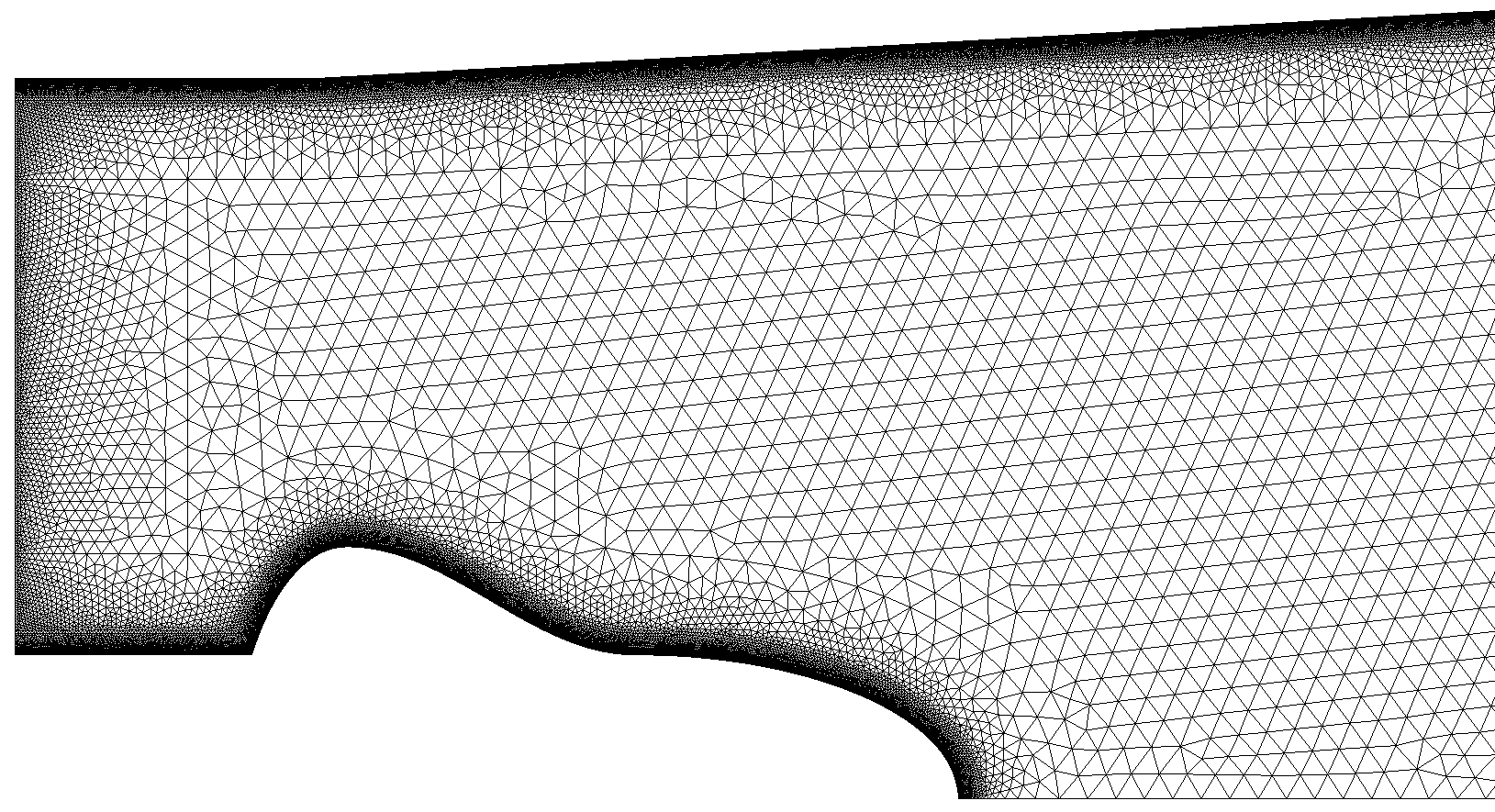}};
                    \begin{scope}[x={(image.south east)},y={(image.north west)}]
                        \draw[draw=paraview_red_light, thick] (0,0) rectangle
                        ++(1,1);
                    \end{scope}
                \end{scope}
            \end{tikzpicture}
            \begin{tikzpicture}
                \begin{scope}
                    \node[anchor=south west,inner sep=0] (image) at (0,0) {\includegraphics[height=0.267\textwidth]{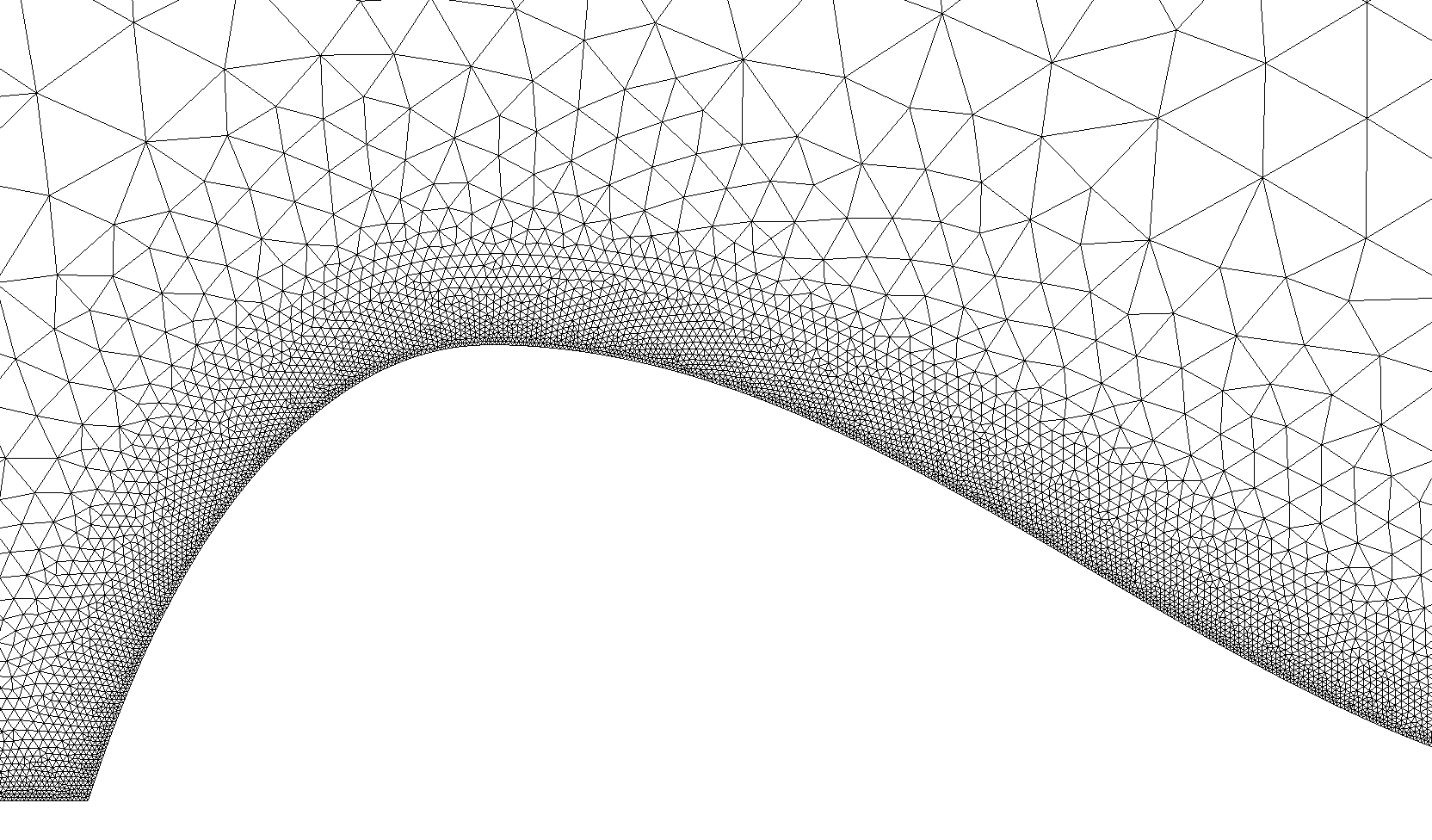}};
                    \begin{scope}[x={(image.south east)},y={(image.north west)}]
                        \draw[draw=paraview_red, thick] (0,0) rectangle
                        ++(1,1);
                    \end{scope}
                \end{scope}
            \end{tikzpicture}
    \end{minipage}
    \caption{Computational mesh of the RANS simulation and LSA.}
    \label{fig:mesh_domains}
\end{figure}
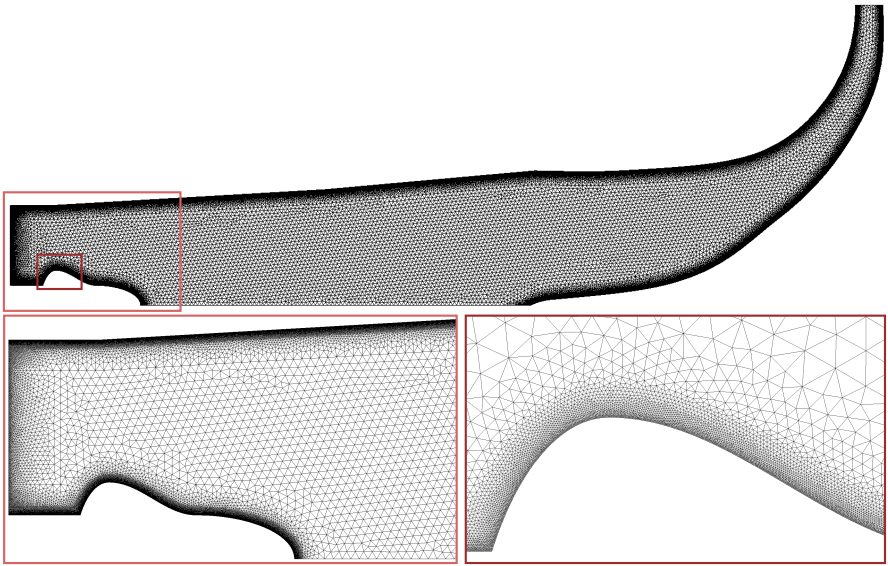

The computational mesh, shown in Figure~\ref{fig:mesh_domains}, is generated using \textit{Gmsh} and consists only of triangular elements that are mostly equilateral. Choosing triangular elements ensures that we can consistently use the same mesh in the subsequent LSA and shape sensitivity computations, which are performed with a finite element code. The high number of equilateral elements ensures a low non-orthogonality for most of the elements, thereby improving numerical stability and convergence in \textit{OpenFOAM}. The main drawback of this approach is the increased number of cells required to adequately resolve the wall boundary layers. However, this is acceptable given the relatively low computational cost of the 2D RANS simulations. The final mesh comprises approximately $2.5\times10^5$ cells, with a resolution of $y^+ < 1$ for the first cell layer. A combined mesh independence study was carried out for the RANS, LSA, and sensitivity analyses by repeating selected cases on a refined mesh containing approximately $5\times10^5$ cells. The qualitative trends remained unchanged, while quantitative variations in the eigenvalue magnitude and its peak sensitivity were limited to about $\numrange{1}{2}$\%.

\subsubsection{Tuning of the RANS inlet conditions} \label{sec:rans_tuning}
So far, we have associated the coherent fluctuations $\ucoh$ with the vortex rope. However, the URANS precursor simulations also reveal other dominant fluctuations contained in $\ucoh$: the wake structures being periodically shed from the runner blades. These wake structures interact with the vortex rope (see the discussion on Figure~\ref{fig:spectrum_crossiwse_Q65} in Appendix~\ref{app:3d_urans_experiments}). They modify both the vortex rope dynamics and the mean flow. When transitioning from the URANS to the combined steady-state RANS and LSA framework, the runner wake structures and their interactions are no longer directly resolved. To account for the increased turbulence intensity at the inlet originating from the runner wake structures, we tune the inlet profiles of $k$, $\eps$ and $\nut$ by a constant scaling factor. This factor is adjusted so that the bifurcation point obtained from the combined RANS and LSA framework matches the bifurcation point of the URANS simulation. Preliminary investigations showed that, without such tuning of the RANS inlet conditions, the base-flow growth rates remain positive and nearly constant across the entire range of flow rates, which does not reflect the physical behavior of the system.

In this RANS tuning approach, we conceptually treat these wake structures as non-resolved fluctuations and lump them into the turbulent fluctuations $\ufull''$. By that, we assume that these non-resolved wake structures act primarily in a dissipative manner and can, therefore, be modeled using the Boussinesq eddy viscosity hypothesis already employed. This addition of the runner wake structures to $\ufull''$ implies an effective increase in the magnitude of the turbulent quantities, which is the motivation for the described approach using the scaling factor.

\subsection{Numerical setup for the LSA and shape sensitivity}
The discretized direct and adjoint eigenvalue problems, as well as the shape sensitivity and its decompositions (described in §~\ref{sec:LSA_theory}-\ref{sec:frozen_vs_linearized_theory}), are solved in variational form using the in-house finite element code \textit{FELiCS} \citep{Kaiser2023b}, a linearized flow solver that utilizes the finite element package \textit{FEniCS} \citep{Baratta2023}. Second-order continuous Galerkin elements are applied for the velocity, turbulent kinetic energy, and turbulent dissipation, whereas first-order elements are used for the pressure. The eigenvalue problem is solved using \textit{SLEPc}. At the inlet, outlet, and wall, homogeneous Dirichlet conditions are imposed for the velocity and turbulent kinetic energy, as well as homogeneous Neumann conditions for the turbulent dissipation and pressure. On the axis, compatibility conditions similar to \cite{Khorrami1989} are set with homogeneous Dirichlet conditions for the axial velocity, pressure, turbulent kinetic energy and turbulent dissipation, and homogeneous Neumann conditions for the radial and azimuthal velocities.

The centerbody geometry is parameterized using $N = 126$ B-spline control points that are uniformly distributed along the axial direction. Each control point is allowed to move in both the axial and radial directions, except for the most upstream point located at the end of the runner axle, which remains fixed. The large number of B-spline nodes ensures a sufficiently high shape resolution and enables the computation of an almost continuous surface distribution of the shape sensitivity. Using Equation~\eqref{eq:shape_sensitivity}, the wall-normal component of the shape sensitivity is then evaluated with
\begin{equation}
    \frac{\mathrm{d} \lambda}{\mathrm{d} \mathbf{n}} = \frac{\mathrm{d} \lambda}{\mathrm{d} \mathbf{a}_i} \cdot \mathbf{n} ,
    \label{eq:shape_sensitivity_normal}
\end{equation}
where $\mathbf{n}$ is the local surface-normal vector with $\left\Vert\mathbf{n} \right\Vert = 1$.

\section{Impact of the perturbed eddy-viscosity model on eigenvalue and eigenmode} \label{sec:results_lsa}

\begin{figure}[tbp]
    \centering
    \sidesubfloat[]{
    \centering
    \includegraphics[height=0.29\columnwidth]{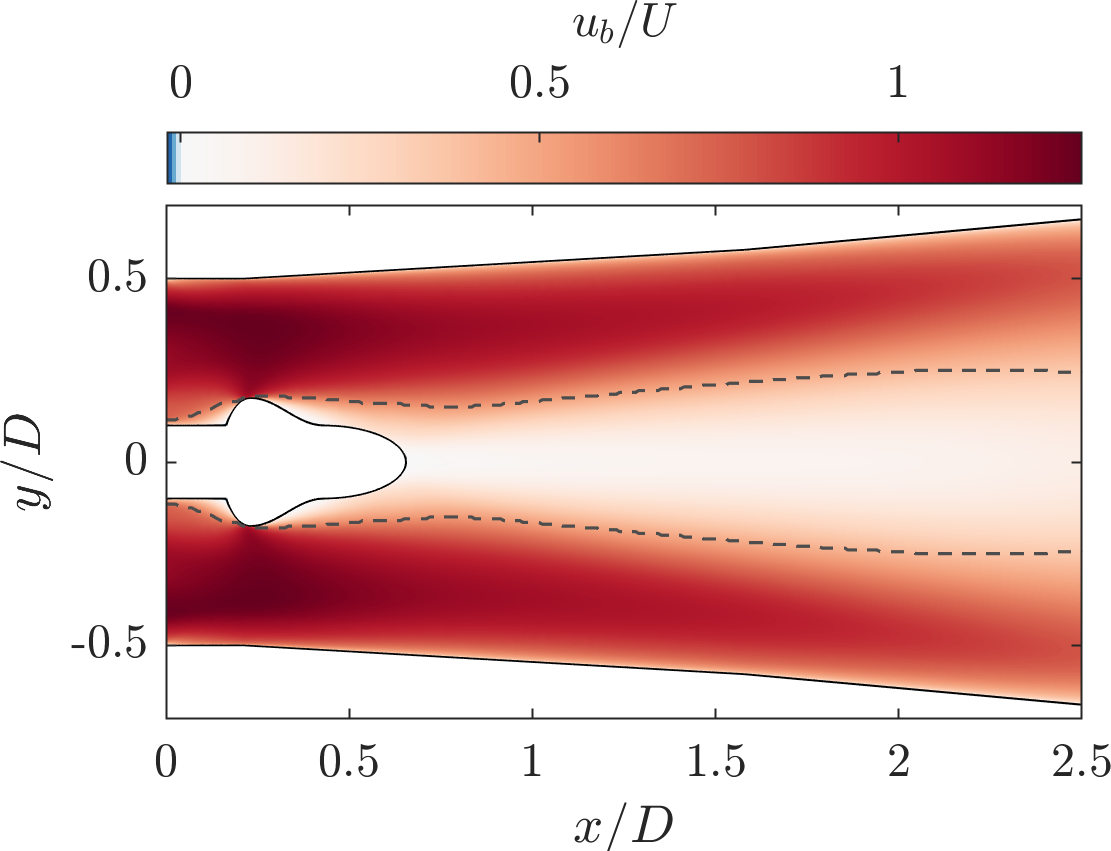}
    }
    \sidesubfloat[]{
    \centering
    \includegraphics[height=0.29\columnwidth]{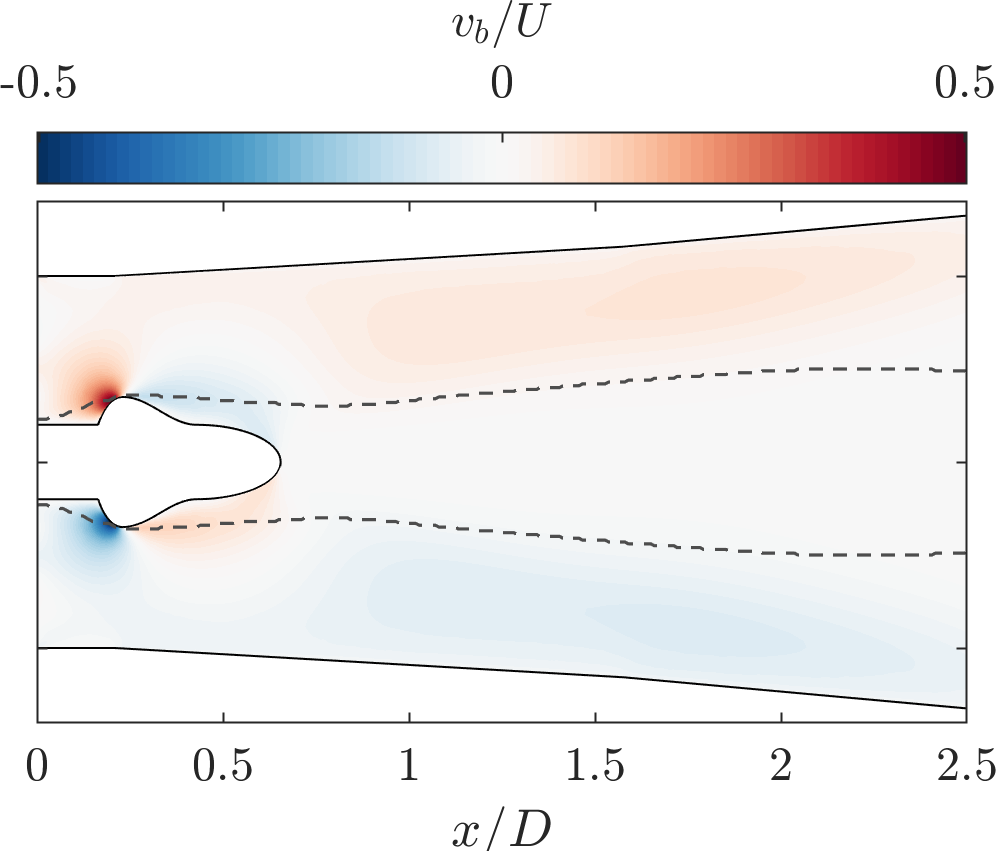}
    }\\
    \vspace{0.2cm}
    \sidesubfloat[]{
    \centering
    \includegraphics[height=0.29\columnwidth]{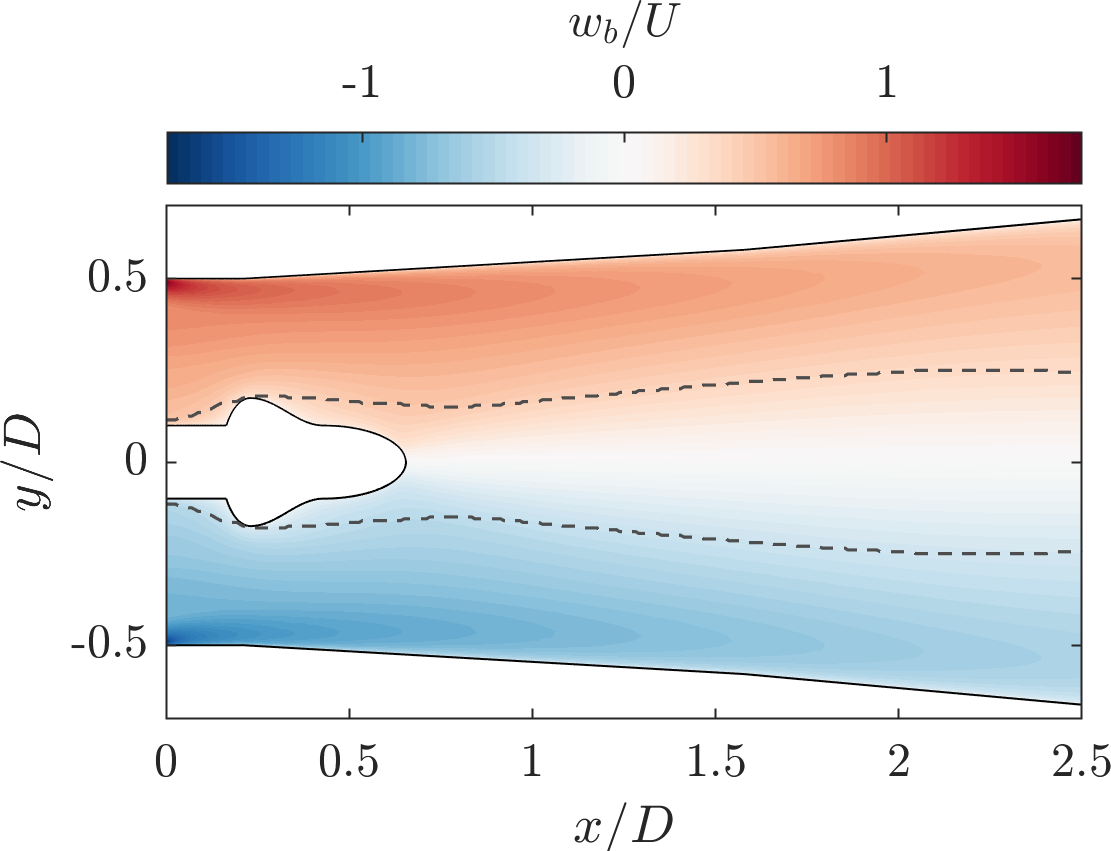}
    }
    \sidesubfloat[]{
    \centering
    \includegraphics[height=0.29\columnwidth]{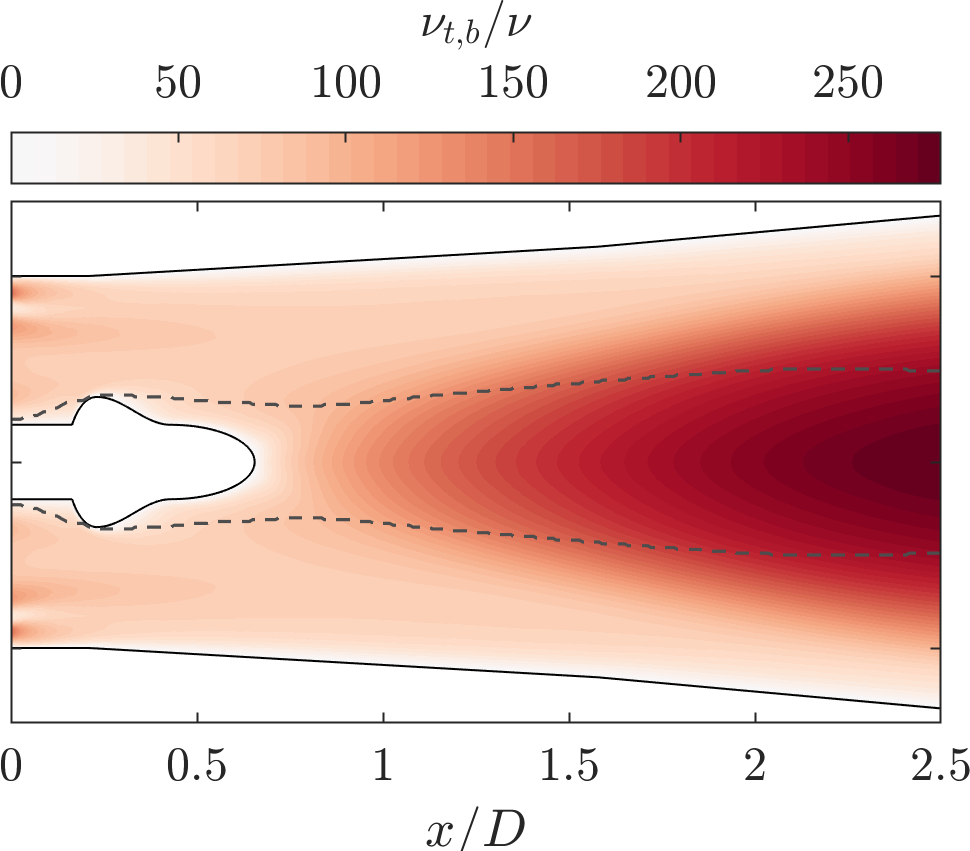}
    }
    \caption{Base-flow fields in a streamwise section for (a) axial velocity, (b) radial velocity, (c) azimuthal velocity, and (d) eddy viscosity at $Q^* = 0.65$. Gray dashed line indicates wake half width $y_{1/2}$.}
    \label{fig:meanflow_contour_Q65_RANS_with_CB_tuned}
\end{figure}

The RANS base flows are calculated for a range of flow rates between $Q^* = 0.50$ and $Q^* = 0.76$. Figure~\ref{fig:meanflow_contour_Q65_RANS_with_CB_tuned} shows the base-flow fields at $Q^* = 0.65$, exemplarily, which is at slightly unstable flow conditions. The axial velocity exhibits a substantial wake deficit region in the center of the draft tube downstream of the centerbody, while velocities remain high near the walls. The wake half width is shown as a gray dashed line in Figure~\ref{fig:meanflow_contour_Q65_RANS_with_CB_tuned}. It is defined as 
\begin{equation}
    y_{1/2}(x) = \underset{y}{\mathrm{arg}} \,
    \begin{cases}
        \pm 1/2[\mathrm{max}(u_b(x,y)) - u_b(x,0)]\textrm{, if } x/D > 0.655\\
        \pm 1/2[\mathrm{max}(u_b(x,y)) - 0]\textrm{, \qquad\quad if } x/D \leq 0.655
    \end{cases}
\end{equation}
and indicates regions of high shear due to large velocity gradients occurring between the wake-like region around the centerline and the jet-like region at the outer walls. The azimuthal velocity shows high values of swirl towards the outer walls, while low values of swirl persist around the centerline due to the vortex breakdown. The eddy viscosity is elevated downstream, reflecting enhanced turbulence production due to shear in the wake. How the flow fields compare with the reference URANS simulations is discussed in Appendix~\ref{app:3d_urans_experiments}.

\begin{figure}[tbp]
    \centering
    \includegraphics[width=0.7\columnwidth]{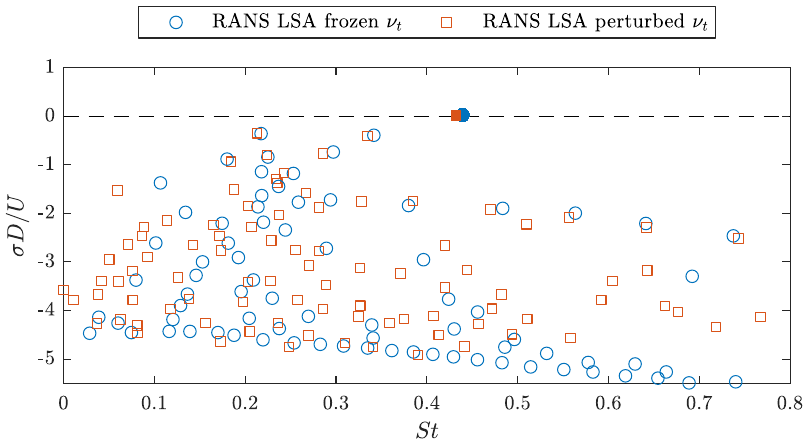}
    \caption{Eigenvalue spectrum for $m=1$ comparing frozen and perturbed eddy-viscosity model at $Q^* = 0.65$. The respective eigenvalues corresponding to the vortex-rope instability are indicated by the filled markers.}
    \label{fig:eigenvalue_spectrum_rans_tuned_frozen_and_kepslin}
\end{figure}

Using the RANS base flow at $Q^* = 0.65$, an LSA is performed for $m=1$. Figure~\ref{fig:eigenvalue_spectrum_rans_tuned_frozen_and_kepslin} shows the spectrum with normalized eigenvalues, defining a Strouhal number $St = fD/U$ and a nondimensional growth rate $\sigma D /U$. An isolated eigenvalue is found close to the stability limit that represents the helical vortex rope instability. Evidently, the perturbed eddy-viscosity model has only a minor impact on the eigenvalue of interest compared to the frozen eddy-viscosity model.

\begin{figure}[tbp]
    \centering
    \sidesubfloat[]{
    \centering
    \includegraphics[height=0.29\columnwidth]{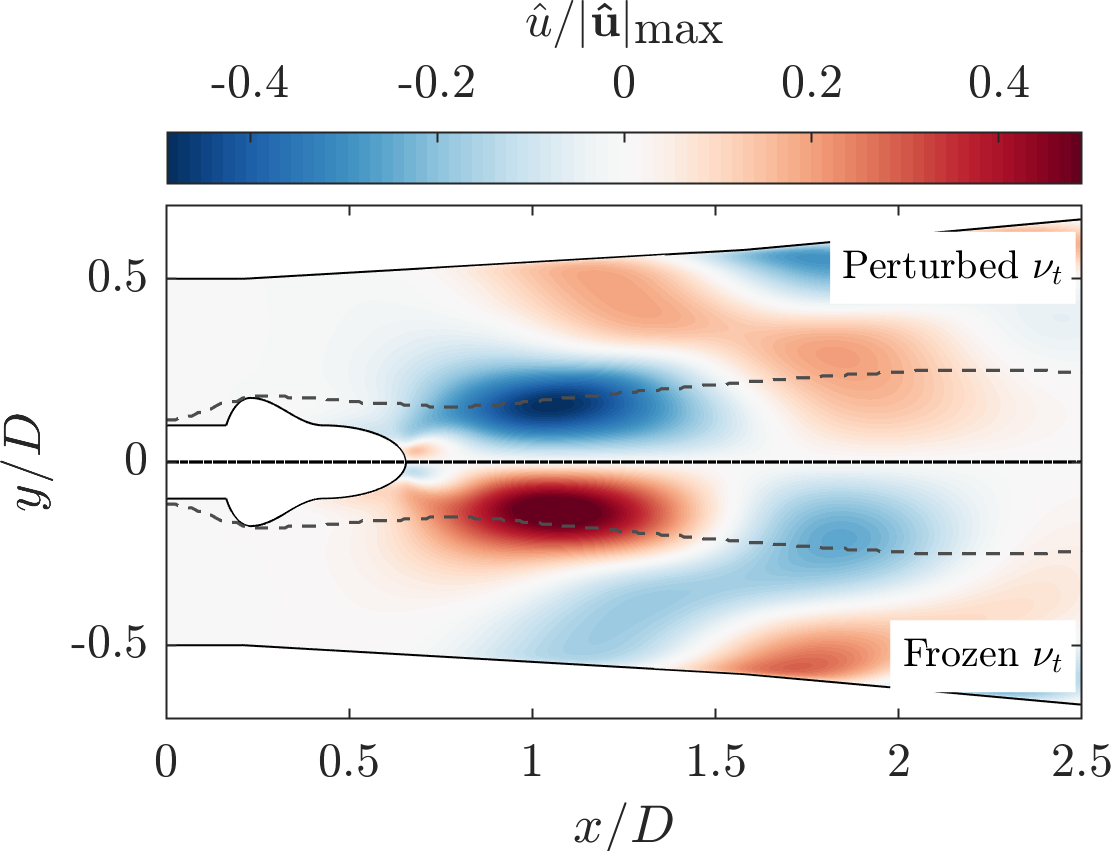}
    }
    \sidesubfloat[]{
    \centering
    \includegraphics[height=0.29\columnwidth]{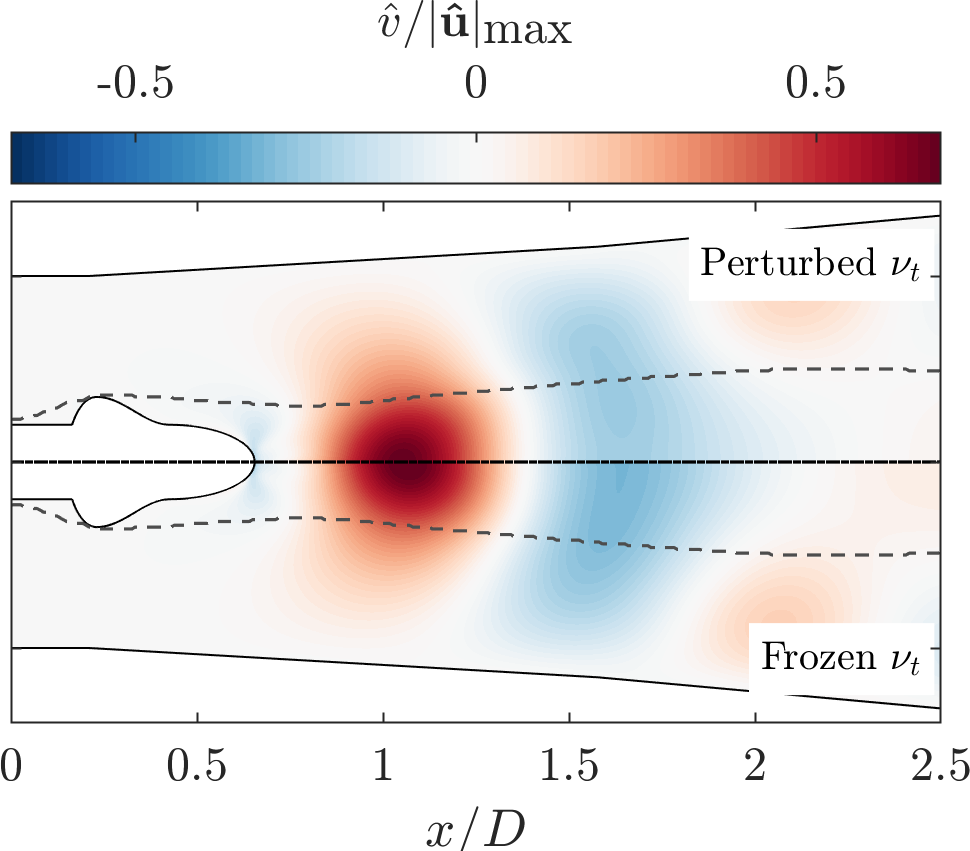}
    }\\
    \vspace{0.2cm}
    \sidesubfloat[]{
    \centering
    \includegraphics[height=0.29\columnwidth]{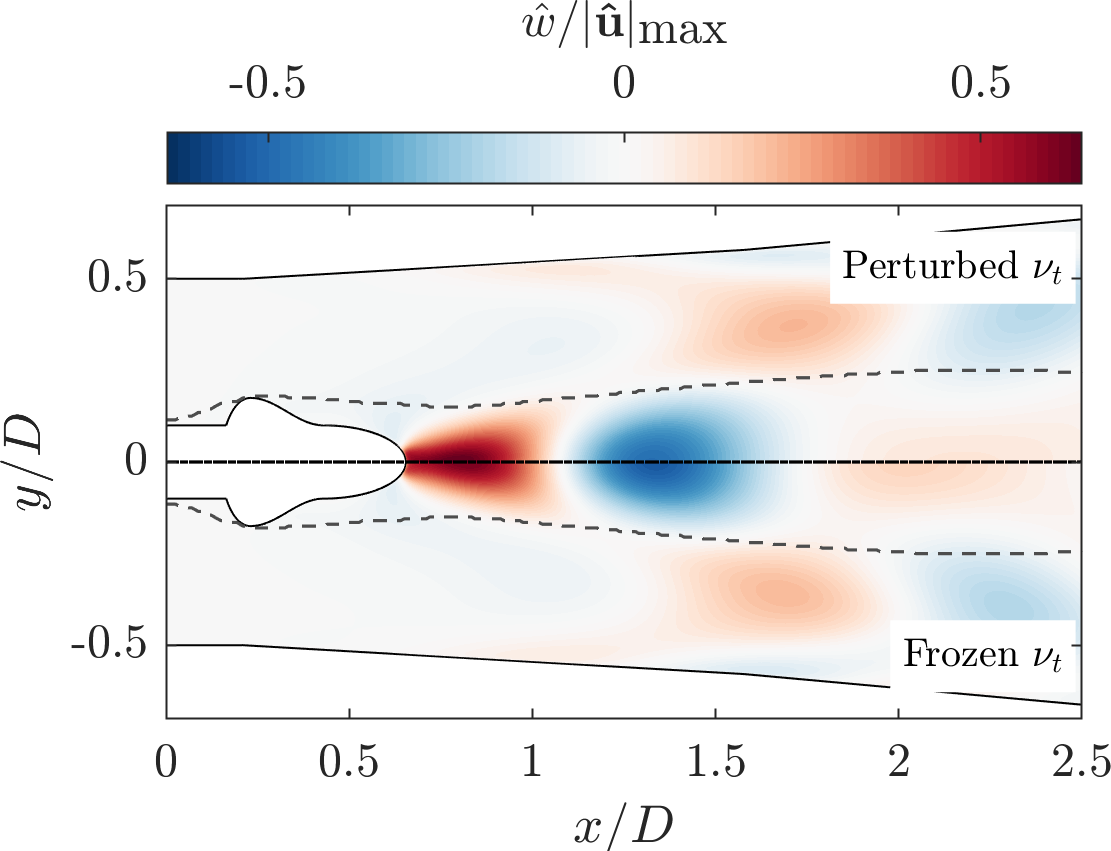}
    }
    \sidesubfloat[]{
    \centering
    \includegraphics[height=0.29\columnwidth]{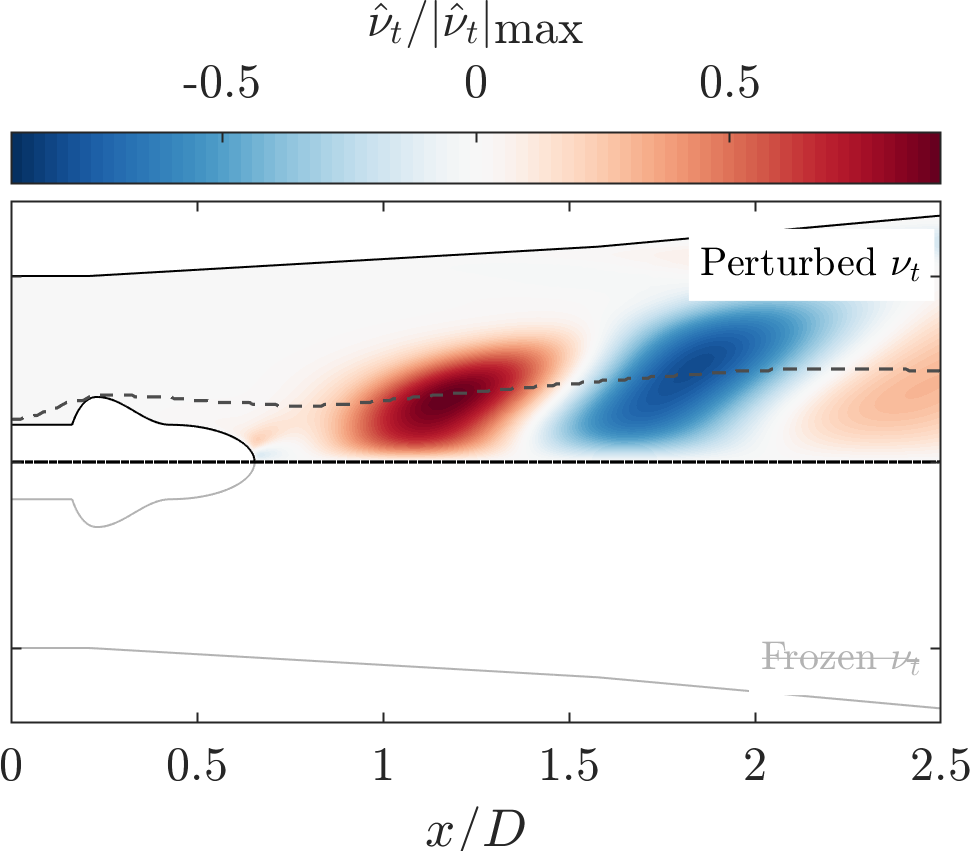}
    }
    \caption{Spatial mode shapes at arbitrary phase angle in a streamwise section comparing (top) perturbed and (bottom) frozen eddy-viscosity model. Displayed are the eigenmodes of (a) axial velocity, (b) radial velocity, (c) azimuthal velocity, and (d) eddy viscosity at $Q^* = 0.65$. Gray dashed line indicates wake half width $y_{1/2}$.}
    \label{fig:mode_streamwise_Q65_RANS_comparison}
\end{figure}

Figure~\ref{fig:mode_streamwise_Q65_RANS_comparison} compares the eigenmodes obtained with the frozen and perturbed eddy-viscosity models at $Q^* = 0.65$. All velocity $\uhat$ and eddy viscosity fluctuations $\nuthat$ are displayed, with $\nuthat$ computed \textit{a posteriori} using the algebraic relation~\eqref{eq:nutcoh}. Consistent with the observations made for the eigenvalue spectrum, the velocity mode shapes are very similar between both eddy viscosity models. The frozen model does not feature $\nuthat$ fluctuations and therefore cannot be compared in this regard. The velocity fluctuations indicate the helical vortex rope structure: the radial component shows the radial deflection from the centerline, while the azimuthal component reveals oscillations in the swirl velocity. The eddy-viscosity fluctuation is strong in the shear layers between the wake and the outer wall flow. While the largest axial velocity and eddy-viscosity fluctuations align with the shear layer (as indicated by the gray dashed line), the largest radial and azimuthal velocity fluctuations are contained within the wake region.

\begin{figure}[tbp]
    \centering
    \sidesubfloat[]{
    \centering
    \includegraphics[width=0.45\columnwidth]{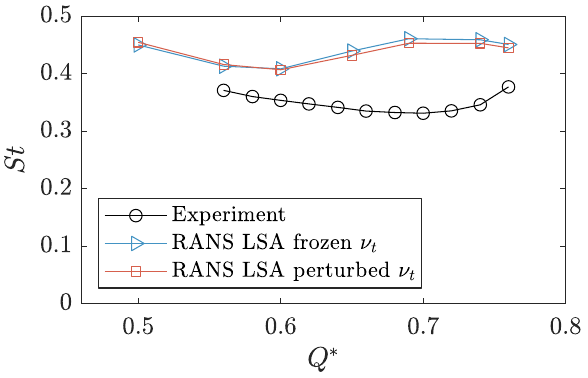}
    }
    \sidesubfloat[]{
    \centering
    \includegraphics[width=0.45\columnwidth]{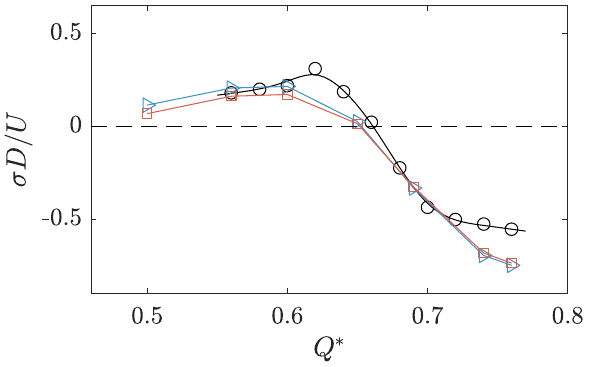}
    }
    \caption{(a) Strouhal number and (b) nondimensional growth rate of the vortex rope mode over normalized flow rate.}
    \label{fig:frequency_growthrate_over_flowrate}
\end{figure}

Repeating the LSA on the RANS base-flow fields for the entire range of considered flow rates, we can compare the Strouhal number and the nondimensional growth rate at other operating conditions. Figure~\ref{fig:frequency_growthrate_over_flowrate}(a) shows the Strouhal number as a function of the normalized flow rate, comparing RANS LSA with frozen and perturbed eddy-viscosity model against experimental measurements. In general, all cases show the same qualitative trend: the Strouhal number is approximately constant over the flow rate. However, there is a clear offset in magnitude: the RANS LSA overestimates the Strouhal number with $St \approx 0.4$ compared to the experiment with $St \approx 0.35$. Notably, in the supercritical regime for $Q^* < 0.65$, a direct match between RANS and experiment is not expected, since the RANS LSA Strouhal number corresponds to the initial oscillations around the base-flow state, whereas the experimental measurements capture the Strouhal number of the nonlinearly saturated mean-flow state at limit cycle \citep{Barkley2006,Sipp2007}. Comparing the Strouhal number of the RANS LSA with and without perturbed eddy-viscosity model, the inclusion of the linearized model has little effect for all flow rates. Figure~\ref{fig:frequency_growthrate_over_flowrate}(b) shows the nondimensional growth rate over the normalized flow rate. The zero crossing of the RANS LSA occurs at $Q^* \approx 0.65$. For $Q^* < 0.65$, the growth rate is positive, indicating global instability, with a maximum growth rate attained at $Q^* \approx 0.60$. For $Q^* > 0.65$, the growth rate becomes negative, corresponding to global stability. Again, the perturbed eddy-viscosity model has minimal impact. In the experiment, the growth rates are determined using a stochastic model \citep{Sieber2021}, which has already been successfully applied to a similar configuration in \cite{Lueckoff2022}, with further details given in Appendix~\ref{app:stochastic_model}. The growth rates determined experimentally match reasonably well with the LSA growth rates. The experimental bifurcation point is predicted at $Q^* \approx 0.66$. Far from the bifurcation point, the results of the stochastic model have to be interpreted with caution. For strong supercriticality, the weakly nonlinear assumption of the Stuart--Landau model is violated, whereas for strong subcriticality, the signal-to-noise ratio becomes very poor. Thus, only close to the bifurcation point is the growth rate estimation expected to be sufficiently valid.

In conclusion, the RANS LSA Strouhal number matches the experiment only approximately, whereas the growth rate is predicted very well, including the identification of the bifurcation point. The excellent agreement of the growth rate might be attributed to the fact that the RANS LSA models the balance between production and dissipation of coherent kinetic energy in close agreement with the experiment. In contrast, the slight mismatch of the Strouhal number might be related to the frequency selection being misrepresented in the RANS LSA. Overall, the impact of the perturbed eddy-viscosity model on the eigenvalues and eigenmodes is marginal. In the next section, we examine the impact of the perturbed eddy-viscosity model on the shape sensitivity.

\section{Impact of the perturbed eddy-viscosity model on shape sensitivity} \label{sec:results_shape_sensitivity}
We will now discuss the growth rate sensitivity with respect to wall-normal shape deformations (or, in short, `shape sensitivity'), $\mathrm{d}\sigma / \mathrm{d}\mathbf{n}$, and how it is influenced by the perturbed eddy-viscosity model. The shape sensitivity is defined with the imaginary part of Equation~\eqref{eq:shape_sensitivity_normal} and quantifies the change in the growth rate with respect to a shape deformation in the direction of the shape-normal vector pointing inward into the flow domain. Positive values, therefore, indicate that increasing the centerbody size has a destabilizing effect (growth rate increases), while decreasing the size is stabilizing (growth rate increases). Negative values, therefore, indicate opposite effects.

\begin{figure}[tbp]
    \centering
    \sidesubfloat[]{
    \centering
    \includegraphics[height=0.25\columnwidth]{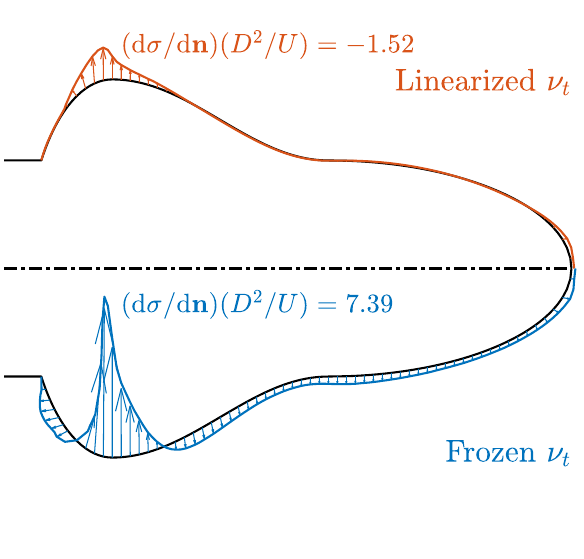}
    \label{fig:optimal_shape_shift_comparison_total}
    }
    \sidesubfloat[]{
    \centering
    \includegraphics[height=0.25\columnwidth]{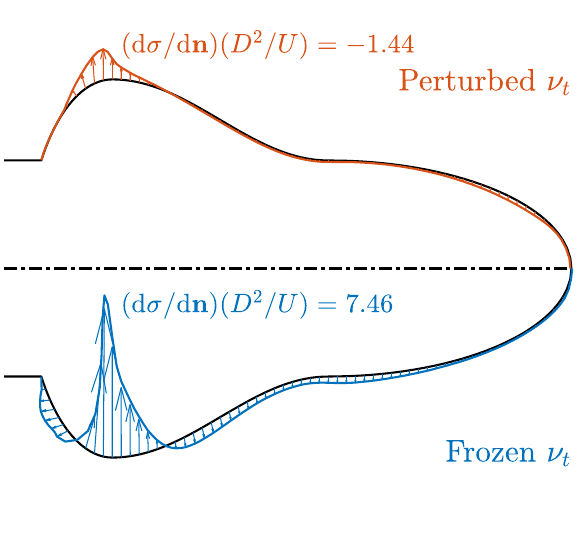}
    \label{fig:optimal_shape_shift_comparison_baseflow}
    }
    \sidesubfloat[]{
    \centering
    \includegraphics[height=0.25\columnwidth]{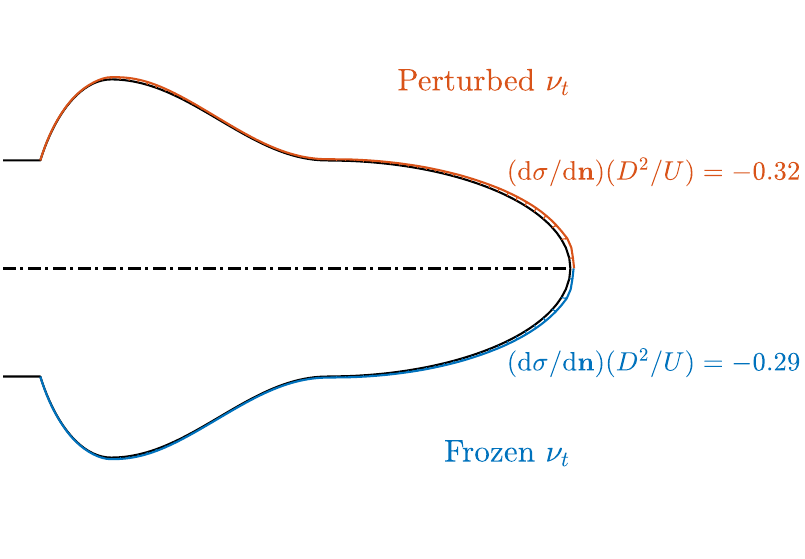}
    \label{fig:optimal_shape_shift_comparison_feedback}
    }
    \caption{Negative growth rate sensitivity with respect to wall-normal shape deformations visualized by a single-step optimal shape inducing a stabilizing effect. Displayed are (a) total sensitivity, decomposed into (b) base-flow and (c) feedback contributions, comparing (top) perturbed and (bottom) frozen eddy-viscosity model at $Q^* = 0.65$.}
    \label{fig:optimal_shape_shift_comparison}
\end{figure}

For comparing the shape sensitivity between frozen and perturbed eddy-viscosity model, we consider `single-step optimal' shape deformations. Figure~\ref{fig:optimal_shape_shift_comparison}(a) shows these optimal changes for reducing the growth rate, obtained from a single step, $\delta \mathbf{n}$, in the direction of steepest descent, $-\mathrm{d}\sigma / \mathrm{d}\mathbf{n}$, with $\delta/D = 5\cdot10^{-5}$ and $\lvert\mathbf{n}\rvert = 1$. For illustration purposes, we plot this deformed shape superposed with the baseline shape and complement it with the normalized peak value of the sensitivity for quantitative comparison. The shape sensitivity differs significantly between the two models. Although both models feature a distinct peak of very high values at the location of maximum centerbody thickness, the signs are opposite: to stabilize the flow, the perturbed eddy-viscosity model requires the centerbody to be thickened at that position, whereas the frozen eddy-viscosity model requires the centerbody to be thinned. Furthermore, the sensitivity peak values are much higher in the frozen than in the perturbed eddy-viscosity model, which would result in much larger growth rate changes when choosing the same step size. In the remaining segments of the centerbody, the shape sensitivity is comparatively low for both models. These results demonstrate that, unlike the eigenvalues and mode shapes, the inclusion of the perturbed eddy-viscosity model has a significant impact on the predicted shape sensitivity.

\subsection{Decomposition of the shape sensitivity into physical mechanisms}
To better understand the origin of the discrepancies between the frozen and perturbed eddy-viscosity models, the shape sensitivity is examined in more detail by decomposing it into base-flow and feedback contributions (see Equations~\eqref{eq:shape_sensitivity_feedback} and \eqref{eq:shape_sensitivity_baseflow}). Figure~\ref{fig:optimal_shape_shift_comparison}(b,c) displays the base-flow and feedback contribution for both eddy-viscosity models, again visualized by a single-step optimal shape. Evidently, at the position of maximum thickness, the peak sensitivity is primarily driven by the base-flow contribution, meaning that the largest values of shape sensitivity are mainly due to sensitivity in the base flow. Like in the total sensitivity of Figure~\ref{fig:optimal_shape_shift_comparison}(a), the signs are opposed between both eddy-viscosity models. The feedback contribution is comparatively low and peaks close to the centerbody's tip, implying that modifications of the feedback mechanism through shape changes are mainly localized in this region. Since the base-flow contribution is responsible for the highest shape sensitivity values, the following discussion will focus primarily on this contribution.

\begin{figure}[tbp]
    \centering
    \sidesubfloat[]{
    \centering
    \includegraphics[height=0.35\columnwidth]{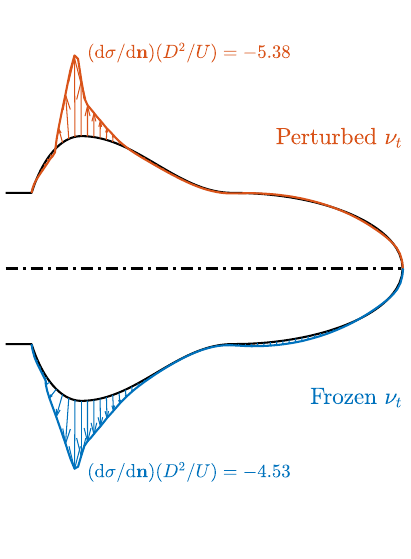}
    }
    \sidesubfloat[]{
    \centering
    \includegraphics[height=0.35\columnwidth]{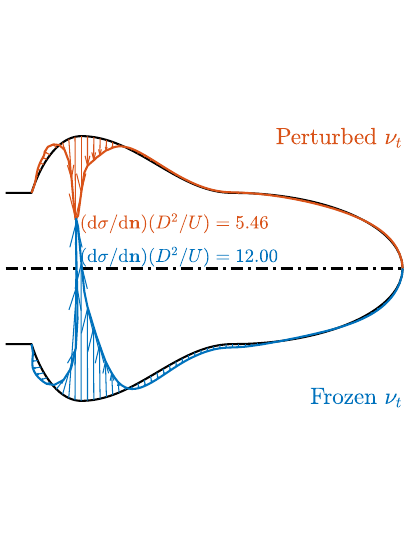}
    }
    \sidesubfloat[]{
    \centering
    \includegraphics[height=0.35\columnwidth]{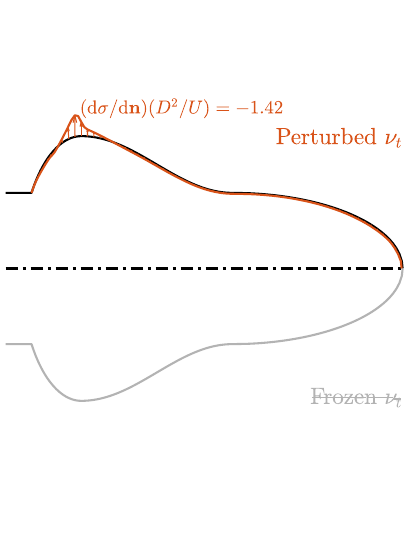}
    }\\
    \sidesubfloat[]{
    \centering
    \includegraphics[height=0.35\columnwidth]{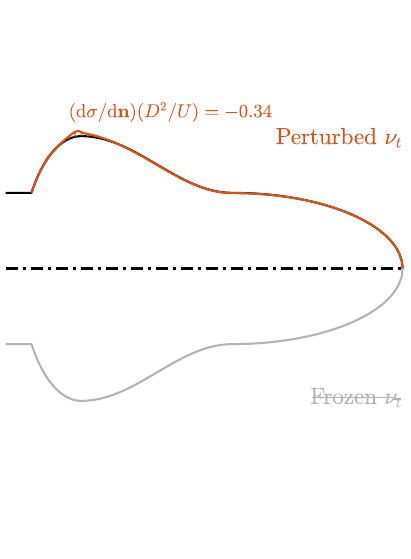}
    }
    \sidesubfloat[]{
    \centering
    \includegraphics[height=0.35\columnwidth]{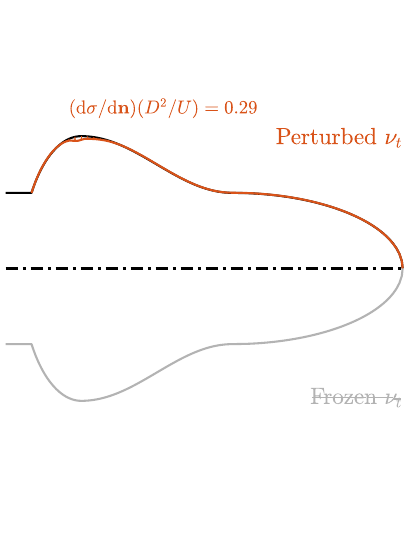}
    }
    \caption{Base-flow contribution of the negative growth rate sensitivity with respect to wall-normal shape deformations, decomposed into individual terms, visualized by a single-step optimal shape inducing a stabilizing effect. Displayed are terms associated with (a) $\uhat$ production, (b) $\uhat$ advection, (c) $\uhat$ diffusion, (d) $\khat$ production, and (e) $\epshat$ production, comparing (top) perturbed and (bottom) frozen eddy-viscosity model at $Q^* = 0.65$.}
    \label{fig:optimal_shape_shift_baseflow_decomposed_comparison}
\end{figure}

The base-flow contribution can be decomposed into individual contributing terms of the linearized RANS equations (see Equation~\eqref{eq:shape_sensitivity_baseflow_decomposed}), each corresponding to a distinct physical mechanism. This decomposition provides a direct link between flow physics and shape sensitivity, allowing us to identify how specific mechanisms respond to shape changes. Figure~\ref{fig:optimal_shape_shift_baseflow_decomposed_comparison} illustrates this decomposition for the five most dominant terms. Using the frozen eddy-viscosity model, only the production and advection terms of $\hat{\mathbf{u}}$ in the momentum equations contribute. The most sensitive regions again lie near the maximum thickness of the centerbody, where shape changes affect production and advection with competing impacts. This implies that thickening yields a stabilizing response through the production term, while thinning yields a stabilizing response through the advection term, with the advection term being much more sensitive. For the perturbed eddy-viscosity model, the sensitivity trends of $\uhat$ production and advection are comparable, although the contribution of the advection term is significantly reduced compared to the frozen eddy-viscosity model. In addition to these two terms, ten further terms arise, introducing ten extra physical mechanisms that can be manipulated by shape changes, with $\uhat$ diffusion, $\khat$ production, and $\epshat$ production being the most significant among the additional terms. As shown in Figure~\ref{fig:optimal_shape_shift_baseflow_decomposed_comparison}, the $\uhat$ diffusion and $\khat$ production terms have a stabilizing effect when thickening at the point of maximum thickness, whereas the $\epshat$ production term has a stabilizing effect when thinning.

Thus, although the main mechanisms remain $\uhat$ advection and production, the additional sensitivity terms in the perturbed eddy-viscosity model shift the balance among competing contributions compared to the frozen eddy-viscosity model, explaining the discrepancies observed between both models. In order to develop a more phenomenological understanding of the mechanisms by which they affect the growth rate, the following section examines the optimal and actual base-flow changes associated with the individual terms.

\subsection{The role of base-flow modifications in stabilizing the flow}
All the shape sensitivities previously examined quantify how the growth rate changes when the shape is deformed. To get a more phenomenological perspective, the base-flow contribution of the shape sensitivity can be expressed as the product of two factors as defined in Equation~\eqref{eq:shape_sensitivity_baseflow_decomposed}. These two factors are the growth rate sensitivity with respect to base-flow modifications, $\partial \sigma / \partial \qfull_b$, and the base-flow sensitivity with respect to shape deformations, $\mathrm{d}\qfull_b / \mathrm{d}\ai$. In the following, we first investigate the growth rate sensitivity to base-flow modifications, which indicates the optimal base-flow modifications required to obtain a stabilizing effect. Again, we restrict the analysis to the dominant terms identified in Figure~\ref{fig:optimal_shape_shift_baseflow_decomposed_comparison}, i.e.\ $\uhat$ production, $\uhat$ advection, $\uhat$ diffusion, $\khat$ production, and $\epshat$ production. While, in principle, the growth rate sensitivity can be evaluated with respect to any base-flow variable, only the terms most relevant in magnitude are considered here.

\begin{figure}[tbp]
    \centering
    \sidesubfloat[]{
    \centering
    \includegraphics[height=0.29\columnwidth]{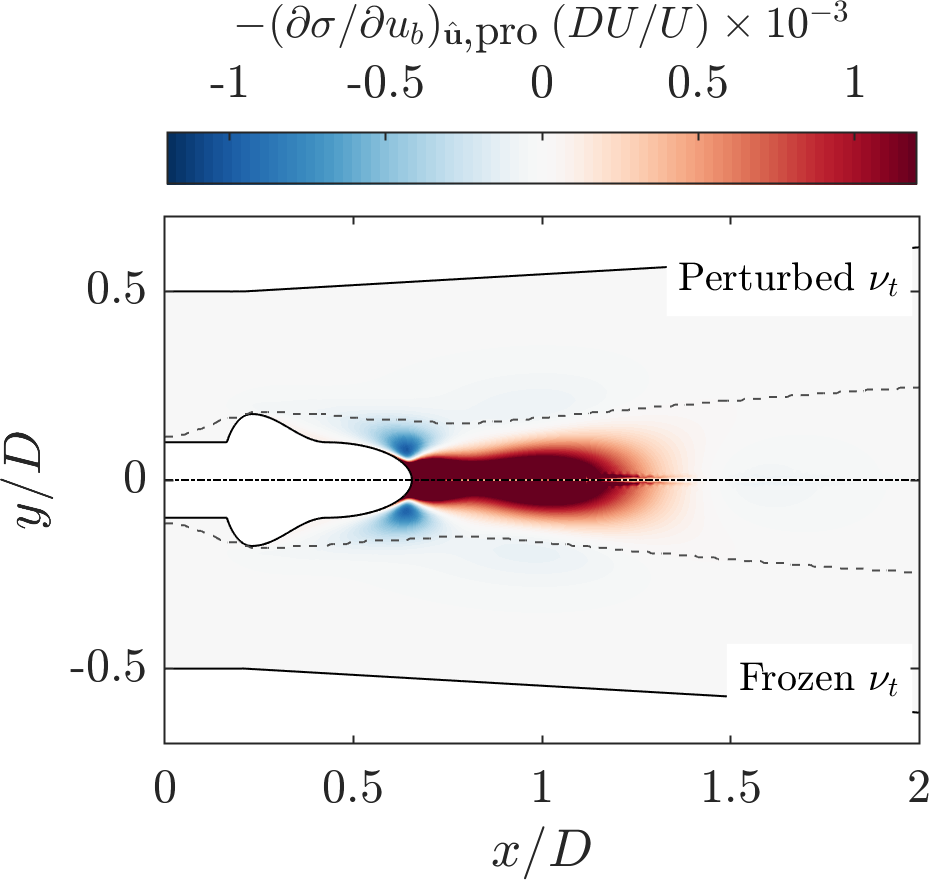}
    }
    \sidesubfloat[]{
    \centering
    \includegraphics[height=0.29\columnwidth]{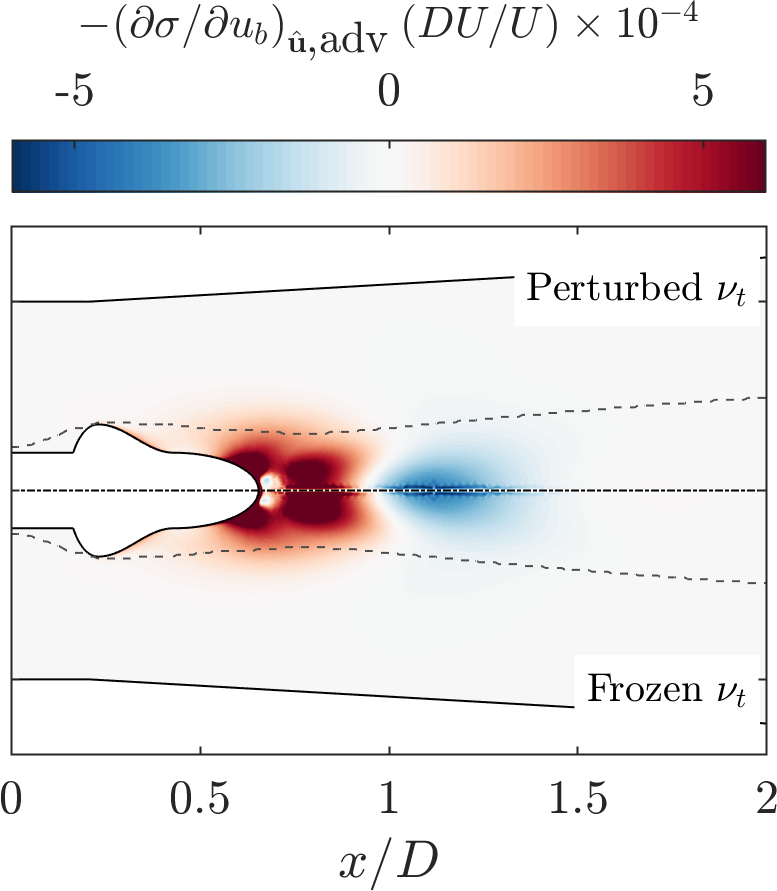}
    }
    \sidesubfloat[]{
    \centering
    \includegraphics[height=0.29\columnwidth]{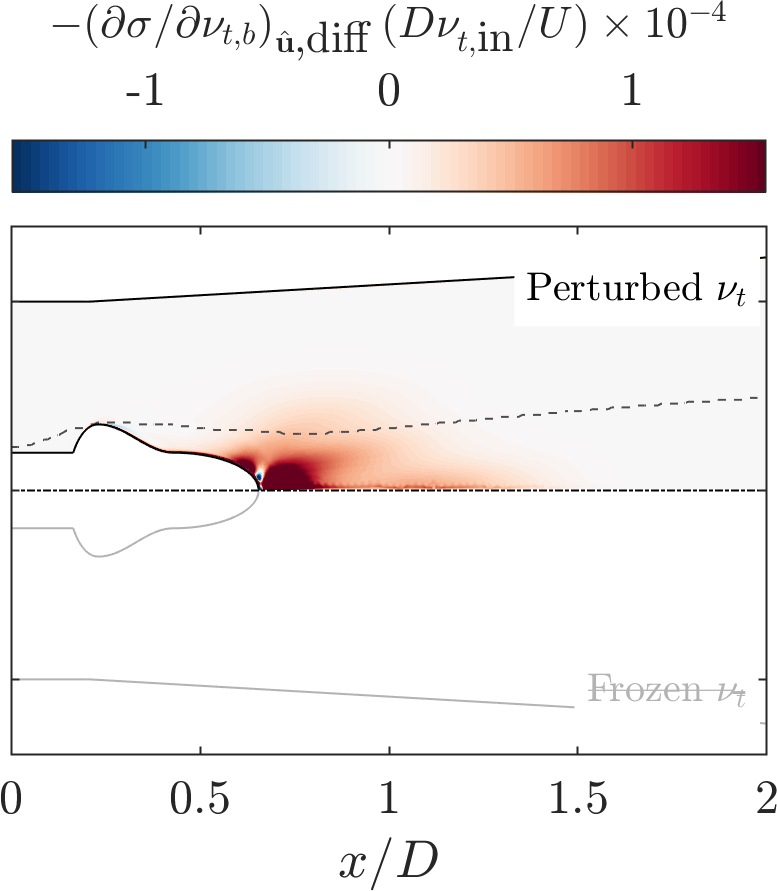}
    }\\
    \vspace{0.2cm}
    \sidesubfloat[]{
    \centering
    \includegraphics[height=0.29\columnwidth]{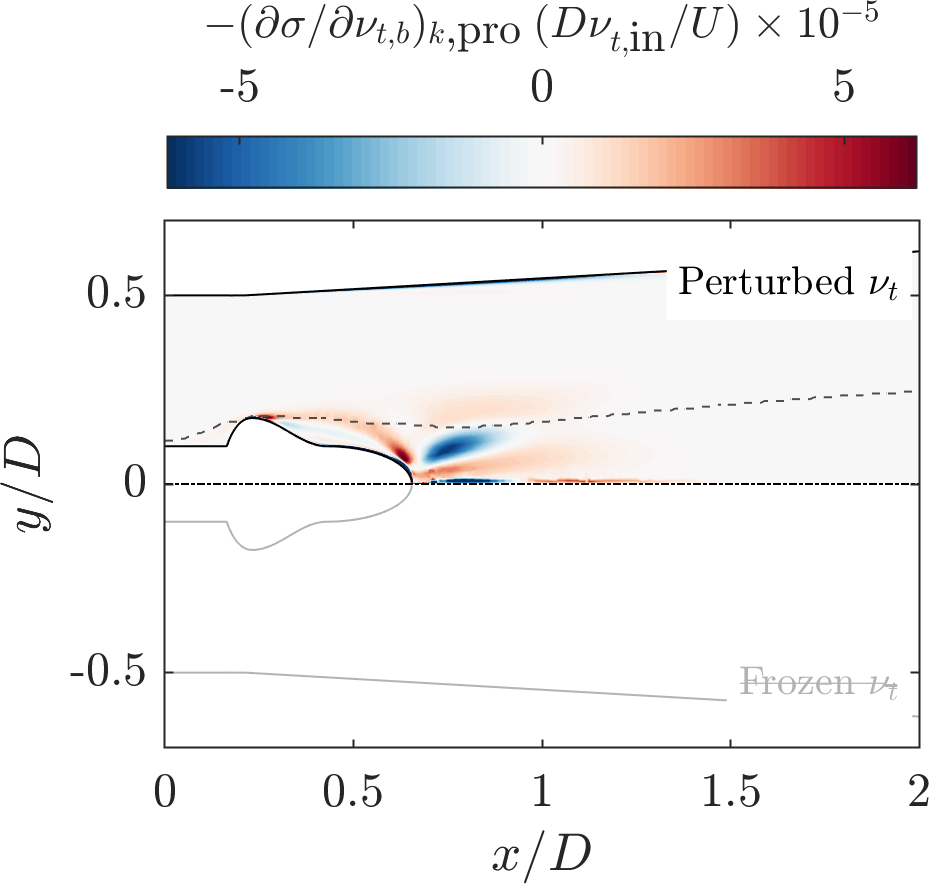}
    }
    \sidesubfloat[]{
    \centering
    \includegraphics[height=0.29\columnwidth]{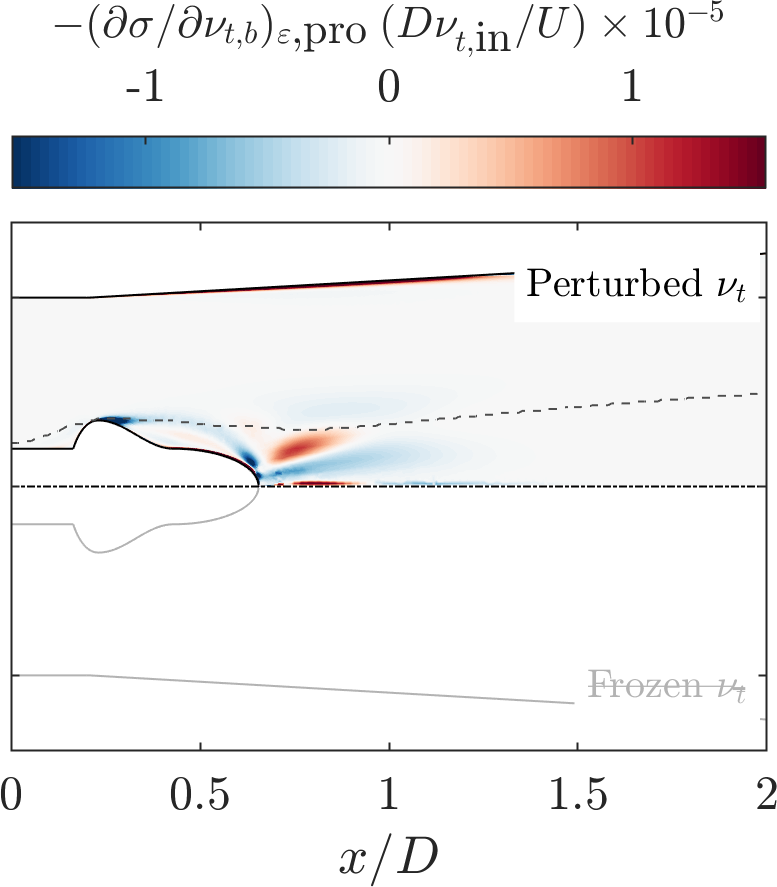}
    }
    \caption{Negative growth rate sensitivity with respect to base-flow modifications. Displayed are terms associated with (a) $\uhat$ production, (b) $\uhat$ advection, (b) $\uhat$ diffusion, (d) $\khat$ production, and (e) $\epshat$ production, comparing (top) perturbed and (bottom) frozen eddy-viscosity model at $Q^* = 0.65$. Gray dashed line indicates wake half width $y_{1/2}$.}
    \label{fig:growth_rate_sensitivity_to_base_flow_change}
\end{figure}

Figure~\ref{fig:growth_rate_sensitivity_to_base_flow_change} presents the five sensitivity terms for both the frozen and perturbed eddy-viscosity models, premultiplied with $-1$. Figure~\ref{fig:growth_rate_sensitivity_to_base_flow_change}(a,b) depicts the sensitivity of the $\uhat$ production and $\uhat$ advection terms for both models, quantifying the growth rate change to modifications in the axial base-flow velocity. The similarity between the two models suggests that the underlying physical mechanisms are captured in a comparable manner. The negative values of the $\uhat$ production sensitivity term around the centerbody indicate that reducing the axial velocity in this region stabilizes the flow. Conversely, positive values downstream of the centerbody’s tip reveal that increasing the axial velocity promotes stabilization. For the advection term, the sensitivity is mainly positive both around the centerbody and downstream of the maximum thickness. Thus, essentially, an increase in axial velocity near the centerbody tip counteracts the wake deficit in this region, presumably weakening the intrinsic feedback sustained by upstream-traveling waves.

Figure~\ref{fig:growth_rate_sensitivity_to_base_flow_change}(c-e) displays the sensitivity of the $\uhat$ diffusion, $\khat$ production, and $\epshat$ production terms for the perturbed eddy-viscosity model, now quantifying the growth rate change with respect to variations in the eddy viscosity. These derivatives are obtained via the chain rule (see Equation~\eqref{eq:egv_sensitivity_due_to_nut}). These terms are absent in the frozen eddy-viscosity model, meaning that the corresponding mechanisms are completely missing in that approach. The $\uhat$ diffusion term in Figure~\ref{fig:growth_rate_sensitivity_to_base_flow_change}(c) indicates that an increase in eddy viscosity around the centerbody tip yields a stabilizing effect. This can be interpreted as an increase of dissipation for the mode. The $\khat$ and $\epshat$ production sensitivities, shown in Figures~\ref{fig:growth_rate_sensitivity_to_base_flow_change}(d,e), are closely related to the eddy viscosity (see Equation~\eqref{eq:nut_baseflow}). While their spatial patterns are very similar, their signs are inverted due to the relations $\nu_{t,b} \propto k_b^2$ and $\nu_{t,b} \propto 1/\eps_b$. Specifically, the $\khat$ production term suggests that increasing eddy viscosity along the concave section of the centerbody stabilizes the flow, whereas the $\epshat$ production term implies a destabilizing effect.

It is important to note that the optimal base-flow modifications identified here are not necessarily realizable by any feasible shape deformation. For this reason, we now turn to the second factor of Equation~\eqref{eq:shape_sensitivity_baseflow_decomposed}, namely the base-flow sensitivity to shape changes, i.e.\ how the base flow itself responds when the shape is deformed.

\begin{figure}[tbp]
    \centering
    \sidesubfloat[]{
    \centering
    \begin{tikzpicture}
        \begin{scope}
            \node[anchor=south west,inner sep=0] (image) at (0,0) {\includegraphics[height=0.29\columnwidth]{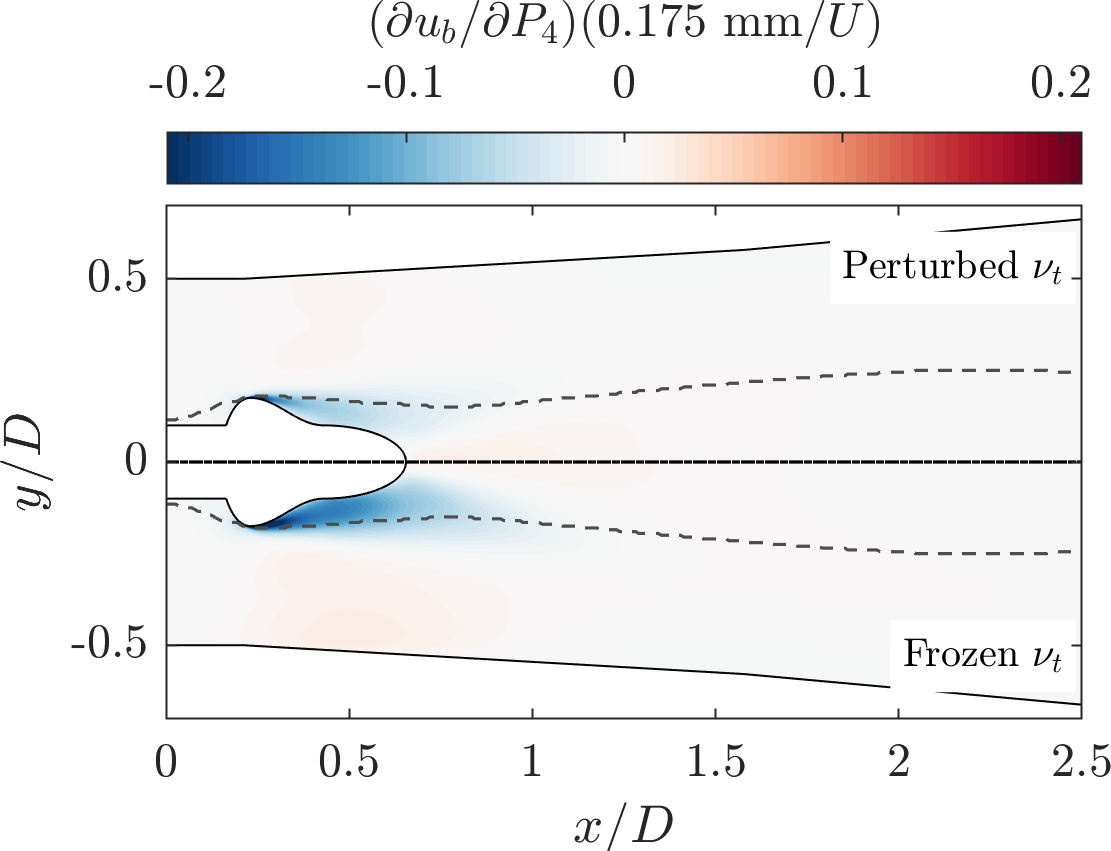}};
            \begin{scope}[x={(image.south east)},y={(image.north west)}]
                \draw [{Stealth[scale=0.8]}-{Stealth[scale=0.8]}, very thin, black] (0.228,0.455) -- ++(0,0.077)
                node [above=2.5pt,midway,black] {\scriptsize $P_4$};
            \end{scope}
        \end{scope}
    \end{tikzpicture}
    }
    \sidesubfloat[]{
    \centering
    \begin{tikzpicture}
        \begin{scope}
            \node[anchor=south west,inner sep=0] (image) at (0,0) {\includegraphics[height=0.2883\columnwidth]{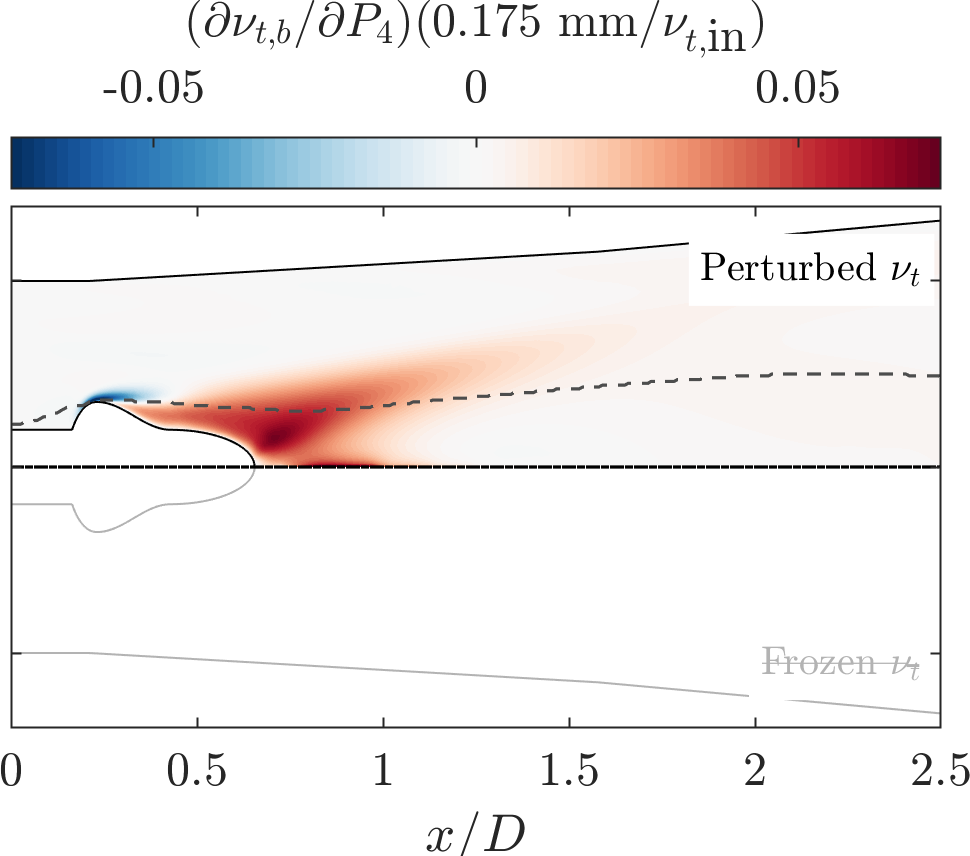}};
            \begin{scope}[x={(image.south east)},y={(image.north west)}]
                \draw [{Stealth[scale=0.8]}-{Stealth[scale=0.8]}, very thin, black] (0.102,0.455) -- ++(0,0.077)
                node [above=2.5pt,midway,black] {\scriptsize $P_4$};
            \end{scope}
        \end{scope}
    \end{tikzpicture}
    }
    \caption{Base-flow sensitivity with respect to changes in the shape parameter $P_4$ that controls the maximum thickness of the centerbody, as defined in \cite{Lueckoff2022}. Displayed are (a) axial velocity, and (b) eddy viscosity, comparing (top) perturbed and (bottom) frozen eddy-viscosity model. Gray dashed line indicates wake half width $y_{1/2}$ at $Q^* = 0.65$.}
    \label{fig:base_flow_sensitivity_decomposed}
\end{figure}

In order to better discuss the base-flow sensitivity to shape changes, we turn to a more `integral' parameterization to better connect shape changes with flow physics. The sensitivities discussed so far were based on a B-spline parameterization with 126 B-spline control points, allowing virtually continuous and highly localized shape deformations. Such localized changes also induce localized base-flow and feedback responses, which are not always straightforward to interpret. To facilitate more global modifications and thus ease interpretation, we follow \cite{Lueckoff2022} and define the centerbody shape by two half-ellipses connected with an arcus cosine function that are described by four parameters, with $P_4$ controlling the maximum thickness. Since this `bulge' has been identified as the most impactful on the growth rate, we focus on changes in $P_4$ to interpret the previously observed results. Figure~\ref{fig:base_flow_sensitivity_decomposed}(a) shows the change in the axial base-flow velocity with respect to a $\SI{0.175}{\mm}$ increase of $P_4$ (which equals 1\% of the baseline value of $P_4$). In the frozen eddy-viscosity model, the axial velocity strongly decreases, indicating an increased wake deficit downstream of the bulge. This increase in wake deficit would have a stabilizing effect through the production term and a destabilizing effect through the advection term according to Figure~\ref{fig:growth_rate_sensitivity_to_base_flow_change}(a,b), which also aligns with the observations made in Figure~\ref{fig:optimal_shape_shift_baseflow_decomposed_comparison}(a,b). This is consistent with the role of wakes in promoting global instabilities through upstream-traveling waves. In the perturbed eddy-viscosity model, an increase in wake deficit is also observed but is much weaker, and thus the destabilizing effect is reduced. At the same time, additional terms introduced by the model lead to further base-flow modifications. Notably, Figure~\ref{fig:base_flow_sensitivity_decomposed}(b) shows that the base-flow eddy viscosity increases substantially with increasing thickness, attributable to enhanced shear and production of $k$, which yields higher $\nu_t$. This additional eddy viscosity has a strongly stabilizing effect by damping coherent fluctuations, as predicted in Figure~\ref{fig:growth_rate_sensitivity_to_base_flow_change}(c), counterbalancing the destabilizing wake contribution and ultimately producing a net stabilizing response in the perturbed eddy-viscosity model in contrast to the frozen model.

With these observations, we can provide a hypothesis why the impact of the perturbed eddy-viscosity model is small on the eigenvalues but large on the sensitivities. The determination of the eigenvalue is governed by the linearized RANS operator, as defined in~\eqref{eq:LRANS_operator} for the perturbed eddy-viscosity model and as defined in~\eqref{eq:LRANS_operator_frozen} for the frozen eddy-viscosity model. From a mathematical point of view, the marginal influence of the additional linearized $k$--$\eps$ equations on the eigenvalue suggests a very weak coupling between the linearized momentum equations and the additional linearized $k$--$\eps$ equations, such that the latter behave almost as quasi-decoupled equations. However, the eigenvalue sensitivity is not determined by the linearized equations, but by perturbations to the linearized equations (see Equation~\eqref{eq:shape_sensitivity}), and part of these perturbations are described by the Hessian operator (see Equation~\eqref{eq:shape_sensitivity_baseflow}). This Hessian operator quantifies the base-flow modifications. As the results demonstrate, it is this strong sensitivity to base-flow modifications that dominate the eigenvalue sensitivity. Therefore, in other words, the linearized equations are sensitive to base-flow modifications (which affect the shape sensitivity), but they are not sensitive to the introduction of additional $k$--$\eps$ equations (which therefore have virtually no effect on the eigenvalue and mode shape).

\subsection{Comparison of shape sensitivity with experimental measurements} \label{sec:shape_sensitivity_comparison_experiment}
\begin{figure}[tbp]
    \centering
    \includegraphics[width=0.45\columnwidth]{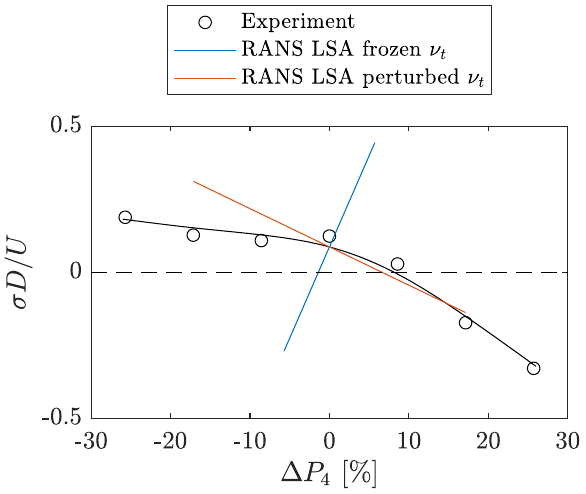}
    \caption{Nondimensional growth rate over shape parameter $P_4$ that controls the maximum thickness of the centerbody, comparing experimentally determined growth rates with predicted linear growth rate gradients using the RANS LSA framework at $Q^* = 0.65$.}
    \label{fig:shape_sensitivity_experimental_comparison_Q65}
\end{figure}

So far, we have viewed the shape sensitivity through the RANS LSA framework only. In the following, we will briefly `validate' the shape sensitivity results with experimental measurements. Figure~\ref{fig:shape_sensitivity_experimental_comparison_Q65} shows the normalized growth rate as a function of the relative change in $P_4$. The experimentally determined growth rates are compared with the predicted linear growth rate changes obtained from both eddy-viscosity approaches. Consistent with the prediction of the perturbed eddy-viscosity model, an increase in $P_4$ in the experiment, corresponding to a local thickening of the centerbody at the bulge, results in a stabilization of the flow. The linear trend predicted by the model reproduces the local experimental slope reasonably well. In contrast, the frozen eddy-viscosity model predicts an entirely opposite effect, suggesting that a thinning of the centerbody at the bulge would stabilize the flow.

These findings indicate that the underlying physical mechanisms governing the instability are reproduced well by the RANS-based linear stability framework, but only when the eddy-viscosity model is linearized to capture the influence of shape deformation on the background turbulent field. The consistency between experimental and linearized predictions is encouraging, as it demonstrates that the combined RANS LSA and shape sensitivity framework provides a physically meaningful basis for designing improved Francis turbine draft-tube geometries aimed at extending the range of stable operating conditions at part-load flow rates.

\section{Conclusion}
\label{sec:conclusion}

In this study, a physics-based framework is laid out to control the global vortex-rope instability in the turbulent draft tube flow of a Francis turbine model via shape modifications. The framework was based on the linearized axisymmetric Reynolds-averaged Navier--Stokes equations equipped with a $k$--$\eps$ turbulence model. The corresponding base flow was tuned by adapting the inlet boundary conditions in order to match the bifurcation point of the vortex-rope instability determined from three-dimensional unsteady simulations. Linear stability and shape sensitivity analyses were then performed on these base flows, comparing a frozen and a perturbed eddy-viscosity model. The frequency, growth rate and shape sensitivity were validated against experiments.

Several conclusions can be drawn from this investigation. First, the influence of the perturbed eddy-viscosity model on the eigenvalue and eigenmode of the vortex-rope instability was shown to be marginal, indicating that effects related solely to the linear operator and the feedback mechanism remain largely unaffected. This is hypothesized to be attributed to quasi-decoupled linearized $k$--$\eps$ equations. In contrast, the shape sensitivities of the growth rate exhibit significant differences between the two modeling approaches. The frozen eddy-viscosity model yields sensitivities that deviate substantially from the perturbed eddy-viscosity model. A closer inspection of the underlying physical mechanisms responsible for the predicted growth rate changes revealed that the base-flow contribution, encoded in the Hessian operator, dominates the total sensitivity. Both eddy-viscosity models consistently indicate that the production and advection terms of the momentum equation are the primary contributors to shape sensitivity. However, the perturbed eddy-viscosity model captures additional physical mechanisms that significantly influence the eddy-viscosity change when the shape is deformed. These mechanisms are entirely neglected in the frozen model. Interestingly, a comparison with experimental measurements demonstrated that only the perturbed eddy-viscosity model reproduces the correct trends in the growth rate sensitivity, whereas the frozen model fails to do so.

Future work should expand on the influence of different turbulence models, as these may emphasize physical mechanisms in different ways \citep{Brewster2019} and thus alter the resulting optimal shape predictions. Moreover, the present analysis should be extended to other flow configurations exhibiting global instabilities, as well as to noise-driven convective instabilities, where the interplay between frozen and perturbed eddy-viscosity models may manifest differently.

Overall, our results highlight the success of predicting shape sensitivity from linearized Reynolds-averaged Navier--Stokes equations, opening new possibilities for gradient-based shape optimization for stability control in fully turbulent environments. The results demonstrate the importance of including a perturbed eddy-viscosity model when aiming for an accurate sensitivity analysis.

\section*{Acknowledgments}
The funding from the \textit{Deutsche Forschungsgemeinschaft} (DFG, German Research Foundation) within project number 429772199 is gratefully acknowledged. The authors express their gratitude to Kai Hildebrandt and Frederic Bergner for their contributions to the shape sensitivity framework.

\section*{Declaration of interests}
The authors report no conflict of interest.

\section*{Declaration of AI assistance}
During the preparation of this manuscript, large-language-model-based AI tools were used to assist with language editing and stylistic refinement. All scientific content and interpretations remain the responsibility of the authors.

\appendix
\section{Mathematical operators}\label{app:operators}
\subsection{Modified $\nabla$ operator in the homogeneous direction}\label{app:operators_nabla}
The spatial derivatives in the $\nabla$ operator are performed analytically in the homogeneous direction of $\theta$. For a scalar quantity $\alpha$, we define the modified gradient operator as
\begin{equation}
    \nabla \alpha = \begin{bmatrix}
        \partial \alpha / \partial x\\
        \partial \alpha / \partial r\\
        -(\imag m/r) \alpha
    \end{bmatrix}.
\end{equation}
For a vector quantity $\mathbf{v} = [v_x, v_r, v_\theta]^T$, we define the modified gradient operator as
\begin{equation}
    \nabla \mathbf{v} = \begin{bmatrix}
        \partial v_x / \partial x & \partial v_x / \partial r & -(\imag m/r) v_x\\
        \partial v_r / \partial x & \partial v_r / \partial r & -(\imag m/r) v_r - v_\theta/r\\
        \partial v_\theta / \partial x & \partial v_\theta / \partial r & -(\imag m/r) v_\theta + v_r/r
    \end{bmatrix} .
\end{equation}
The modified divergence operator we define as
\begin{equation}
    \nabla \cdot \mathbf{v} = \frac{\partial v_x}{\partial x} + \frac{1}{r}\frac{\partial (rv_r)}{\partial r} - \frac{\imag m}{r} v_\theta .
\end{equation}
For a matrix quantity $\mathbf{T}$, we define the modified divergence operator as
\begin{equation}
    \nabla \cdot \mathbf{T} = \begin{bmatrix}
        \partial T_{xx}/\partial x + \partial T_{xr}/\partial r  -(\imag m/r) T_{x\theta} + T_{xr}/r \\
        \partial T_{rx}/\partial x + \partial T_{rr}/\partial r - (\imag m/r) T_{r\theta} + (T_{rr} - T_{\theta\theta})/r \\
        \partial T_{\theta x}/\partial x + \partial T_{\theta r}/\partial r - (\imag m/r) T_{\theta\theta} + (T_{\theta r} + T_{r\theta})/r
    \end{bmatrix} .
\end{equation}
In all the spatial derivatives given in this work, the $\nabla$ operator is implicitly defined as stated above. When acting on an eigenmode $\hat{(\cdot)}$, the azimuthal wavenumber is set to $m=1$. When acting on a base-flow field $(\cdot)_b$, the azimuthal wavenumber is set to $m=0$.

\subsection{Hessian operator}\label{app:operators_hessian}
For the perturbed eddy-viscosity model, the Hessian operator is defined as $\mathcal{H}(\qfull_b) = \partial \mathcal{L}(\qfull_b) / \partial \qfull_b$ with $\mathcal{L}(\qfull_b) = (\partial \mathcal{N} / \partial \qfull)_{\qfull=\qfull_b}$. Thus, the Hessian operator can also be written as
\begin{equation}
    \mathcal{H}(\qfull_b) = \frac{\partial^2 \mathcal{N}(\qfull)_{\qfull=\qfull_b}}{\partial\qfull\partial\qfull_b} \, ,
\end{equation}
meaning that the Hessian operator is a second-order derivative along two different directions that acts on two different fields. The Hessian operator in this work is thus a bilinear operator and is almost always contracted with two different fields (except in Equation~\eqref{eq:shape_sensitivity_with_respect_to_base_flow}). In that light, we define the Hessian operator contracted with an eigenmode $\qhat$ and a perturbed base-flow field $\qfull_b'$. Due to the homogeneous direction in $\theta$, we again work with a modified $\nabla$ operator as given in \ref{app:operators_nabla}. The modified Hessian operator reads
\begin{equation}
    \mathcal{H}(\qfull_b,m)\qhat\qfull_b' = \frac{\partial \mathcal{L}(\qfull_b,m)}{\partial \qfull_b}\qhat\qfull_b' = \begin{bmatrix}
        \mathcal{H}_\ufull(\qfull_b)\qhat\qfull_b' \\
        0 \\
        \mathcal{H}_k(\qfull_b)\qhat\qfull_b' \\
        \mathcal{H}_\eps(\qfull_b)\qhat\qfull_b'
    \end{bmatrix},
\end{equation}
with
\begin{subequations}
    \begin{equation}
        \mathcal{H}_\ufull(\qfull_b,m) \qhat \qfull_b'
        =
        \underbrace{(\uhat \cdot \nabla)\ufull_b'}_{\textrm{(a)}}
        + \underbrace{(\ufull_b' \cdot \nabla)\uhat}_{\textrm{(b)}} \\
        \underbrace{- 2 \nabla \cdot (\nuthat \strain_b' + \nuthat' \strain_b
        + \nu_{t,b}' \strainhat)}_{\textrm{(c)}} ,
    \end{equation}
    \begin{equation}
        \begin{split}
            \mathcal{H}_k(\qfull_b,m) \qhat \qfull_b'
            =
            \uhat \cdot \nabla k_b'
            + \ufull_b' \cdot \nabla \khat
            - \nabla \cdot \left( \frac{\nuthat'}{\sigma_k} \nabla k_b + \frac{\nuthat}{\sigma_k} \nabla k_b' + \frac{\nu_{t,b}'}{\sigma_k} \nabla \khat \right) \\
            \underbrace{- 2 \nuthat' \strain_b : \strain_b
            - 4 \nuthat \strain_b : \strain_b'
            - 4 \nu_{t,b}' \strainhat : \strain_b
            - 4 \nu_{t,b} \strainhat : \strain_b'}_{\textrm{(d)}} ,
        \end{split}
    \end{equation}
    \begin{equation}
        \begin{split}
            \mathcal{H}_\eps(\qfull_b,m)\qhat\qfull_b'
            =
            \uhat \cdot \nabla \eps_b'
            + \ufull_b' \cdot \nabla \epshat
            - \nabla \cdot \left( \frac{\nuthat'}{\sigma_\eps} \nabla \eps_b + \frac{\nuthat}{\sigma_\eps} \nabla \eps_b' + \frac{\nu_{t,b}'}{\sigma_\eps} \nabla \epshat \right) \\
            \underbrace{- 4 C_{1\eps} \Biggl(
            \frac{\epshat}{k_b} \nu_{t,b} \strain_b : \strain_b'
            -  \frac{\eps_b}{k_b^2} \khat \nu_{t,b} \strain_b : \strain_b'
            + \frac{\eps_b}{k_b} \nuthat \strain_b : \strain_b'}_{\textrm{(e)}}\\
            \underbrace{+ \frac{\eps_b'}{k_b} \nu_{t,b} \strainhat : \strain_b
            - \frac{\eps_b}{k_b^2} k_b' \nu_{t,b} \strainhat : \strain_b
            + \frac{\eps_b}{k_b} \nu_{t,b}' \strainhat : \strain_b
            + \frac{\eps_b}{k_b} \nu_{t,b} \strainhat : \strain_b'
            \Biggr)}_{\textrm{(e)}} \\
            + 2C_{2\eps} \left(
            \frac{\eps_b'\epshat}{k_b}
            - \frac{\eps_b\epshat}{k_b^2}k_b'
            - \frac{\eps_b\eps_b'}{k_b^2}\khat
            + \frac{\eps_b^2}{k_b^3}k_b'\khat
            \right) ,
        \end{split}
    \end{equation}
\end{subequations}
with the eddy-viscosity mode being
\begin{equation}
    \nuthat = C_\mu \left( \frac{2 k_b \khat}{\eps_b} - \frac{k_b^2}{\eps_b^2}\epshat \right) ,
\end{equation}
the perturbation of the eddy-viscosity mode being
\begin{equation}
    \nuthat' = C_\mu \left( \frac{2k_b'\khat}{\eps_b} - \frac{2k_b\khat}{\eps_b^2}\eps_b' - \frac{2k_b k_b'}{\eps_b^2}\epshat + \frac{2k_b^2}{\eps_b^3}\epshat \eps_b' \right) ,
\end{equation}
and the perturbation of the base-flow eddy viscosity being
\begin{equation}
    \nu_{t,b}' = C_\mu \left( \frac{2k_b k_b'}{\eps_b} - \frac{k_b^2}{\eps_b^2}\eps_b' \right) .
\end{equation}
The terms denoted with an underbrace are the terms that correspond to the most dominant physical mechanisms discussed in Figure~\ref{fig:optimal_shape_shift_baseflow_decomposed_comparison} and \ref{fig:growth_rate_sensitivity_to_base_flow_change}. These are the terms associated with a change in (a) the production of $\uhat$ fluctuations, (b) the advection of $\uhat$ fluctuations, (c) the diffusion of $\uhat$ fluctuations, (d) the production of $\khat$ fluctuations, and (e) the production of $\epshat$ fluctuations.

For the frozen eddy-viscosity model, the Hessian operator with a twofold contraction is defined as
\begin{equation}
    \breve{\mathcal{H}}(\qfull_b,m)\qhat\qfull_b' = \frac{\partial \breve{\mathcal{L}}(\qfull_b,m)}{\partial \qfull_b}\qhat\qfull_b' = \begin{bmatrix}
        \breve{\mathcal{H}}_\ufull(\qfull_b,m)\qhat\qfull_b' \\
        0
    \end{bmatrix},
\end{equation}
with
\begin{equation}
    \mathcal{H}_\ufull(\qfull_b,m) \qhat \qfull_b'
    =
    \underbrace{(\uhat \cdot \nabla)\ufull_b'}_{\textrm{(a)}}
    + \underbrace{(\ufull_b' \cdot \nabla)\uhat}_{\textrm{(b)}} .
\end{equation}

\section{Stochastic model for data-driven growth rate estimation} \label{app:stochastic_model}
A stochastic amplitude model offers a robust, data-driven methodology to estimate the linear growth rate of a global oscillator directly from (experimental) measurements \citep{Noiray2013,Bonciolini2017,Lee2019}. This framework has recently been adapted to turbulent swirling flows in \cite{Sieber2021}, to which the reader is referred for further details. In that work, the dynamics of a global vortex-rope-type instability is obscured by the superposed broadband turbulence. The key idea is that, sufficiently close to the Hopf bifurcation, the slow dynamics of the dominant coherent structure can be reduced to a Stuart--Landau amplitude equation. To account for stochastic forcing by the background turbulence, the deterministic amplitude evolution is complemented by a stochastic forcing term. Using a Fokker--Planck equation, a model for the stationary probability density function of the oscillation amplitude is found. Fitting the analytical Fokker--Planck solution to the measured probability density function allows the linear growth rate (and other dynamical parameters) to be inferred.

In practice, a scalar observable is used as data input, which is bandpass-filtered around the mode frequency, and long time series are used to construct this stationary probability density function. In our work, we take an azimuthally decomposed wall pressure signal as also used in Appendix~\ref{app:urans_experiment_comparison}. We choose a bandpass filter width of $[f_1 / 1.2, 1.2 f_1]$ at $Q^* = 0.50$ and a filter width of $[f_1 / 1.7, 1.7 f_1]$ at $Q^* = 0.76$, with a linear interpolation of the filter width for the flow rates between and with $f_1$ being the mean frequency of the vortex rope. The filter width is chosen such that (a) the Kulback--Leibler divergence is minimized, and (b) the bifurcation point estimated by the stochastic model approximately agrees with the bifurcation point found by the zero crossing of the squared amplitude curve fit (see Appendix~\ref{app:urans_experiment_comparison}).

\section{3D URANS simulations and experimental measurements} \label{app:3d_urans_experiments}
\begin{figure}
    \centering
    \begin{tikzpicture}
        \begin{scope}
            \node[anchor=south west,inner sep=0] (image) at (0,0) {\includegraphics[height=0.9\textwidth,angle=90,trim={30.4cm 3.85cm 25.6cm 2.6cm},clip]{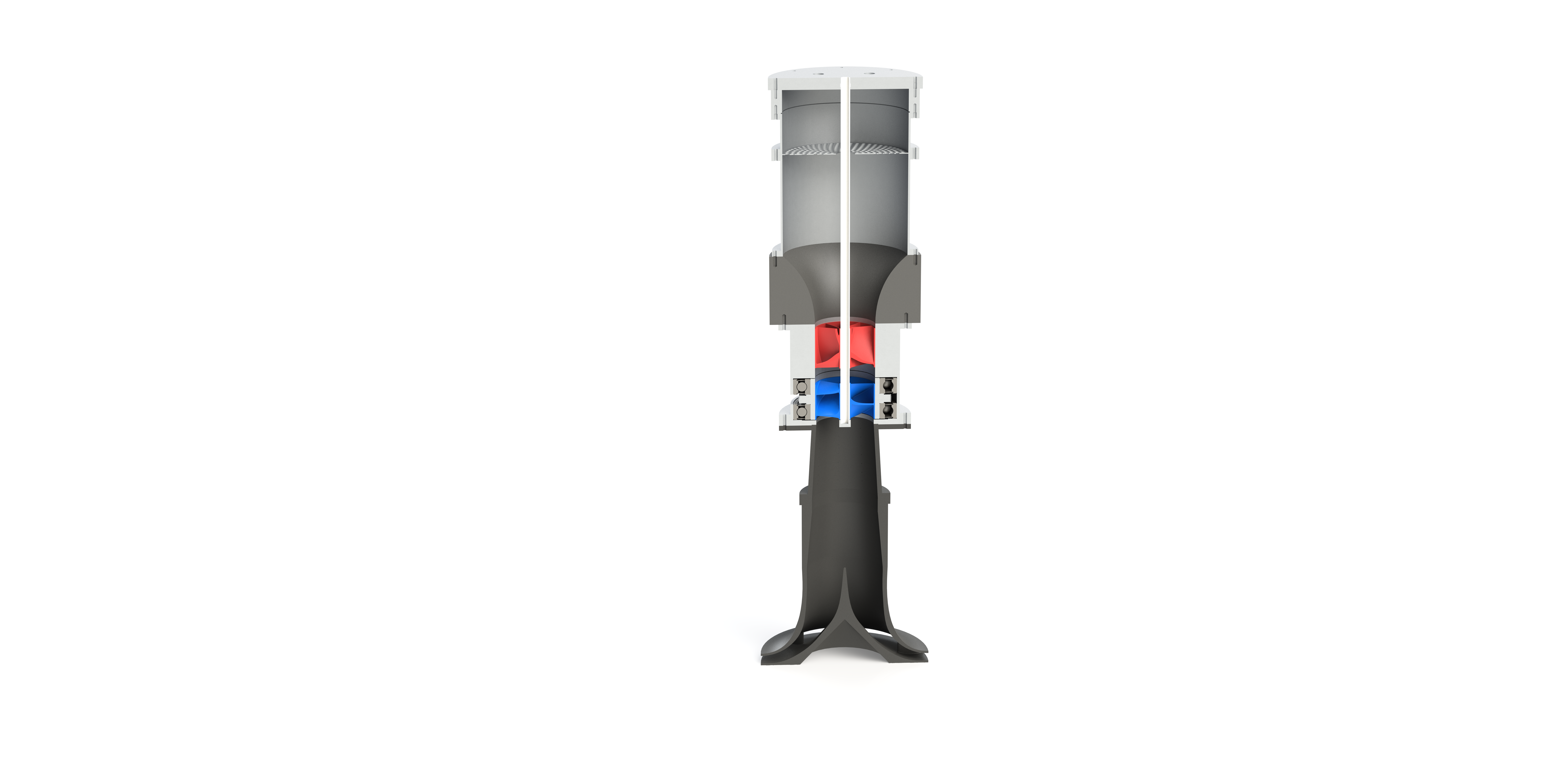}}; 
            \begin{scope}[x={(image.south east)},y={(image.north west)}]
                \draw[draw=gn, thick] (0.41,-0.055) rectangle
                ++(0.61,1.11);
                \node [anchor=north west,fill=gn,text=white] at (0.41,1.055) {\footnotesize URANS};
                \draw[draw=oj, thick] (0.58,-0.02) rectangle
                ++(0.43,1.04);
                \node [anchor=north west,fill=oj,text=white] at (0.58,1.02) {\footnotesize RANS};
                \draw [latex-, thick, guidevanes_red] (0.475,0.33) -- ++(0,-0.45)
                node [below,guidevanes_red] {\footnotesize Guide vanes};
                \draw [latex-, thick, runner_blue] (0.547,0.67) -- ++(0,+0.45)
                node [above,runner_blue] {\footnotesize Runner};
                \draw [latex-, thick, white] (0.605,0.505) -- ++(0.04,0)
                node [right,white] {\footnotesize Runner axle};
                \draw [latex-, thick, black] (0.65,0.3) -- ++(0,-0.42)
                node [below,black] {\footnotesize Draft tube};
                \draw [latex-, thick, black] (0.89,0.255) -- ++(0,-0.375)
                node [below,black] {\footnotesize Moody tube};
                \draw [latex-, thick, white] (0.28,0.28) -- ++(-0.1,0);
                \draw [latex-, thick, white] (0.28,0.38) -- ++(-0.1,0);
                \draw [latex-, thick, white] (0.28,0.62) -- ++(-0.1,0);
                \draw [latex-, thick, white] (0.28,0.72) -- ++(-0.1,0);
                \node[white] at (0.228,0.21) {\footnotesize Flow};
            \end{scope}
        \end{scope}
         
    \end{tikzpicture}

    \caption{Setup rendering of the experimental rig with considered URANS and RANS domains indicated by the colored rectangles. Centerbody is not shown and would be mounted at the end of the runner axle.}
    \label{fig:experimental_rig}
\end{figure}
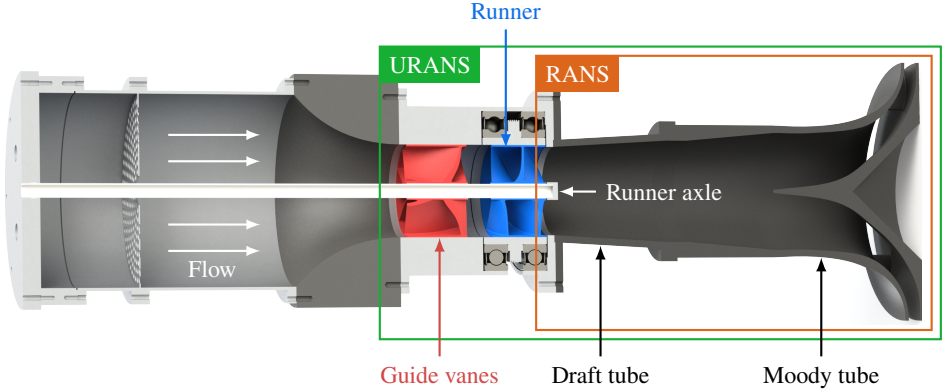

The numerical URANS and the experimental setup described here are based on earlier studies of \cite{Mueller2022b,Lueckoff2022}, where further details are provided. A schematic of the experimental setup is shown in Figure~\ref{fig:experimental_rig}. The experimental rig is operated under atmospheric conditions. Pressurized air first enters an axisymmetric plenum and is homogenized by perforated plates. Subsequently, the flow is accelerated in a nozzle before passing through the guide vanes and the rotating runner stage. The runner is mounted on a central axle and is driven by an external motor with a fixed rotation rate of $n = \SI{2432}{\rpm}$. At the downstream end of the axle, the centerbody is installed (not shown in the figure). The flow then enters the draft tube, followed by the Moody tube. In the URANS precursor simulations, the plenum, the perforated plates, and the nozzle are omitted, and only the downstream flow is resolved. The URANS simulations are performed using the $k$--$\omega$ SST turbulence model and non-conformal, arbitrary mesh interfaces between the stationary domains and the rotating runner domain.

\subsection{Time-averaged URANS inlet profiles}
\begin{figure}
    \centering
    \includegraphics[width=0.9\linewidth]{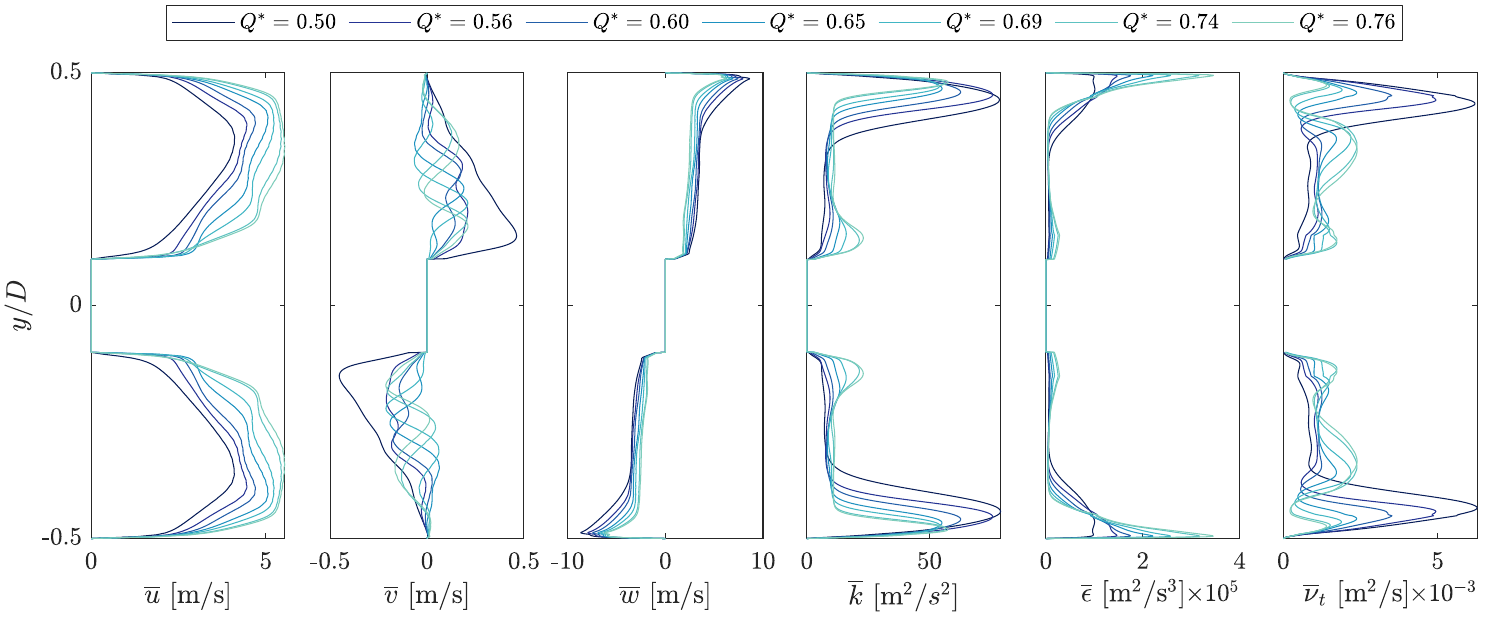}
    \caption{Temporally and azimuthally averaged flow profiles of the 3D URANS simulations extracted at $x/D = 0$, which have been tuned \textit{a posteriori} for bifurcation point matching. These profiles serve as inlet conditions for the 2D RANS simulations.}
    \label{fig:inlet_profiles_2d_rans}
\end{figure}

Figure~\ref{fig:inlet_profiles_2d_rans} presents the inlet profiles extracted from the URANS simulations between the runner outlet and the draft tube inlet at $x/D = 0$. All flow quantities are temporally and azimuthally averaged. The 2D RANS simulations are performed using these flow profiles as inlet conditions, with the turbulent quantities tuned by a constant factor, as described in §\ref{sec:rans_tuning}. The axial velocity distribution exhibits a parabolic-like shape at low flow rates and gradually evolves towards a more block-like, homogeneous profile as $Q^*$ increases, particularly in the region close to the centerbody. In contrast, the azimuthal velocity decreases with increasing flow rate, indicating a reduction in the swirl number. The mean turbulence kinetic energy, $\kmean$, and the mean eddy viscosity, $\nutmean$, both decrease with increasing $Q^*$, reflecting reduced shear and rotational effects, and hence diminished turbulence production. The mean dissipation rate, $\epsmean$, becomes more concentrated near the outer walls at higher flow rates, with an increasing magnitude in this region.

\subsection{Comparison of 3D URANS with 2D RANS simulation}
\begin{figure}[tbp]
    \centering
    \sidesubfloat[]{
    \centering
    \includegraphics[height=0.29\columnwidth]{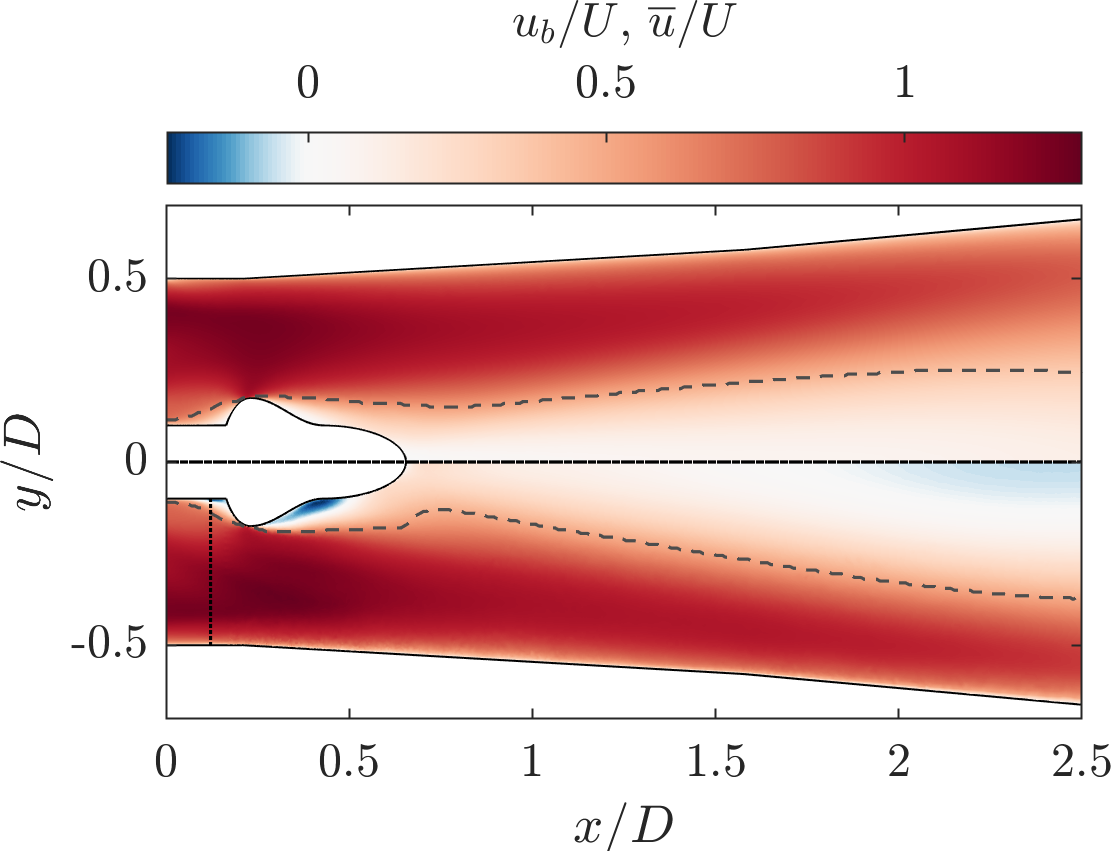}
    }
    \sidesubfloat[]{
    \centering
    \includegraphics[height=0.29\columnwidth]{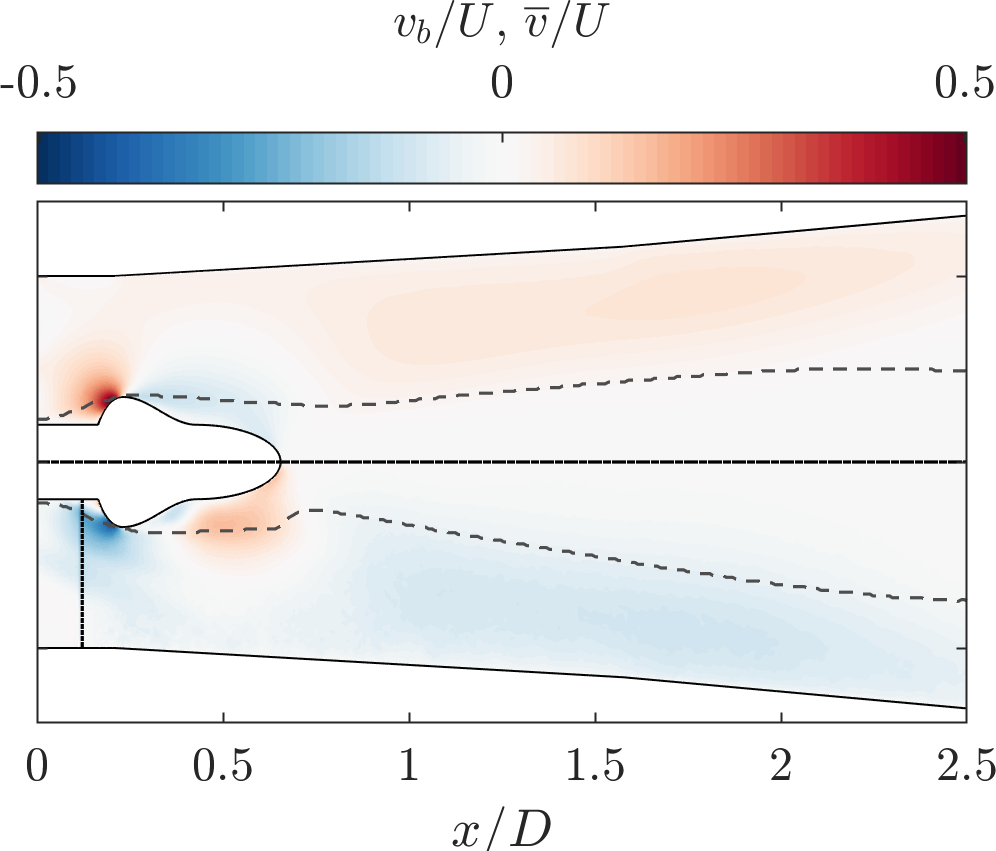}
    }\\
    \vspace{0.2cm}
    \sidesubfloat[]{
    \centering
    \includegraphics[height=0.29\columnwidth]{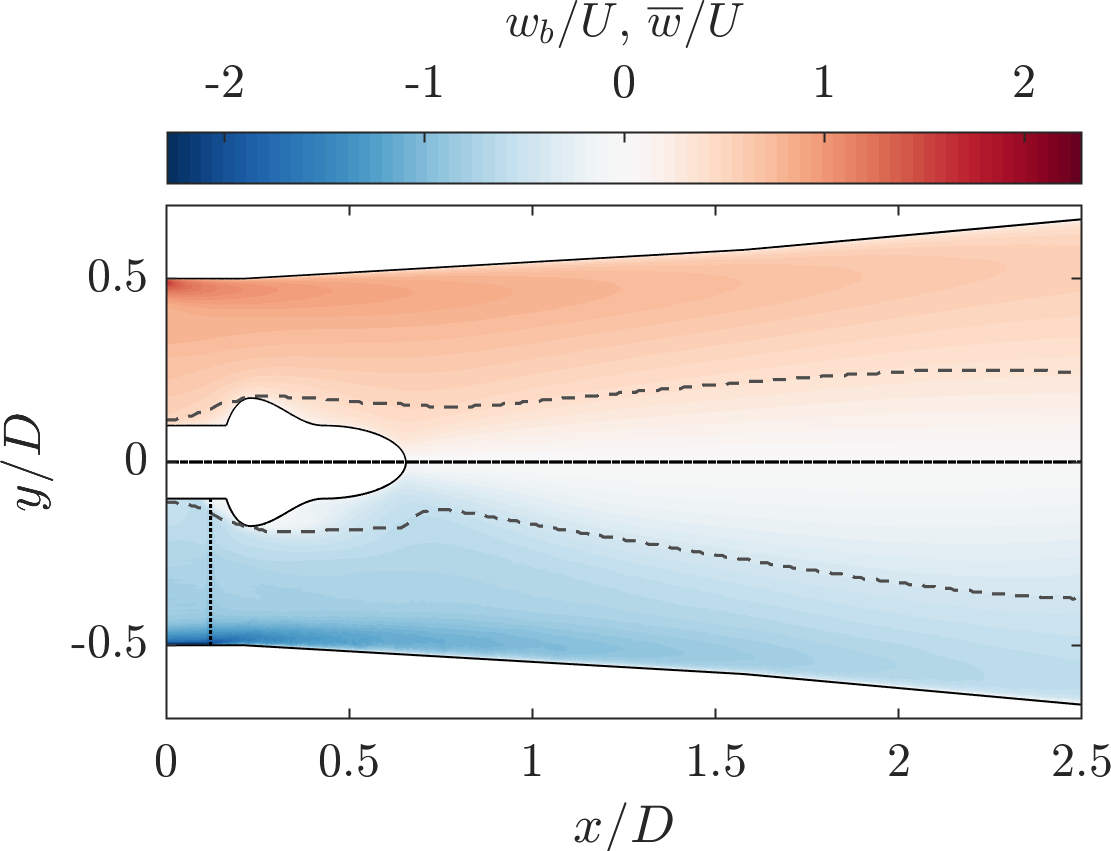}
    }
    \sidesubfloat[]{
    \centering
    \includegraphics[height=0.29\columnwidth]{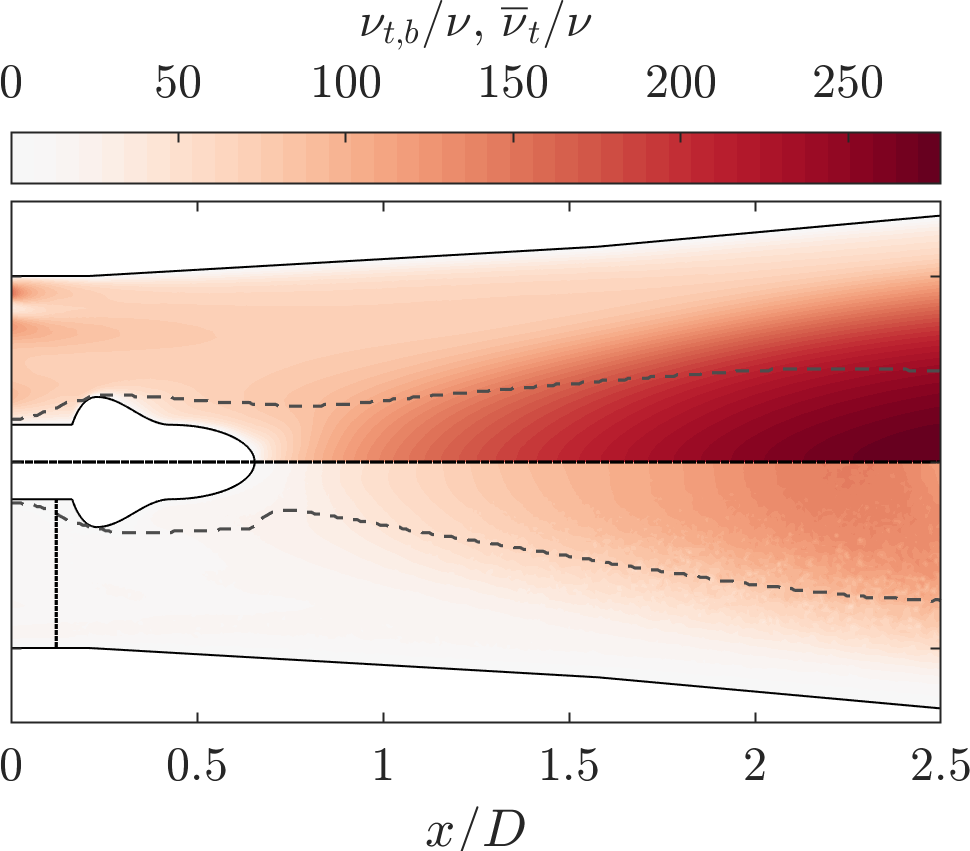}
    }
    \caption{Flow fields in a streamwise section, comparing (top) base flow of 2D RANS simulation with (bottom) mean flow of 3D URANS simulation. Displayed are (a) axial velocity, (b) radial velocity, (c) azimuthal velocity, and (d) eddy viscosity at $Q^* = 0.65$. Gray dashed line indicates wake half width $y_{1/2}$, black dotted line indicates the arbitrary mesh interface of the rotating domain in the URANS case.}
    \label{fig:meanflow_contour_Q65_RANS_URANS_comparison}
\end{figure}

Figure~\ref{fig:meanflow_contour_Q65_RANS_URANS_comparison} compares the RANS base flow with the (time-averaged) URANS mean-flow fields of the velocity vector and the eddy viscosity at $Q^* = 0.65$. The wake region predicted by the RANS simulation is of comparable extent to the URANS results. However, in contrast to URANS, the RANS simulations do not predict regions of reverse flow at the concave part of the centerbody. The azimuthal velocity is slightly higher close to the outer walls at the domain inlet of the URANS simulation, indicating that angular momentum is better conserved. The eddy viscosity magnitude is generally higher in the RANS simulation, which is mainly due to the higher base level of eddy viscosity convected from the inlet, as motivated in §\ref{sec:rans_tuning}.

A key challenge in all cases is the prediction of flow separation and reverse flow, which is known to be difficult for simulations using Boussinesq-type turbulence models. The $k$--$\varepsilon$ model, in particular, is well known for its poor performance in separated flows, while the $k$--$\omega$ SST model typically behaves better \citep{Menter1993}. In this work, however, the $k$--$\varepsilon$ model was chosen for the RANS simulation since precursor studies showed better agreement with the URANS mean flow at subcritical conditions.

Note that an exact agreement between 2D RANS and time-averaged 3D URANS fields is not expected due to the following reasons. First, the 2D RANS simulation uses a scaled-up inlet $k$, $\varepsilon$, and $\nu_t$, as described in §\ref{sec:rans_tuning}, to mimic the additional dissipation caused by the rotating wake structures. Second, the RANS solution represents a steady base flow, which is not necessarily the same as the time average of the URANS solution. At least for unstable conditions, a mismatch is expected due to the  mean-field modification induced by the vortex rope dynamics.

\subsection{Comparison of 3D URANS simulation with experimental measurements} \label{app:urans_experiment_comparison}
In this section, we validate the reference URANS simulations against experimental measurements, focusing on both mean-flow quantities and dynamic quantities. Because only the pressure along the outer draft-tube walls was measured experimentally, this is the only flow quantity available for direct comparison of the URANS fields with the experimental measurements.

\begin{figure}
    \centering
    \includegraphics[width=0.7\linewidth]{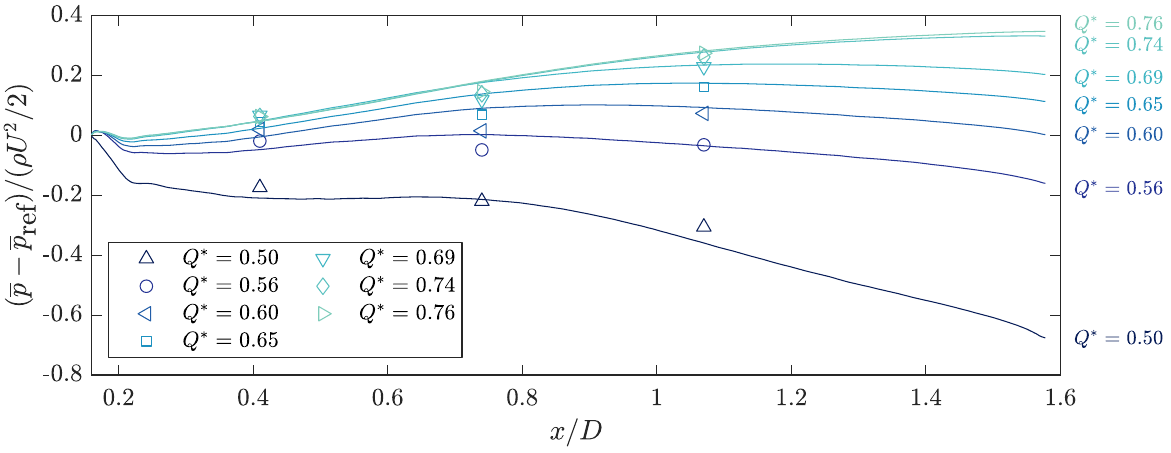}
    \caption{Mean wall pressure distribution for all considered flow rates, comparing 3D URANS simulation values (curves) with experimental measurements (markers).}
    \label{fig:mean_pressure}
\end{figure}

First, we compare the mean-flow behavior considering the mean wall-pressure distribution along the axial coordinate (temporally and azimuthally averaged) for all flow rates considered, comparing URANS and experimental measurements in Figure~\ref{fig:mean_pressure}. The pressure is expressed as a difference relative to an upstream reference pressure $\overline{p}_\textrm{ref}$ at $x/D=0.16$, normalized with the bulk-flow dynamic pressure $\rho U^2/2$. The comparison with experimental wall-pressure measurements shows that, although there are small discrepancies in detail, the general trends are well reproduced across all flow rates. Hence, the URANS mean pressure losses and recovery can be regarded as valid, indicating that the integral behavior of the mean flow is captured reasonably well.

\begin{figure}[tbp]
    \centering
    \sidesubfloat[]{
    \centering
    \includegraphics[width=0.4\columnwidth]{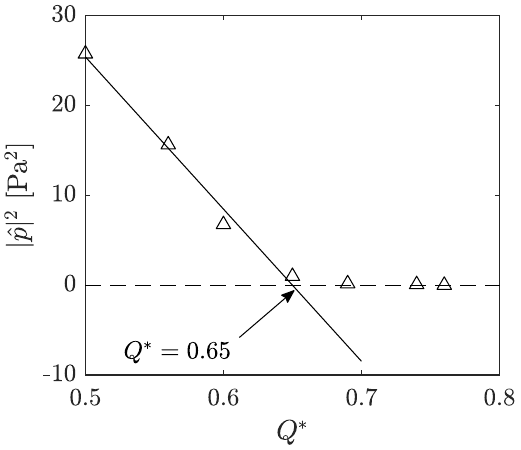}
    }
    \sidesubfloat[]{
    \centering
    \includegraphics[width=0.445\columnwidth]{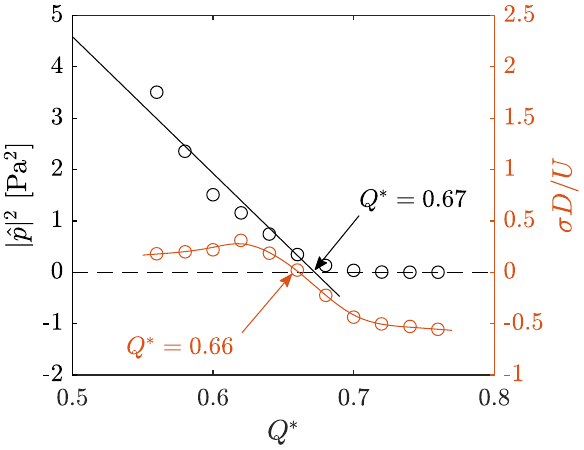}
    }
    \caption{Squared pressure amplitude over normalized flow rate to identify the supercritical Hopf bifurcation point in the (a) URANS simulation and (b) experiment. The experimentally determined growth rate is superposed in red to compare with the bifurcation point obtained with the stochastic model.}
    \label{fig:amplitudes_hopf_bifurcation_over_flowrate}
\end{figure}

Secondly, we compare the dynamic behavior of the flow with respect to the growth rates. In contrast to the experiment, we cannot utilize the data-driven stochastic model to determine the growth rates in the URANS simulations due to the prohibitively high demand for data samples required for statistical convergence. We therefore revert to a `graphical approach'. For a supercritical Hopf bifurcation, the squared amplitude follows a linear slope in the supercritical regime, e.g.\ $|\hat{p}|^2 \propto Q^*_\textrm{crit} - Q^*$ for $Q^*_\textrm{crit} - Q^* > 0$, such that the zero crossing of this line can provide an estimate of the bifurcation point \citep{Huerre1990}. At subcritical conditions ($Q^*_\textrm{crit} - Q^* < 0$), amplitudes are not necessarily zero since residual forcing can intermittently excite the vortex rope instability, e.g.\ through background turbulence or coherent structures originating from the rotating wake of the runner \citep{Sieber2021}. This leads to a finite amplitude that no longer follows the proportionality law of a supercritical Hopf bifurcation. Figure~\ref{fig:amplitudes_hopf_bifurcation_over_flowrate} shows the squared amplitude of the pressure fluctuations quantified with the spectral peak of the identified vortex-rope frequency for $m=1$ at $x/D = 1.1$, comparing experimental measurements with URANS simulations. In the experiment, the zero crossing of the linear curve-fit following the quadratic proportionality occurs at $Q^* = 0.67$, in good agreement with the bifurcation point predicted at $Q^* = 0.66$ using the data-driven stochastic model. In the URANS simulation, the zero crossing occurs at a slightly lower flow rate $Q^* = 0.65$, and the predicted squared amplitudes are generally higher than those in the experiment. Still, the Hopf bifurcation point is reproduced reasonably well.

Thirdly, we examine the dynamic behavior of the flow with regard to the vortex-rope frequency. For that, a two-dimensional discrete Fourier transform (DFT) is performed in the temporal and azimuthal directions according to the convention
\begin{equation}
    \uhat(x,r,m,\omega) = \frac{1}{N_\theta/N_t} \sum_t \sum_\theta \ufull(x,r,\theta,t) \mathrm{e}^{-\mathrm{i}(m\theta - \omega t)} .
\end{equation}
With this phase-angle convention, co-rotating helical modes are identified by positive azimuthal wavenumbers $m > 0$ and positive frequencies $f > 0$. Although a spectral density estimation would provide a more rigorous approach (e.g. by using Welch's method), it would significantly deteriorate the spectral resolution, being $\Delta f \approx \SI{1}{\Hz}$ with a simulation time of $\approx \SI{1}{\s}$. Thus, the higher spectral leakage of the DFT is accepted in favor of a better spectral resolution, as the focus lies on the location of the spectral peaks rather than on their precise amplitudes. Furthermore, the URANS simulation models the turbulent fluctuations, rendering the signal as virtually non-stochastic, and thus not necessarily requiring a spectral density estimator.

\begin{figure}
    \centering
    \includegraphics[width=0.7\linewidth]{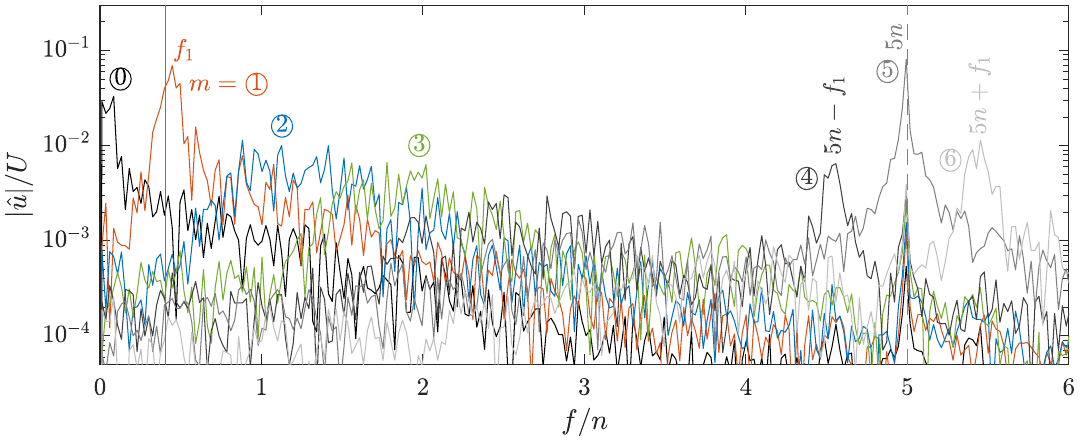}
    \caption{DFT spectrum of the axial velocity fluctuation at $x/D = 0.7$, $r/D = 0.43$, $Q^* = 0.65$. The circled numbers denote the azimuthal wavenumbers, the vortex-rope frequency is $f_1$, the runner rotation rate is $n$. The solid vertical line denotes the experimentally measured vortex-rope frequency, the solid dashed line denotes the blade passing frequency.}
    \label{fig:spectrum_crossiwse_Q65}
\end{figure}

Figure~\ref{fig:spectrum_crossiwse_Q65} presents the DFT spectrum evaluated at a representative location of $x/D = 0.7$, $r/D = 0.43$ for $Q^* = 0.65$. The frequencies are normalized by the runner rotation rate $n = \SI{2432}{\rpm}$. A dominant single-helical mode with $m = 1$ and $f/n = 0.45$ (corresponding to a Strouhal number $St = 0.37$) can be identified that represents the vortex rope. This value is close to the experimentally measured frequency of $f/n = 0.41$ ($St = 0.34$), indicated by the vertical solid line. Hence, the URANS simulation reproduces the experimentally observed vortex-rope frequency quite accurately. No distinct higher harmonics are observed, which is expected given that the operating condition is very close to the bifurcation point.

Three additional peaks appear at higher azimuthal wavenumbers. For $m = 5$, a peak at the blade passing frequency $f = 5n$ (marked by the vertical dashed line) is visible, corresponding to the runner wake structures discussed in §\ref{sec:rans_tuning}. Additional peaks occur below and above the blade passing frequency that satisfy the triad condition $m_a \pm m_b = m_c$ and $f_a \pm f_b = f_c$, indicating significant nonlinear interactions between the runner wake structures ($m = 5$) and the vortex rope ($m = 1$ and its complex conjugate $m = -1$). These interactions generate new flow structures with $m_c = 5 \pm 1$ and $f_c = 5n \pm f_1$. The presence of these nonlinear interactions motivates the tuning of the RANS inlet conditions as described in §\ref{sec:rans_tuning}.

\bibliographystyle{jfm}

\begin{thebibliography}{87}
\expandafter\ifx\csname natexlab\endcsname\relax\def\natexlab#1{#1}\fi
\def\au#1{#1} \def\ed#1{#1} \def\yr#1{#1}\def\at#1{#1}\def\jt#1{\textit{#1}} \def\bt#1{#1}\def\bvol#1{\textbf{#1}} \def\vol#1{#1} \def\pg#1{#1} \def\publ#1{#1}\def\arxiv#1{#1}\def\org#1{#1}\def\st#1{\textit{#1}}

\bibitem[Amoignon {\em et~al.\/}(2006)Amoignon, Pralits, Hanifi, Berggren \& Henningson]{Amoignon2006}
{\sc \au{Amoignon, O.~G.}, \au{Pralits, J.~O.}, \au{Hanifi, A.}, \au{Berggren, M.} \& \au{Henningson, D.~S.}} \yr{2006}  \at{{Shape Optimization for Delay of Laminar-Turbulent Transition}}.  \jt{AIAA Journal}  \bvol{44}~(5),  \pg{1009--1024}.

\bibitem[Baratta {\em et~al.\/}(2023)Baratta, Dean, Dokken, Habera, Hale, Richardson, Rognes, Scroggs, Sime \& Wells]{Baratta2023}
{\sc \au{Baratta, I.~A.}, \au{Dean, J.~P.}, \au{Dokken, J.~S.}, \au{Habera, M.}, \au{Hale, J.~S.}, \au{Richardson, C.~N.}, \au{Rognes, M.~E.}, \au{Scroggs, M.~W.}, \au{Sime, N.} \& \au{Wells, G.~N.}} \yr{2023}  \at{{DOLFINx: The next generation FEniCS problem solving environment}}.  \jt{Zenodo (preprint)} .

\bibitem[Barkley(2006)]{Barkley2006}
{\sc \au{Barkley, D.}} \yr{2006}  \at{{Linear analysis of the cylinder wake mean flow}}.  \jt{Europhysics Letters}  \bvol{75}~(5),  \pg{750--756}.

\bibitem[Bonciolini {\em et~al.\/}(2017)Bonciolini, Boujo \& Noiray]{Bonciolini2017}
{\sc \au{Bonciolini, G.}, \au{Boujo, E.} \& \au{Noiray, N.}} \yr{2017}  \at{{Output-only parameter identification of a colored-noise-driven Van-der-Pol oscillator: Thermoacoustic instabilities as an example}}.  \jt{Physical Review E}  \bvol{95}~(6),  \pg{062217}.

\bibitem[Bosioc {\em et~al.\/}(2012)Bosioc, Susan-Resiga, Muntean \& Tanasa]{Bosioc2012}
{\sc \au{Bosioc, A.~I.}, \au{Susan-Resiga, R.}, \au{Muntean, S.} \& \au{Tanasa, C.}} \yr{2012}  \at{{Unsteady Pressure Analysis of a Swirling Flow With Vortex Rope and Axial Water Injection in a Discharge Cone}}.  \jt{Journal of Fluids Engineering}  \bvol{134}~(8),  \pg{81104}.

\bibitem[Boujo {\em et~al.\/}(2019)Boujo, Fani \& Gallaire]{Boujo2019}
{\sc \au{Boujo, E.}, \au{Fani, A.} \& \au{Gallaire, F.}} \yr{2019}  \at{{Second-order sensitivity in the cylinder wake: Optimal spanwise-periodic wall actuation and wall deformation}}.  \jt{Physical Review Fluids}  \bvol{4}~(5),  \pg{53901}.

\bibitem[Brewster(2019)]{Brewster2019}
{\sc \au{Brewster, J.}} \yr{2019}  \at{{Shape Optimisation for Hydrodynamic Stability and its Application to Cyclone Separators}}. PhD thesis, University of Cambridge.

\bibitem[Brewster \& Juniper(2020)]{Brewster2020}
{\sc \au{Brewster, J.} \& \au{Juniper, M.~P.}} \yr{2020}  \at{{Shape sensitivity of eigenvalues in hydrodynamic stability, with physical interpretation for the flow around a cylinder}}.  \jt{European Journal of Mechanics, B/Fluids}  \bvol{80},  \pg{80--91}.

\bibitem[Browne {\em et~al.\/}(2014)Browne, Rubio, Ferrer \& Valero]{Browne2014}
{\sc \au{Browne, O.~M.}, \au{Rubio, G.}, \au{Ferrer, E.} \& \au{Valero, E.}} \yr{2014}  \at{{Sensitivity analysis to unsteady perturbations of complex flows: a discrete approach}}.  \jt{International Journal for Numerical Methods in Fluids}  \bvol{76}~(12),  \pg{1088--1110}.

\bibitem[Camarri(2015)]{Camarri2015}
{\sc \au{Camarri, S.}} \yr{2015}  \at{{Flow control design inspired by linear stability analysis}}.  \jt{Acta Mechanica}  \bvol{226}~(4),  \pg{979--1010}.

\bibitem[Crouch {\em et~al.\/}(2007)Crouch, Garbaruk \& Magidov]{Crouch2007}
{\sc \au{Crouch, J.~D.}, \au{Garbaruk, A.} \& \au{Magidov, D.}} \yr{2007}  \at{{Predicting the onset of flow unsteadiness based on global instability}}.  \jt{Journal of Computational Physics}  \bvol{224}~(2),  \pg{924--940}.

\bibitem[Delfour \& Zol{\'{e}}sio(2011)]{Delfour2011}
{\sc \au{Delfour, M.~C.} \& \au{Zol{\'{e}}sio, J.-P.}} \yr{2011} {\em {Shapes and geometries: metrics, analysis, differential calculus, and optimization}\/}, 2nd edn.  \publ{Society for Industrial and Applied Mathematics}.

\bibitem[Fan {\em et~al.\/}(2024)Fan, Kozul, Li \& Sandberg]{Fan2024}
{\sc \au{Fan, Y.}, \au{Kozul, M.}, \au{Li, W.} \& \au{Sandberg, R.~D.}} \yr{2024}  \at{{Eddy-viscosity-improved resolvent analysis of compressible turbulent boundary layers}}.  \jt{Journal of Fluid Mechanics}  \bvol{983},  \pg{A46}.

\bibitem[Giannetti \& Luchini(2007)]{Giannetti2007}
{\sc \au{Giannetti, F.} \& \au{Luchini, P.}} \yr{2007}  \at{{Structural sensitivity of the first instability of the cylinder wake}}.  \jt{Journal of Fluid Mechanics}  \bvol{581},  \pg{167--197}.

\bibitem[Goyal \& Gandhi(2018)]{Goyal2018a}
{\sc \au{Goyal, R.} \& \au{Gandhi, B.~K.}} \yr{2018}  \at{{Review of hydrodynamics instabilities in Francis turbine during off-design and transient operations}}.  \jt{Renewable Energy}  \bvol{116},  \pg{697--709}.

\bibitem[Grinfeld(2010)]{Grinfeld2010}
{\sc \au{Grinfeld, P.}} \yr{2010}  \at{{Hadamard's Formula Inside and Out}}.  \jt{Journal of Optimization Theory and Applications}  \bvol{146}~(3),  \pg{654--690}.

\bibitem[Heuveline \& Strau{\ss}(2009)]{Heuveline2009}
{\sc \au{Heuveline, V.} \& \au{Strau{\ss}, F.}} \yr{2009}  \at{{Shape optimization towards stability in constrained hydrodynamic systems}}.  \jt{Journal of Computational Physics}  \bvol{228}~(4),  \pg{938--951}.

\bibitem[Hinch(1991)]{Hinch1991}
{\sc \au{Hinch, E.}} \yr{1991} {\em {Perturbation Methods}\/}.  \publ{Cambridge University Press}.

\bibitem[Holmstr{\"{o}}m {\em et~al.\/}(2021)Holmstr{\"{o}}m, Sundstr{\"{o}}m \& Cervantes]{Holmstrom2021}
{\sc \au{Holmstr{\"{o}}m, H.}, \au{Sundstr{\"{o}}m, J.} \& \au{Cervantes, M.~J.}} \yr{2021}  \at{{Vortex rope mitigation with azimuthal perturbations: A numerical study}}.  \jt{IOP Conference Series: Earth and Environmental Science}  \bvol{774}~(1),  \pg{012144}.

\bibitem[Holmstr{\"{o}}m {\em et~al.\/}(2022)Holmstr{\"{o}}m, Sundstr{\"{o}}m \& Cervantes]{Holmstrom2022}
{\sc \au{Holmstr{\"{o}}m, H.}, \au{Sundstr{\"{o}}m, J.} \& \au{Cervantes, M.~J.}} \yr{2022}  \at{{Vortex rope interaction with radially protruded solid bodies in an axial turbine: A numerical study}}.  \jt{IOP Conference Series: Earth and Environmental Science}  \bvol{1079}~(1),  \pg{012055}.

\bibitem[Huerre \& Monkewitz(1990)]{Huerre1990}
{\sc \au{Huerre, P.} \& \au{Monkewitz, P.~A.}} \yr{1990}  \at{{Instabilities in Spatially Developing Flows}}.  \jt{Annual Review of Fluid Mechanics}  \bvol{22},  \pg{473--537}.

\bibitem[Javadi \& Nilsson(2017)]{Javadi2017}
{\sc \au{Javadi, A.} \& \au{Nilsson, H.}} \yr{2017}  \at{{Active flow control of the vortex rope and pressure pulsations in a swirl generator}}.  \jt{Engineering Applications of Computational Fluid Mechanics}  \bvol{11}~(1),  \pg{30--41}.

\bibitem[Kaiser {\em et~al.\/}(2023)Kaiser, Demange, M{\"{u}}ller, Knechtel \& Oberleithner]{Kaiser2023b}
{\sc \au{Kaiser, T.~L.}, \au{Demange, S.}, \au{M{\"{u}}ller, J.~S.}, \au{Knechtel, S.} \& \au{Oberleithner, K.}} \yr{2023} {FELiCS: A versatile linearized solver addressing dynamics in multi-physics flows}.  \bt{In {\em AIAA Aviation 2023 Forum\/}},  \pg{p. 3434}.

\bibitem[Kaiser {\em et~al.\/}(2018)Kaiser, Poinsot \& Oberleithner]{Kaiser2018}
{\sc \au{Kaiser, T.~L.}, \au{Poinsot, T.} \& \au{Oberleithner, K.}} \yr{2018}  \at{{Stability and Sensitivity Analysis of Hydrodynamic Instabilities in Industrial Swirled Injection Systems}}.  \jt{Journal of Engineering for Gas Turbines and Power}  \bvol{140}~(5),  \pg{051506}.

\bibitem[Khorrami {\em et~al.\/}(1989)Khorrami, Malik \& Ash]{Khorrami1989}
{\sc \au{Khorrami, M.~R.}, \au{Malik, M.~R.} \& \au{Ash, R.~L.}} \yr{1989}  \at{{Application of spectral collocation techniques to the stability of swirling flows}}.  \jt{Journal of Computational Physics}  \bvol{81}~(1),  \pg{206--229}.

\bibitem[Khullar {\em et~al.\/}(2022)Khullar, Singh, Cervantes \& Gandhi]{Khullar2022}
{\sc \au{Khullar, S.}, \au{Singh, K.~M.}, \au{Cervantes, M.~J.} \& \au{Gandhi, B.~K.}} \yr{2022}  \at{{Influence of runner cone profile and axial water jet injection in a low head Francis turbine at part load}}.  \jt{Sustainable Energy Technologies and Assessments}  \bvol{50},  \pg{101810}.

\bibitem[Kiriyama {\em et~al.\/}(2018)Kiriyama, Katamine \& Azegami]{Kiriyama2018}
{\sc \au{Kiriyama, Y.}, \au{Katamine, E.} \& \au{Azegami, H.}} \yr{2018}  \at{{Shape optimisation problem for stability of Navier–Stokes flow field}}.  \jt{International Journal of Computational Fluid Dynamics}  \bvol{32}~(2-3),  \pg{68--87}.

\bibitem[Knechtel {\em et~al.\/}(2024)Knechtel, Kaiser, Orchini \& Oberleithner]{Knechtel2024}
{\sc \au{Knechtel, S.~J.}, \au{Kaiser, T.~L.}, \au{Orchini, A.} \& \au{Oberleithner, K.}} \yr{2024}  \at{{Arbitrary-order sensitivities of the incompressible base flow and its eigenproblem}}.  \jt{Journal of Fluid Mechanics}  \bvol{985},  \pg{A32}.

\bibitem[Kuhn {\em et~al.\/}(2022)Kuhn, M{\"{u}}ller, Oberleithner, Knechtel \& Soria]{Kuhn2022}
{\sc \au{Kuhn, P.}, \au{M{\"{u}}ller, J.~S.}, \au{Oberleithner, K.}, \au{Knechtel, S.} \& \au{Soria, J.}} \yr{2022} {Influence of eddy viscosity on linear modeling of self-similar coherent structures in the jet far field}.  \bt{In {\em AIAA SciTech Forum 2022\/}},  \pg{p. 0460}.

\bibitem[Kumar {\em et~al.\/}(2021)Kumar, Cervantes \& Gandhi]{Kumar2021}
{\sc \au{Kumar, S.}, \au{Cervantes, M.~J.} \& \au{Gandhi, B.~K.}} \yr{2021}  \at{{Rotating vortex rope formation and mitigation in draft tube of hydro turbines – A review from experimental perspective}}.  \jt{Renewable and Sustainable Energy Reviews}  \bvol{136}~(March 2020),  \pg{110354}.

\bibitem[Kungurtsev \& Juniper(2019)]{Kungurtsev2019}
{\sc \au{Kungurtsev, P.~V.} \& \au{Juniper, M.~P.}} \yr{2019}  \at{{Adjoint-based shape optimization of the microchannels in an inkjet printhead}}.  \jt{Journal of Fluid Mechanics}  \bvol{871},  \pg{113--138}.

\bibitem[Kurokawa {\em et~al.\/}(2010)Kurokawa, Imamura \& Choi]{Kurokawa2010}
{\sc \au{Kurokawa, J.}, \au{Imamura, H.} \& \au{Choi, Y.~D.}} \yr{2010}  \at{{Effect of J-Groove on the Suppression of Swirl Flow in a Conical Diffuser}}.  \jt{Journal of Fluids Engineering, Transactions of the ASME}  \bvol{132}~(7),  \pg{071101}.

\bibitem[Launder \& Spalding(1974)]{Launder1974}
{\sc \au{Launder, B.~E.} \& \au{Spalding, D.~B.}} \yr{1974}  \at{{The numerical computation of turbulent flows}}.  \jt{Computer Methods in Applied Mechanics and Engineering}  \bvol{3}~(2),  \pg{269--289}.

\bibitem[Lee {\em et~al.\/}(2019)Lee, Zhu, Li \& Gupta]{Lee2019}
{\sc \au{Lee, M.}, \au{Zhu, Y.}, \au{Li, L.~K.} \& \au{Gupta, V.}} \yr{2019}  \at{{System identification of a low-density jet via its noise-induced dynamics}}.  \jt{Journal of Fluid Mechanics}  \bvol{862},  \pg{200--215}.

\bibitem[Li {\em et~al.\/}(2021)Li, Yu, Yan, Wang, Shi \& Wei]{Li2021}
{\sc \au{Li, D.~Y.}, \au{Yu, L.}, \au{Yan, X.~Y.}, \au{Wang, H.~J.}, \au{Shi, Q.} \& \au{Wei, X.~Z.}} \yr{2021}  \at{{Runner cone optimization to reduce vortex rope-induced pressure fluctuations in a Francis turbine}}.  \jt{Science China Technological Sciences}  \bvol{64}~(9),  \pg{1953--1970}.

\bibitem[Litvinov {\em et~al.\/}(2018)Litvinov, Shtork, Gorelikov, Mitryakov \& Hanjalic]{Litvinov2018}
{\sc \au{Litvinov, I.}, \au{Shtork, S.}, \au{Gorelikov, E.}, \au{Mitryakov, A.} \& \au{Hanjalic, K.}} \yr{2018}  \at{{Unsteady regimes and pressure pulsations in draft tube of a model hydro turbine in a range of off-design conditions}}.  \jt{Experimental Thermal and Fluid Science}  \bvol{91},  \pg{410--422}.

\bibitem[Lucca-Negro \& O'Doherty(2001)]{LuccaNegro2001}
{\sc \au{Lucca-Negro, O.} \& \au{O'Doherty, T.}} \yr{2001}  \at{{Vortex breakdown: a review}}.  \jt{Prog. Energy Combust. Sci.}  \bvol{27}~(4),  \pg{431--481}.

\bibitem[Luchini \& Bottaro(2014)]{Luchini2014}
{\sc \au{Luchini, P.} \& \au{Bottaro, A.}} \yr{2014}  \at{{Adjoint Equations in Stability Analysis}}.  \jt{Annual Review of Fluid Mechanics}  \bvol{46}~(1),  \pg{493--517}.

\bibitem[L{\"{u}}ckoff {\em et~al.\/}(2022)L{\"{u}}ckoff, Naster, M{\"{u}}ller, Sieber, Litvinov \& Oberleithner]{Lueckoff2022}
{\sc \au{L{\"{u}}ckoff, F.}, \au{Naster, M.}, \au{M{\"{u}}ller, J.~S.}, \au{Sieber, M.}, \au{Litvinov, I.} \& \au{Oberleithner, K.}} \yr{2022}  \at{{Impact of runner crown shape modifications on the onset of the precessing vortex core}}.  \jt{IOP Conference Series: Earth and Environmental Science}  \bvol{1079}~(1),  \pg{12051}.

\bibitem[Marquet {\em et~al.\/}(2008)Marquet, Sipp \& Jacquin]{Marquet2008}
{\sc \au{Marquet, O.}, \au{Sipp, D.} \& \au{Jacquin, L.}} \yr{2008}  \at{{Sensitivity analysis and passive control of cylinder flow}}.  \jt{Journal of Fluid Mechanics}  \bvol{615},  \pg{221--252}.

\bibitem[Martinez-Cava {\em et~al.\/}(2020)Martinez-Cava, Ch{\'{a}}vez-Modena, Valero, {De Vicente} \& Ferrer]{Martinez-Cava2020}
{\sc \au{Martinez-Cava, A.}, \au{Ch{\'{a}}vez-Modena, M.}, \au{Valero, E.}, \au{{De Vicente}, J.} \& \au{Ferrer, E.}} \yr{2020}  \at{{Sensitivity gradients of surface geometry modifications based on stability analysis of compressible flows}}.  \jt{Physical Review Fluids}  \bvol{5}~(6),  \pg{63902}.

\bibitem[Meliga {\em et~al.\/}(2016)Meliga, Cadot \& Serre]{Meliga2016}
{\sc \au{Meliga, P.}, \au{Cadot, O.} \& \au{Serre, E.}} \yr{2016}  \at{{Experimental and Theoretical Sensitivity Analysis of Turbulent Flow Past a Square Cylinder}}.  \jt{Flow, Turbulence and Combustion}  \bvol{97}~(4),  \pg{987--1015}.

\bibitem[Meliga {\em et~al.\/}(2012)Meliga, Pujals \& Serre]{Meliga2012b}
{\sc \au{Meliga, P.}, \au{Pujals, G.} \& \au{Serre, {\'{E}}.}} \yr{2012}  \at{{Sensitivity of 2-D turbulent flow past a D-shaped cylinder using global stability}}.  \jt{Physics of Fluids}  \bvol{24}~(6),  \pg{061701}.

\bibitem[Menter(1993)]{Menter1993}
{\sc \au{Menter, F.~R.}} \yr{1993} {Zonal two equation k-omega turbulence models for aerodynamic flows}.  \bt{In {\em AIAA 24th Fluid Dynamics Conference\/}},  \pg{p. 2906}.

\bibitem[Mettot {\em et~al.\/}(2014{\natexlab{{\em a\/}}})Mettot, Renac \& Sipp]{Mettot2014}
{\sc \au{Mettot, C.}, \au{Renac, F.} \& \au{Sipp, D.}} \yr{2014{\natexlab{{\em a\/}}}}  \at{{Computation of eigenvalue sensitivity to base flow modifications in a discrete framework: Application to open-loop control}}.  \jt{Journal of Computational Physics}  \bvol{269},  \pg{234--258}.

\bibitem[Mettot {\em et~al.\/}(2014{\natexlab{{\em b\/}}})Mettot, Sipp \& B{\'{e}}zard]{Mettot2014a}
{\sc \au{Mettot, C.}, \au{Sipp, D.} \& \au{B{\'{e}}zard, H.}} \yr{2014{\natexlab{{\em b\/}}}}  \at{{Quasi-laminar stability and sensitivity analyses for turbulent flows: Prediction of low-frequency unsteadiness and passive control}}.  \jt{Physics of Fluids}  \bvol{26}~(4),  \pg{045112}.

\bibitem[Mitruţ {\em et~al.\/}(2022)Mitruţ, Bucur, Dunca \& Cervantes]{Mitrut2022}
{\sc \au{Mitruţ, R.}, \au{Bucur, D.~M.}, \au{Dunca, G.} \& \au{Cervantes, M.~J.}} \yr{2022}  \at{{Global linear stability analysis of the flow inside a conical draft tube}}.  \jt{IOP Conference Series: Earth and Environmental Science}  \bvol{1079}~(1),  \pg{012049}.

\bibitem[Mons {\em et~al.\/}(2024)Mons, Vervynck \& Marquet]{Mons2024}
{\sc \au{Mons, V.}, \au{Vervynck, A.} \& \au{Marquet, O.}} \yr{2024}  \at{{Data assimilation and linear analysis with turbulence modelling: application to airfoil stall flows with PIV measurements}}.  \jt{Theoretical and Computational Fluid Dynamics}  \bvol{38}~(3),  \pg{403--429}.

\bibitem[Morra {\em et~al.\/}(2019)Morra, Semeraro, Henningson \& Cossu]{Morra2019}
{\sc \au{Morra, P.}, \au{Semeraro, O.}, \au{Henningson, D.~S.} \& \au{Cossu, C.}} \yr{2019}  \at{{On the relevance of Reynolds stresses in resolvent analyses of turbulent wall-bounded flows}}.  \jt{Journal of Fluid Mechanics}  \bvol{867},  \pg{969--984}.

\bibitem[M{\"{u}}ller {\em et~al.\/}(2022)M{\"{u}}ller, L{\"{u}}ckoff, Kaiser \& Oberleithner]{Mueller2022b}
{\sc \au{M{\"{u}}ller, J.~S.}, \au{L{\"{u}}ckoff, F.}, \au{Kaiser, T.~L.} \& \au{Oberleithner, K.}} \yr{2022}  \at{{On the relevance of the runner crown for flow instabilities in a Francis turbine}}.  \jt{IOP Conference Series: Earth and Environmental Science}  \bvol{1079}~(1),  \pg{012053}.

\bibitem[M{\"{u}}ller {\em et~al.\/}(2020)M{\"{u}}ller, L{\"{u}}ckoff, Paredes, Theofilis \& Oberleithner]{Mueller2020}
{\sc \au{M{\"{u}}ller, J.~S.}, \au{L{\"{u}}ckoff, F.}, \au{Paredes, P.}, \au{Theofilis, V.} \& \au{Oberleithner, K.}} \yr{2020}  \at{{Receptivity of the turbulent precessing vortex core: Synchronization experiments and global adjoint linear stability analysis}}.  \jt{Journal of Fluid Mechanics}  \bvol{888},  \pg{A3}.

\bibitem[M{\"{u}}ller {\em et~al.\/}(2024{\natexlab{{\em a\/}}})M{\"{u}}ller, Reumsch{\"{u}}ssel, Kaiser, Knechtel \& Oberleithner]{Mueller2024}
{\sc \au{M{\"{u}}ller, J.~S.}, \au{Reumsch{\"{u}}ssel, J.~M.}, \au{Kaiser, T.~L.}, \au{Knechtel, S.~J.} \& \au{Oberleithner, K.}} \yr{2024{\natexlab{{\em a\/}}}} {Combining Bayesian optimization with adjoint-based gradients for efficient control of flow instabilities}.  \bt{In {\em Center for Turbulence Research Proceedings of the Summer Program\/}},  \pg{pp. 143--152}. Stanford University.

\bibitem[M{\"{u}}ller {\em et~al.\/}(2024{\natexlab{{\em b\/}}})M{\"{u}}ller, von Saldern, Kaiser \& Oberleithner]{Mueller2024b}
{\sc \au{M{\"{u}}ller, J.~S.}, \au{von Saldern, J.~G.}, \au{Kaiser, T.~L.} \& \au{Oberleithner, K.}} \yr{2024{\natexlab{{\em b\/}}}}  \at{{Linear amplification of inertial-wave-driven swirl fluctuations in turbulent swirling pipe flows: a resolvent analysis approach}}.  \jt{Journal of Fluid Mechanics}  \bvol{1000},  \pg{A91}.

\bibitem[Muntean {\em et~al.\/}(2014)Muntean, Susan-Resiga, Cǎmpian, Dumbrava \& Cuzmoş]{Muntean2014}
{\sc \au{Muntean, S.}, \au{Susan-Resiga, R.~F.}, \au{Cǎmpian, V.~C.}, \au{Dumbrava, C.} \& \au{Cuzmoş, A.}} \yr{2014}  \at{{In situ unsteady pressure measurements on the draft tube cone of the francis turbine with air injection over an extended operating range}}.  \jt{UPB Scientific Bulletin, Series D: Mechanical Engineering}  \bvol{76}~(3),  \pg{173--180}.

\bibitem[Nakazawa \& Azegami(2016)]{Nakazawa2016}
{\sc \au{Nakazawa, T.} \& \au{Azegami, H.}} \yr{2016}  \at{{Shape optimization of flow field improving hydrodynamic stability}}.  \jt{Japan Journal of Industrial and Applied Mathematics}  \bvol{33}~(1),  \pg{167--181}.

\bibitem[Noiray \& Schuermans(2013)]{Noiray2013}
{\sc \au{Noiray, N.} \& \au{Schuermans, B.}} \yr{2013}  \at{{Deterministic quantities characterizing noise driven Hopf bifurcations in gas turbine combustors}}.  \jt{International Journal of Non-Linear Mechanics}  \bvol{50},  \pg{152--163}.

\bibitem[Pasche {\em et~al.\/}(2017)Pasche, Avellan \& Gallaire]{Pasche2017}
{\sc \au{Pasche, S.}, \au{Avellan, F.} \& \au{Gallaire, F.}} \yr{2017}  \at{{Part Load Vortex Rope as a global unstable mode}}.  \jt{Journal of Fluids Engineering, Transactions of the ASME}  \bvol{139}~(5),  \pg{051102}.

\bibitem[Pasche {\em et~al.\/}(2019)Pasche, Avellan \& Gallaire]{Pasche2019}
{\sc \au{Pasche, S.}, \au{Avellan, F.} \& \au{Gallaire, F.}} \yr{2019}  \at{{Optimal Control of Part Load Vortex Rope in Francis Turbines}}.  \jt{Journal of Fluids Engineering, Transactions of the ASME}  \bvol{141}~(8),  \pg{1--12}.

\bibitem[Pickering {\em et~al.\/}(2021)Pickering, Rigas, Schmidt, Sipp \& Colonius]{Pickering2021}
{\sc \au{Pickering, E.}, \au{Rigas, G.}, \au{Schmidt, O.~T.}, \au{Sipp, D.} \& \au{Colonius, T.}} \yr{2021}  \at{{Optimal eddy viscosity for resolvent-based models of coherent structures in turbulent jets}}.  \jt{Journal of Fluid Mechanics}  \bvol{917},  \pg{A29}.

\bibitem[Pironneau(1974)]{Pironneau1974}
{\sc \au{Pironneau, O.}} \yr{1974}  \at{{On optimum design in fluid mechanics}}.  \jt{Journal of Fluid Mechanics}  \bvol{64}~(1),  \pg{97--110}.

\bibitem[Qadri {\em et~al.\/}(2013)Qadri, Mistry \& Juniper]{Qadri2013}
{\sc \au{Qadri, U.~A.}, \au{Mistry, D.} \& \au{Juniper, M.~P.}} \yr{2013}  \at{{Structural sensitivity of spiral vortex breakdown}}.  \jt{Journal of Fluid Mechanics}  \bvol{720},  \pg{558--581}.

\bibitem[Reau \& Tumin(2002)]{Reau2002}
{\sc \au{Reau, N.} \& \au{Tumin, A.}} \yr{2002}  \at{{Harmonic perturbations in turbulent wakes}}.  \jt{AIAA Journal}  \bvol{40}~(3),  \pg{526--530}.

\bibitem[Reynolds \& Hussain(1972)]{Reynolds1972}
{\sc \au{Reynolds, W.~C.} \& \au{Hussain, A. K. M.~F.}} \yr{1972}  \at{{The mechanics of an organized wave in turbulent shear flow. Part 3. Theoretical models and comparisons with experiments}}.  \jt{Journal of Fluid Mechanics}  \bvol{54}~(2),  \pg{263--288}.

\bibitem[Rukes {\em et~al.\/}(2016)Rukes, Paschereit \& Oberleithner]{Rukes2016}
{\sc \au{Rukes, L.}, \au{Paschereit, C.~O.} \& \au{Oberleithner, K.}} \yr{2016}  \at{{An assessment of turbulence models for linear hydrodynamic stability analysis of strongly swirling jets}}.  \jt{European Journal of Mechanics, B/Fluids}  \bvol{59},  \pg{205--218}.

\bibitem[Sarras {\em et~al.\/}(2024)Sarras, Tayeh, Mons \& Marquet]{Sarras2024}
{\sc \au{Sarras, K.}, \au{Tayeh, C.}, \au{Mons, V.} \& \au{Marquet, O.}} \yr{2024}  \at{{Linear stability analysis of turbulent mean flows based on a data-consistent Reynolds-averaged Navier–Stokes model: prediction of three-dimensional stall cells around an airfoil}}.  \jt{Journal of Fluid Mechanics}  \bvol{1001},  \pg{A41}.

\bibitem[Seifi {\em et~al.\/}(2023)Seifi, Raisee \& Cervantes]{Seifi2023}
{\sc \au{Seifi, Z.}, \au{Raisee, M.} \& \au{Cervantes, M.~J.}} \yr{2023}  \at{{Global linear stability analysis of flow inside an axial swirl generator with a rotating vortex rope}}.  \jt{Journal of Hydraulic Research}  \bvol{61}~(1),  \pg{34--50}.

\bibitem[Shiraghaee {\em et~al.\/}(2024)Shiraghaee, Sundstrom, Raisee \& Cervantes]{Shiraghaee2024}
{\sc \au{Shiraghaee, S.}, \au{Sundstrom, J.}, \au{Raisee, M.} \& \au{Cervantes, M.~J.}} \yr{2024}  \at{{Experimental Investigation of Part Load Vortex Rope Mitigation With Rod Protrusion in an Axial Turbine}}.  \jt{Journal of Fluids Engineering, Transactions of the ASME}  \bvol{146}~(8),  \pg{081205}.

\bibitem[Sieber {\em et~al.\/}(2021)Sieber, Paschereit \& Oberleithner]{Sieber2021}
{\sc \au{Sieber, M.}, \au{Paschereit, C.~O.} \& \au{Oberleithner, K.}} \yr{2021}  \at{{Stochastic modelling of a noise-driven global instability in a turbulent swirling jet}}.  \jt{Journal of Fluid Mechanics}  \bvol{916},  \pg{A7}.

\bibitem[Sipp \& Lebedev(2007)]{Sipp2007}
{\sc \au{Sipp, D.} \& \au{Lebedev, A.}} \yr{2007}  \at{{Global stability of base and mean flows: A general approach and its applications to cylinder and open cavity flows}}.  \jt{Journal of Fluid Mechanics}  \bvol{593},  \pg{333--358}.

\bibitem[Sipp {\em et~al.\/}(2010)Sipp, Marquet, Meliga \& Barbagallo]{Sipp2010}
{\sc \au{Sipp, D.}, \au{Marquet, O.}, \au{Meliga, P.} \& \au{Barbagallo, A.}} \yr{2010}  \at{{Dynamics and Control of Global Instabilities in Open-Flows: A Linearized Approach}}.  \jt{Applied Mechanics Reviews}  \bvol{63}~(3),  \pg{030801}.

\bibitem[Skripkin {\em et~al.\/}(2022)Skripkin, Suslov, Litvinov, Gorelikov, Tsoy \& Shtork]{Skripkin2022}
{\sc \au{Skripkin, S.~G.}, \au{Suslov, D.~A.}, \au{Litvinov, I.~V.}, \au{Gorelikov, E.~U.}, \au{Tsoy, M.~A.} \& \au{Shtork, S.~I.}} \yr{2022}  \at{{Comparative analysis of air and water flows in simplified hydraulic turbine models}}.  \jt{Journal of Physics: Conference Series}  \bvol{2150}~(1),  \pg{12001}.

\bibitem[Susan-Resiga {\em et~al.\/}(2007)Susan-Resiga, Muntean, Hasmatuchi, Ruprecht \& Sandor]{Susan-Resiga2007}
{\sc \au{Susan-Resiga, R.}, \au{Muntean, S.}, \au{Hasmatuchi, V.}, \au{Ruprecht, A.} \& \au{Sandor, B.}} \yr{2007} {Development of a swirling flow control technique for Francis turbines operated at partial discharge}.  \bt{In {\em 3rd Workshop on Vortex Dominated Flows\/}},  \pg{pp. 1--12}. Timisoara, Romania.

\bibitem[Susan-Resiga {\em et~al.\/}(2021)Susan-Resiga, Muntean, Bosioc \& Stuparu]{Susan-Resiga2021}
{\sc \au{Susan-Resiga, R.~F.}, \au{Muntean, S.}, \au{Bosioc, A.} \& \au{Stuparu, A.}} \yr{2021}  \at{{Stabilisation of the swirl exiting a Francis runner far from the best efficiency point}}.  \jt{IOP Conference Series: Earth and Environmental Science}  \bvol{774}~(1),  \pg{012113}.

\bibitem[Sweby(1984)]{Sweby1984}
{\sc \au{Sweby, P.~K.}} \yr{1984}  \at{{High resolution schemes using flux limiters for hyperbolic conservation laws}}.  \jt{SIAM Journal on Numerical Analysis}  \bvol{21}~(5),  \pg{995--1011}.

\bibitem[Symon {\em et~al.\/}(2023)Symon, Madhusudanan, Illingworth \& Marusic]{Symon2023}
{\sc \au{Symon, S.}, \au{Madhusudanan, A.}, \au{Illingworth, S.~J.} \& \au{Marusic, I.}} \yr{2023}  \at{{Use of eddy viscosity in resolvent analysis of turbulent channel flow}}.  \jt{Physical Review Fluids}  \bvol{8}~(6),  \pg{064601}.

\bibitem[Taira {\em et~al.\/}(2020)Taira, Hemati, Brunton, Sun, Duraisamy, Bagheri, Dawson \& Yeh]{Taira2020}
{\sc \au{Taira, K.}, \au{Hemati, M.~S.}, \au{Brunton, S.~L.}, \au{Sun, Y.}, \au{Duraisamy, K.}, \au{Bagheri, S.}, \au{Dawson, S. T.~M.} \& \au{Yeh, C.~A.}} \yr{2020}  \at{{Modal analysis of fluid flows: Applications and outlook}}.  \jt{AIAA Journal}  \bvol{58}~(3),  \pg{998--1022}.

\bibitem[Tammisola(2017)]{Tammisola2017}
{\sc \au{Tammisola, O.}} \yr{2017}  \at{{Optimal wavy surface to suppress vortex shedding using second-order sensitivity to shape changes}}.  \jt{European Journal of Mechanics, B/Fluids}  \bvol{62},  \pg{139--148}.

\bibitem[Tammisola \& Juniper(2016)]{Tammisola2016}
{\sc \au{Tammisola, O.} \& \au{Juniper, M.~P.}} \yr{2016}  \at{{Coherent structures in a swirl injector at Re = 4800 by nonlinear simulations and linear global modes}}.  \jt{Journal of Fluid Mechanics}  \bvol{792},  \pg{620--657}.

\bibitem[Tanasa {\em et~al.\/}(2019)Tanasa, Bosioc, Muntean \& Susan-Resiga]{Tanasa2019}
{\sc \au{Tanasa, C.}, \au{Bosioc, A.}, \au{Muntean, S.} \& \au{Susan-Resiga, R.}} \yr{2019}  \at{{A novel passive method to control the Swirling Flow with Vortex Rope from the conical diffuser of hydraulic turbines with fixed blades}}.  \jt{Applied Sciences}  \bvol{9}~(22),  \pg{4910}.

\bibitem[Trivedi {\em et~al.\/}(2016)Trivedi, Cervantes \& Dahlhaug]{Trivedi2016}
{\sc \au{Trivedi, C.}, \au{Cervantes, M.~J.} \& \au{Dahlhaug, O.~G.}} \yr{2016}  \at{{Experimental and numerical studies of a high-head Francis turbine: A review of the Francis-99 test case}}.  \jt{Energies}  \bvol{9}~(2),  \pg{74}.

\bibitem[Urban {\em et~al.\/}(2022)Urban, Kurkov{\'{a}} \& Pochyl{\'{y}}]{Urban2022}
{\sc \au{Urban, O.}, \au{Kurkov{\'{a}}, M.} \& \au{Pochyl{\'{y}}, F.}} \yr{2022}  \at{{Mitigation of swirling flow with a vortex rope by passive installations - theory, simulations, and experiments}}.  \jt{Physics of Fluids}  \bvol{34}~(12),  \pg{124111}.

\bibitem[Valent{\'{i}}n {\em et~al.\/}(2017)Valent{\'{i}}n, Presas, Egusquiza, Valero, Egusquiza \& Bossio]{Valentin2017}
{\sc \au{Valent{\'{i}}n, D.}, \au{Presas, A.}, \au{Egusquiza, E.}, \au{Valero, C.}, \au{Egusquiza, M.} \& \au{Bossio, M.}} \yr{2017}  \at{{Power Swing Generated in Francis Turbines by Part Load and Overload Instabilities}}.  \jt{Energies}  \bvol{10}~(12),  \pg{2124}.

\bibitem[{Van Doormaal} \& Raithby(1984)]{VanDoormaal1984}
{\sc \au{{Van Doormaal}, J.~P.} \& \au{Raithby, G.~D.}} \yr{1984}  \at{{Enhancements of the SIMPLE method for predicting incompressible fluid flows}}.  \jt{Numerical Heat Transfer}  \bvol{7}~(2),  \pg{147--163}.

\bibitem[Viola {\em et~al.\/}(2014)Viola, Iungo, Camarri, Port{\'{e}}-Agel \& Gallaire]{Viola2014}
{\sc \au{Viola, F.}, \au{Iungo, G.~V.}, \au{Camarri, S.}, \au{Port{\'{e}}-Agel, F.} \& \au{Gallaire, F.}} \yr{2014}  \at{{Prediction of the hub vortex instability in a wind turbine wake: Stability analysis with eddy-viscosity models calibrated on wind tunnel data}}.  \jt{Journal of Fluid Mechanics}  \bvol{750},  \pg{R1}.

\bibitem[{Von Saldern} {\em et~al.\/}(2024){Von Saldern}, Schmidt, Jordan \& Oberleithner]{VonSaldern2024}
{\sc \au{{Von Saldern}, J.~G.}, \au{Schmidt, O.~T.}, \au{Jordan, P.} \& \au{Oberleithner, K.}} \yr{2024}  \at{{On the role of eddy viscosity in resolvent analysis of turbulent jets}}.  \jt{Journal of Fluid Mechanics}  \bvol{1000},  \pg{A51}.

\bibitem[Wang {\em et~al.\/}(2019)Wang, Ferrer, Mart{\'{i}}nez-Cava, Zheng \& Valero]{Wang2019}
{\sc \au{Wang, Y.}, \au{Ferrer, E.}, \au{Mart{\'{i}}nez-Cava, A.}, \au{Zheng, Y.} \& \au{Valero, E.}} \yr{2019}  \at{{Enhanced stability of flows through contraction channels: Combining shape optimization and linear stability analysis}}.  \jt{Physics of Fluids}  \bvol{31}~(7),  \pg{074109}.

\bibitem[Zhou {\em et~al.\/}(2024)Zhou, Hu, Huang, Wu, Tang \& Cervantes]{Zhou2024}
{\sc \au{Zhou, X.}, \au{Hu, X.}, \au{Huang, Q.}, \au{Wu, H.}, \au{Tang, X.} \& \au{Cervantes, M.~J.}} \yr{2024}  \at{{Optimization design of an innovative francis draft tube: Insight into improving operational flexibility}}.  \jt{Energy}  \bvol{299},  \pg{131489}.

\end{thebibliography}

\end{document}